\documentclass[12pt,titlepage]{utarticle}
\usepackage{latexsym}
\usepackage{graphicx}
\usepackage{mathtools}
\usepackage{hyperref}

\numberwithin{table}{section}
\newcommand{\Pf}{\mathrm{Pf}}

\begin{document}

\preprint{
UTTG-01-13 \\
TCC-001-13 \\
}
\title{Three Dimensional Mirror Symmetry and Partition Function on $S^3$}

\author{Anindya Dey and Jacques Distler
     \oneaddress{
      Theory Group and\\
      Texas Cosmology Center\\
      Department of Physics,\\
      University of Texas at Austin,\\
      Austin, TX 78712, USA \\
      {~}\\
      \email{anindya@physics.utexas.edu}\\
      \email{distler@golem.ph.utexas.edu}\\
      }
}

\date{January 5, 2013}

\Abstract{
We provide non-trivial checks of $\mathcal{N}=4,\; D=3$ mirror symmetry in a large class of quiver gauge theories whose Type IIB (Hanany-Witten) descriptions involve D3 branes ending on orbifold/orientifold 5-planes at the boundary. From the M-theory perspective, such theories can be understood in terms of coincident M2 branes sitting at the origin of a product of an A-type and a D-type ALE (Asymtotically Locally Euclidean) space with G-fluxes. Families of mirror dual pairs, which  arise in this fashion, can be labeled as $(A_{m-1},D_n)$, where $m$ and $n$ are integers. For a large subset of such infinite families of dual theories, corresponding to generic values of $n\geq 4$, arbitrary ranks of the gauge groups and varying $m$, we test the conjectured duality by proving the precise equality of the $S^3$ partition functions for dual gauge theories in the IR as functions of masses and FI parameters. The mirror map for a given pair of mirror dual theories can be read off at the end of this computation and we explicitly present these for the aforementioned examples. The computation uses non-trivial identities of hyperbolic functions including certain generalizations of Cauchy determinant identity and Schur's Pfaffian identity, which are discussed in the paper.
}

\maketitle

\thispagestyle{empty}
\tableofcontents
\vfill
\newpage
\setcounter{page}{1}

\section{Introduction and Main Results}\label{introduction}
Localization techniques in supersymmetric/superconformal field theories in $D\leq 6$ dimensions have emerged as an extremely efficient toolbox for non-perturbative/exact computation of various supersymmetric observables. For a large class of supersymmetric gauge theories, localization on a compact manifold reduces supersymmetric observables (partition function, Wilson loops, superconformal indices \emph{etc.}) to fairly simple finite dimensional matrix integrals.

Following the initial examples involving $\mathcal{N}=2$ theories on four and three dimensional spheres \cite{Pestun:2007rz, Kapustin:2009kz}, a systematic framework for the computation was presented in \cite{Festuccia:2011ws}. This progress has led to a unique opportunity to directly test various conjectured dualities for $D\leq6$-dimensional field theories by computing and matching appropriate field theory observables on both sides of the duality. For example, a large class of dualities for $\mathcal{N}\geq 2$ theories in three dimensions, including various cases of mirror symmetry and Seiberg duality, has been studied in recent times in this fashion \cite{Kapustin:2010mh,Kapustin:2010xq}.

In this note, we are concerned with $\mathcal{N}=4$ mirror symmetry \cite{deBoer:1996mp, Intriligator:1996ex} in three dimensions for a certain class of ``elliptic" quivers, which we will specify presently. Mirror symmetry for a large class of quiver gauge theories can be understood in terms of coincident M2 branes sitting at the singularity of a product of two ALE (Asymptotically Locally Euclidean) spaces with certain G-fluxes turned on \cite{Porrati:1996xi,Dey:2011pt}. From the ADE classification of ALE spaces, families of mirror dual theories can be labeled as $(\Gamma_1,\Gamma_2)$ (see the discussion in \S\ref{M-theory}) where $\Gamma_{1,2}$ are finite subgroups of $SU(2)$ associated to the orbifold limits of the respective ALE spaces. A complete classification of the infinite families of dual theories arising from A and D type ALE spaces was obtained in \cite{Dey:2011pt}. A particularly interesting feature of these supersymmetric gauge theories is that their Higgs branches  can be identified with the moduli spaces of instantons on ALE spaces of A and D type obtained via the ADHM construction. The theories corresponding to $(\Gamma_1=\mathbb{Z}_m, \Gamma_2=\mathbb{Z}_n)\, (\forall m,n)$ in the aforementioned M-theory picture have been analyzed in detail in \cite{deBoer:1996mp,Intriligator:1996ex} as prototypes of three dimensional mirror symmetry. Discussions of the Type IIB realization of theories corresponding to $(\Gamma_1=\mathbb{Z}_n, \Gamma_2=D_m) (m\geq 4)$ and $(\Gamma_1=D_m, \Gamma_2=D_n) (n\geq 4)$ can be found in \cite{Gaiotto:2008ak,Hanany:1996ie}.

\begin{figure}[htbp]
\begin{center}
\includegraphics[height=3.0in]{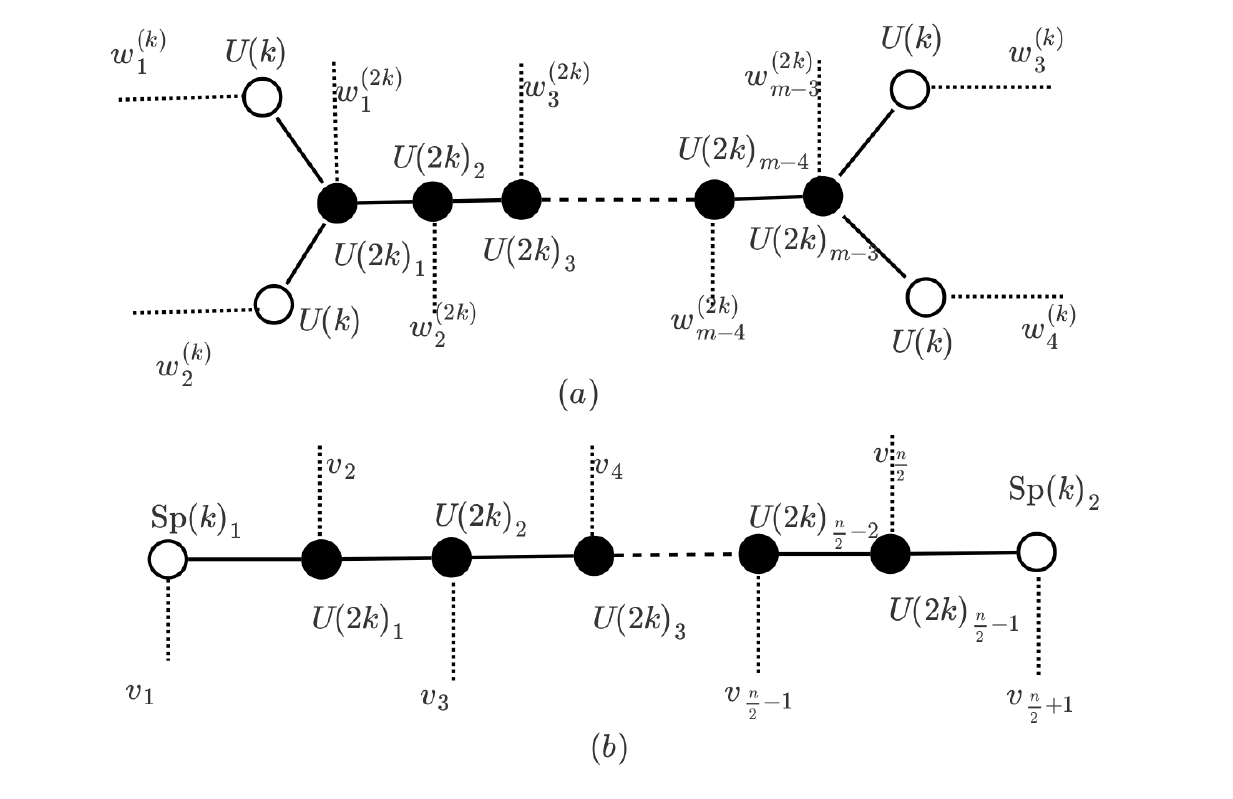}
\caption{Mirror pairs for $\Gamma_1=\mathbb{Z}_n$,$\Gamma_2={D}_{m}$, with even $n\; (\geq 4)$ and generic $m\;(\geq 4)$. Fundamental hypers are simply denoted as dotted lines connected to the appropriate black/white node.}
\label{fig4l}
\end{center}
\end{figure}

\begin{figure}[htbp]
\begin{center}
\includegraphics[height=3.0in]{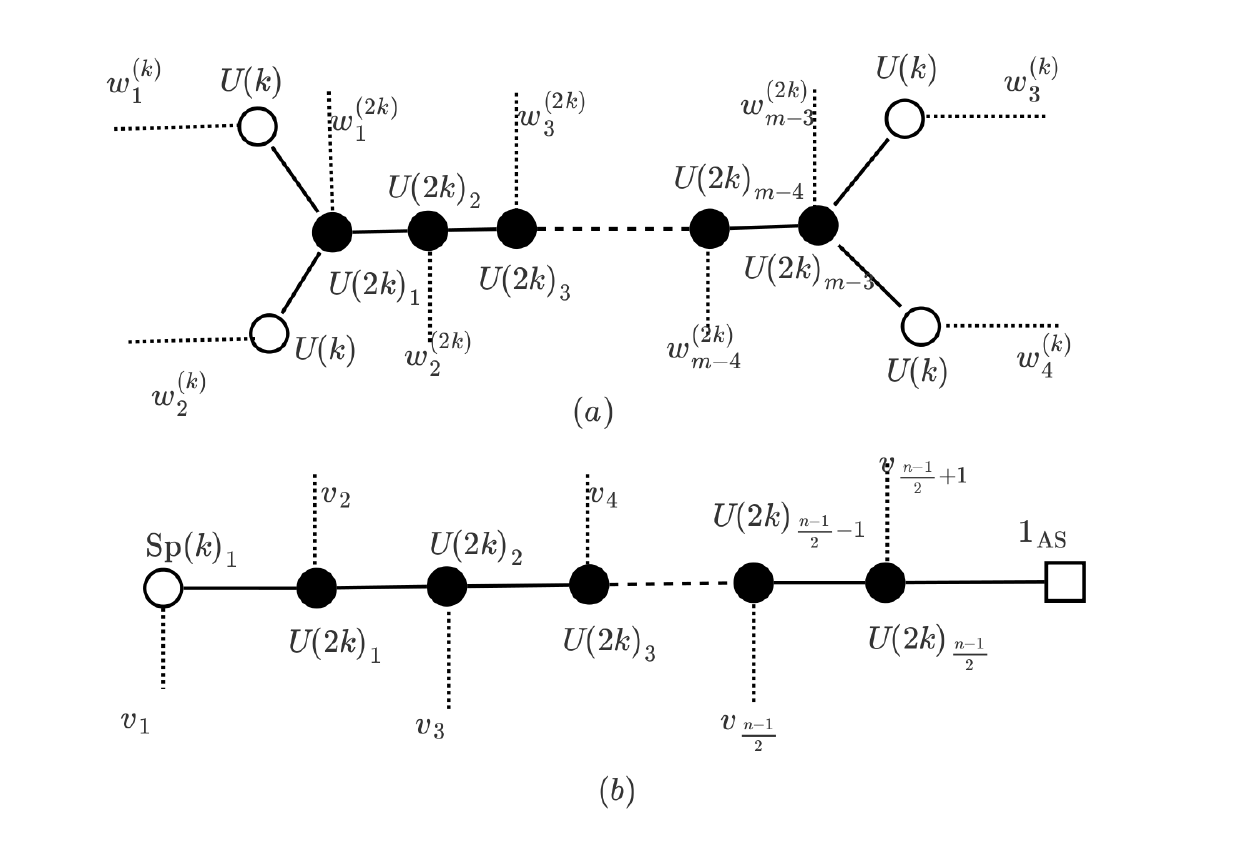}
\caption{Mirror Duals for $\Gamma_1={Z}_n$,$\Gamma_2={D}_{m}$, with odd $n\;(\geq 3)$ and generic $m\;(\geq 4)$}
\label{fig5l}
\end{center}
\end{figure}

Our focus, in this paper, will be the category of dual theories labeled as $(\Gamma_1=\mathbb{Z}_n, \Gamma_2=D_m) (m\geq 4)$ --- generic examples of which are shown in figures \ref{fig4l} and \ref{fig5l}, for even and odd $m(>2)$ respectively. In addition to NS5 and D5 branes, Type IIB realizations of these theories involve D3 branes ending on orbifold/orientifold 5-planes \cite{Gaiotto:2008ak,Hanany:1996ie}. 

The central theme of this paper is to perform some extremely non-trivial checks of mirror symmetry for the aforementioned class of theories. The strategy, outlined first in  \cite{Kapustin:2010xq}, is to compute the $S^3$ partition functions for a given dual pair as functions of the independent mass and FI parameters of the respective theories and prove that they are exactly equal to each other, provided
parameters on side of the duality are related to parameters on the other side by a certain linear map --- known as the ``mirror map". The precise linear map which relate masses and FI parameters of a given theory with the FI parameters and masses respectively in the dual theory is read off at the end of the computation. The main result of this work is the derivation of these mirror maps for the various examples of $(\Gamma_1=\mathbb{Z}_m, \Gamma_2=D_n) (n\geq 4)$-type dual theories for generic values of $n\geq 4$ and arbitrary rank of the gauge group, as we vary the order of the A-type singularity.

As mentioned above, $S^3$ partition functions were first used to study mirror symmetry in \cite{Kapustin:2010xq}  for quiver gauge theories corresponding to $(\Gamma_1=\mathbb{Z}_m, \Gamma_2=\mathbb{Z}_n)$. The two key ingredients required for appropriately manipulating the partition functions in this case were the Fourier transform of the hyperbolic secant and Cauchy's determinant identity:
\begin{equation}
\int \frac{e^{2\pi i x z}}{\cosh{\pi z}} dx = \frac{1}{\cosh{\pi x}}
\end{equation}

\begin{equation}
\sum _{\rho} (-1)^{\rho} \frac{1}{\prod_{i}\cosh{(x_i -y_{\rho(i)})}}=\frac{\prod_{i <j} \sinh{(x_i-x_j)} \sinh{(y_i -y_j)}}{\prod_{i,j}\cosh{(x_i -y_j)}} \label{Cauchycosh}
\end{equation}
\noindent These two identities allowed one to write the matrix integral as a product of NS5 and D5 contributions, which in turn made the action of IIB S-duality (and therefore of 3D mirror symmetry) on the integrand obvious.

For theories whose Type IIB descriptions involve boundary orbifold/orientifold 5-planes, one needs a further set of important identities to be able to prove the equality of partition functions of dual theories. One of the essential ingredients is the Fourier transform of hyperbolic cosecant:
\begin{equation}
\int \frac{e^{2\pi i x z}}{\sinh{\pi z}} dx = i\tanh{\pi x}
\end{equation}
Another important identity is a slight variant of Cauchy's determinant identity, which we may call Cauchy's hyperbolic sine identity:
\begin{equation}
\sum _{\rho} (-1)^{\rho} \frac{1}{\prod_{i}\sinh{(x_i +y_{\rho(i)})}}=\frac{\prod_{i <j} \sinh{(x_i-x_j)} \sinh{(y_i -y_j)}}{\prod_{i,j}\sinh{(x_i +y_j)}} 
\end{equation}
Finally, the most important identity for analyzing this class of theories is obtained (see Appendix \ref{SchurPfaffian}) from a certain generalization of Schur's Pfaffian identity \cite{Okada:2004Pf}:
\begin{equation}
\begin{split}
&\sum _{\rho} (-1)^{\rho} \frac{1}{\prod_{i} \cosh{(x_{\rho(i)}-m)}\cosh{(x_{\rho(k+i)}-m^{'})}\cosh{(x_{\rho(i)}+x_{\rho(k+i)} -M)}}  \\
=&\left(\frac{\kappa \;k! \; \sinh^k{(m-m^{'})}}{\prod_i \cosh{(x_i-m)} \cosh{(x_i-m^{'})} \cosh{(x_{k+i}-m)}\cosh{(x_{k+i}-m^{'})}}\right) \times \Pf\left[\frac{\sinh{(x_p - x_l)}}{\cosh{(x_p + x_l-M)}} \right] \\ 
=&\left(\frac{\kappa \; k!\; \sinh^k{(m-m^{'})}}{\prod_i \cosh{(x_i-m)} \cosh{(x_i-m^{'})} \cosh{(x_{k+i}-m)}\cosh{(x_{k+i}-m^{'})}}\right) \times \left(\prod_{p<l} \frac{\sinh{(x_p - x_l)}}{\cosh{(x_p + x_l-M)}}\right) 
\end{split}
\end{equation}
where $\rho$ denotes the set of all permutations of the integers $\{1,2,....,k,k+1,...,2k-1,2k\}$ with $p,l =1,2,..,2k$ and $i,j = 1,2,...,k$. The constant $\kappa$ depends on $k$ : $\kappa=1$ for even $k=4m$ and odd $k=4m+1$ ($m=1,2,3,...$) and $\kappa=-1$ otherwise.

The paper is organized as follows: \S\ref{M-theory} briefly describes the M-theory interpretation of $\mathcal{N}=4$ mirror symmetry in three dimensional quiver gauge theories. \S\ref{BuildingBlocks} summarizes the rules for constructing the matrix integral corresponding to the partition function of any $\mathcal{N}=4$
theory for a given gauge group and a given matter content. \S4--7 deal with the computation of partition functions for dual theories and deriving the corresponding mirror maps. \S\ref{D-trivial} discusses the family of dual theories labeled by $\Gamma_1=\mathbb{Z}_1,\Gamma_2=D_n$ . \S\ref{D-A} describes the different cases corresponding to $\Gamma_1=\mathbb{Z}_2,\Gamma_2=D_n$. There are four infinite families of mirror dual theories in this category which are labeled by the manifest flavor symmetry groups of the A and B model quiver diagrams - $G^A_{flavor}$ and $G^B_{flavor}$.
Finally, \S\ref{D-AgenEven} and \S\ref{D-AgenOdd} present examples of theories in the category $\Gamma_1=\mathbb{Z}_m,\Gamma_2=D_n$, with generic $m >2$, where $m$ is even and odd respectively.

The Appendix summarizes (and occasionally proves) the important identities required at various stages of the computation.

\section{M-theory Description of Mirror Symmetry}\label{M-theory}
In this section, we discuss briefly the M-theory description of mirror symmetry in $\mathcal{N}=4$ theories in three dimensions. Consider a solution of $11D$ supergravity with the following geometry, $\mathcal{M}=\mathbb{R}^{2,1}\times \text{ALE}_1 \times \text{ALE}_2$, where $\text{ALE}$ is an ``Asymptotically Locally Euclidean'' space  --- a smooth hyper-Kahler resolution of an orbifold singularity of the form $\mathbb{C}^2/\Gamma$ ($\Gamma$ being a finite subgroup of $SU(2)$). The \text{ALE} spaces have a well-known A-D-E classification, which is related to the A-D-E classification of the discrete subgroups of $SU(2)$ appearing in the orbifold limit of these spaces. In what follows, we will restrict ourselves to  $A$ and $D$ type \text{ALE} spaces, which can be further deformed to ``Asymptotically Locally Flat"  or \text{ALF} spaces. The \text{ALF} spaces are four-dimensional hyper-Kahler manifolds, locally asymptotic to $\mathbb{R}^3 \times S^1$ at infinity.

For example, let us review  a simpler version of the aforementioned  $11D$ background, namely, $\mathcal{M}=\mathbb{R}^{2,1}\times \mathbb{C}^2 \times \text{ALE}$, where the $\text{ALE}$ space has an $A_{n-1}$ singularity, i.e. this ALE space is the hyper-Kahler resolution of the orbifold singularity $\mathbb{C}^2/\mathbb{Z}_n$ .The $\text{ALE}$ metric can be  explicitly written as follows:

\begin{equation}
ds_{\text{ALE}}^2=H d\vec{r}^2 + H^{-1}(dx_{11}+\vec{C}.d\vec{r})^2
\end{equation}
where $H=\frac{1}{2}\sum_{i=1}^n \frac{1}{|\vec{r}-\vec{r_i}|}$ and $\nabla \times \vec{C}=-\nabla H$.

On compactifying the coordinate $x_{11}$ as $x_{11} \cong x_{11}+ 2 \pi g_{s} \sqrt{\alpha^{'}}$ and deforming $H \to H^{'}$ such that

\begin{equation}
H^{'}=\frac{1}{g_{\text{YM}}^2 \alpha^{'}}+ \frac{1}{2}\sum_{i=1}^n\frac{1}{|\vec{r}-\vec{r_i}|}   \label{deform}
\end{equation}
with $g_{\text{YM}}^2=g_s \frac{1}{\sqrt{\alpha'}}$, one obtains a muti-centered Taub-Nut space, which is locally asymptotic to $\mathbb{R}^3\times S^1$ at infinity (and hence ``asymptotically locally flat''). At a generic point, the manifold is circle bundle over $\mathbb{R}^3$ and at the centers of the Taub-Nut $\vec{r}=\vec{r}_i$ the circle fiber shrinks to zero radius. Note that the deformed background approaches the original background in the limit $g_{\text{YM}}\to \infty$.

This deformed M-theory background can be understood as the M-theory lift of a Type IIA supergravity background of $n$ $D6$ branes at transverse positions $\vec{r_i}$, wrapping the $\mathbb{C}^2$. In addition, if we have $k$ $M2$ branes wrapping the $\mathbb{R}^{2,1}$ in the M-theory picture, then these become $k$ $D2$ probe branes in the IIA picture. This Type IIA background preserves 8 real supercharges and as a result the world-volume gauge theory on the D2 branes has $\mathcal{N}=4$ supersymmetry. The gauge couplings are all proportional to $g_{\text{YM}}$ as defined above.

One can similarly deform a $D$-type ALE  to an ALF  --- the corresponding space turns out to be a generalization of the Atiyah-Hitchin space \cite{Sen:1997kz}, which appears as the resolution of the  orbifold singularity $\mathbb{C}^2/D_n$. The resulting Type IIA background consists of $n$ $D6$ branes at transverse positions $\vec{r_i}$ parallel to an $O6$ plane which wraps the $\mathbb{C}^2$.

Now, consider the manifold we are interested in, namely, $\mathcal{M}: \mathbb{R}^{2,1}\times \text{ALE}_1 \times \text{ALE}_2$, for the special case in which the $\text{ALE}$ spaces are associated with $A_{n-1}$ and $A_{m-1}$ singularities respectively. Deforming $\text{ALE}_1 \to \text{ALF}_1$, the resulting M-theory background $\mathcal{M}: \mathbb{R}^{2,1}\times \text{ALF}_1 \times \text{ALE}_2$ can easily be seen to correspond to a Type IIA background with $n$ D6 branes wrapping $\text{ALE}_2$. The $k$ M2 branes become $k$ D2 branes in the Type IIA picture. The three-dimensional world-volume gauge theory given by the spectrum of D2-D2 and D2-D6 open strings is an $A_{m-1}$ quiver gauge theory with $n$ fundamental hypers associated with one of the $m$ $U(k)$ gauge groups. The gauge couplings are proportional to $g_{\text{YM}}$ which, in turn, is proportional to the radius of the circle fiber for $\text{ALF}_1$ at infinity (equation \eqref{deform}).

One can alternatively consider the deformed M-theory background $\mathcal{M}^{'}: \mathbb{R}^{2,1}\times \text{ALE}_1 \times \text{ALF}_2$, which leads to a Type IIA background with $m$ D6 branes wrapping $\text{ALE}_1$. The three-dimensional world-volume gauge theory on the D2 branes, in this case, is an $A_{n-1}$ quiver gauge theory with $m$ fundamental hypers associated with one of the $n$ $U(k)$ gauge groups. The gauge couplings are proportional to $g^{'}_{YM}$ which is proportional to the radius of the circle fiber for $\text{ALF}_2$ at infinity.

In the IR limit, when $g_{\text{YM}},g'_{\text{YM}} \to \infty$, the two quiver gauge theories, in question, are described by the same M-theory background $\mathcal{M}: \mathbb{R}^{2,1}\times \text{ALE}_1 \times \text{ALE}_2$ and therefore have equivalent IR dynamics. The exchange of hypermultiplet masses and FI parameters can also be clearly seen. In the A-model (obtained by deforming $\text{ALE}_1$ to $\text{ALF}_1$), the parameters $\vec{r}^{(n)}_i$ appear as masses of the fundamental hypers while the parameters $\vec{r}^{(m)}_j$ are the FI parameters. In the B-model, their roles are reversed.
The M-theory background also has G-fluxes, which determine the flavor symmetries of gauge theories on both sides of the duality. G-fluxes in backgrounds with only A-type ALE spaces have been discussed in \cite{Dey:2011pt}. The related Type IIA/IIB picture has been discussed in detail in \cite{Witten:2009xu}.

The M-theory interpretation of mirror symmetry provides a neat way to catalogue a large class of mirror dual quiver gauge theories in terms of singularities appearing in the orbifold limit of the ALE spaces. A generic M-theory background $\mathcal{M}: \mathbb{R}^{2,1}\times \text{ALE}_1 \times \text{ALE}_2$ reduces to the orbifold $\mathbb{R}^{2,1}\times \mathbb{C}^2/\Gamma_1 \times \mathbb{C}^2/\Gamma_2$ in this limit, with the G-fluxes collapsing at the $\Gamma_1 \times \Gamma_2$ singularity . Therefore, one can classify mirror duals in terms of the different choices for $\Gamma_1$ and $\Gamma_2$, restricting to A and D type of singularities only. A complete catalogue of theories that can be generated in this fashion, was presented in 
\cite{Dey:2011pt}. For the rest of this paper, we will focus on theories which arise from the aforementioned M-theory background with $\Gamma_1=D_n$ ($n \geq 4$) and $\Gamma_2=\mathbb{Z}_m$.

\section{$S^3$ Partition Function for a YM-Matter Theory: Building Blocks}\label{BuildingBlocks}
In \cite{Kapustin:2010xq,Kapustin:2009kz}, the rules for computing supersymmetric observables like the partition function for $\mathcal{N} \geq 2$ Chern Simmons-Yang Mills -Matter theories on $S^3$ were derived using localization on the Coulomb branch. As explained in \cite{Kapustin:2009kz}, the zero locus of the localizing functional requires all bosonic fields in the matter hypermultiplets to vanish. For the vector multiplets, the only non-vanishing fields are the adjoint scalar $\sigma$ and the auxiliary field $D$ of the $\mathcal{N}=2$ vector multiplet (which is part of the $\mathcal{N}=4$ vector multiplet), such that $\sigma = -D = \sigma_0$, where $\sigma_0$ is a constant on $S^3$. One can gauge-fix $\sigma_0$ to an element of the Cartan sub-algebra of the gauge group, introducing a Vandermonde determinant for each factor in the gauge group into the integration measure of the path-integral.

The localized path integral, therefore, can be written as a finite dimensional matrix integral over a single matrix variable $\sigma_0$, while the integrand consists of contributions from the classical action as well as the fluctuation determinant around the zero locus, i.e.
\begin{equation}
Z=\int  d\sigma_0 e^{S_{\text{cl}}[\sigma_0]} Z_{\text{1-loop}}[\sigma_0]
\end{equation}

For an $\mathcal{N}=4$ theory with no CS terms, the classical and 1-loop contributions for the gauge and matter fields can be summarized as follows \cite{Kapustin:2010xq}:

\paragraph*{\bf Classical Action:} For the $\mathcal{N}=4$ theories with no CS terms, $S_{\text{cl}}[\sigma_0]$ contributes a term linear in $\sigma_0$:
\begin{itemize}%
\item $S^{\text{FI}}_{\text{cl}}= 2\pi i \eta Tr(\sigma_0)$
\end{itemize}
for every $U(1)$ factor in the gauge group, with $\eta$ being the corresponding FI parameter.

\paragraph*{\bf 1-loop Determinant:}

\begin{itemize}%
\item Each $\mathcal{N}=4$ vector multiplet contributes

\begin{equation}
Z^v_{\text{1-loop}}=\prod_{\alpha} \frac{\sinh{\pi \alpha(\sigma_0)}}{(\pi \alpha(\sigma_0))^2}
\end{equation}
where the product extends over all the roots of the Lie algebra of G. The Vandermonte factor in the measure exactly cancels with the denominator of the 1-loop contribution of the vector multiplet for each factor in the gauge group.

\item Each $\mathcal{N}=4$ hypermultiplet contributes

\begin{equation}
Z^h_{\text{1-loop}}=\prod_{\rho} \frac{1}{\cosh{\pi \rho(\sigma_0+m)}}
\end{equation}
where the product extends over all the weights of the representation R of the gauge group G and $m$ is a real mass parameter.
\end{itemize}
Finally, the integrand obtained from the above set of rules has to be divided by the order of the Weyl group to account for the residual gauge symmetry.

The above rules completely specify the partition function on $S^3$ for any $\mathcal{N}=4$ gauge theory in three dimensions with a given gauge group and a given matter content in some representation(s) of the gauge group (with the same set of restrictions as noted in \cite{Kapustin:2010xq}).

In the following sections, we perform a non-trivial check of mirror symmetry for dual theories arising from the M-theory background described in the previous section with $\Gamma_1=D_n$ (with $n \geq 4$) and $\Gamma_2=\mathbb{Z}_m$, by computing partition functions of the dual theories and showing that they agree as functions of masses and FI parameters. This naturally implies that the masses and FI parameters on one side respectively map onto FI parameters and masses on the other side. These maps, usually referred to as ``mirror maps", have been explicitly computed in each case.

\section{$\Gamma_1= D_{n}$, $\Gamma_2=\mbox{Trivial}$ ($n \geq 4$)}\label{D-trivial}
We start with the case where $\Gamma_1= D_{n}$, $\Gamma_2=\mbox{Trivial}$.
For this family of dual theories, we perform the computation for the $k=1$ case first and then present the case of generic $k >1$. As we demonstrate below, proving the equality of partition functions for dual theories in the latter case requires using the identity in equation \eqref{Id2} as described in the Appendix.
\subsection{$k=1$ Case}
The dual theories, in this case, are as follows: 

\paragraph*{\bf A-model:} $U(1)^4 \times U(2)^{n-3}$ gauge theory with the matter content given by an extended ${D}_{n}$ quiver diagram. One of the $U(1)$ factors (labeled as $U(1)_1$) has a single fundamental hyper.

\paragraph*{\bf B-model:} $Sp(1)$ gauge theory with $n$ fundamental hypers and one additional hyper that transforms as a singlet of $Sp(1)$.

The partition functions of the two quiver gauge theories can be directly computed as follows:
\begin{equation}
\begin{split}
Z_A&=\frac{1}{(2!)^{n-3}}\int  \prod^{4}_{\alpha=1}d\sigma_{\alpha} e^{2\pi i \eta_{\alpha}\sigma_{\alpha}}\prod^{n-3}_{\beta=1}d^2\tilde{\sigma}_{\beta}  \prod^{2}_{i=1}e^{2\pi i \tilde{\eta}_{\beta}\tilde{\sigma}^i_{\beta}}\\
&\times\frac{1}{\prod_{i}\cosh{\pi(\sigma_1-\tilde{\sigma}^i_{1}+m_1)}\cosh{\pi(\sigma_2-\tilde{\sigma}^i_{1}+m_2)} } \frac{\prod^{n-3}_{\beta=1}\sinh^2{\pi(\tilde{\sigma}^1_{\beta}-\tilde{\sigma}^2_{\beta})}}{\prod^{n-4}_{\beta=1}\prod_{i,j} \cosh{\pi(\tilde{\sigma}^i_{\beta}-\tilde{\sigma}^j_{\beta+1}+M_{\beta})} }\\
&\times \frac{1}{\prod_{i}\cosh{\pi(\sigma_3-\tilde{\sigma}^i_{n-3}+m_3)}\cosh{\pi(\sigma_4-\tilde{\sigma}^i_{n-3}+m_4)} }\frac{1}{\cosh{\pi(\sigma_3+m_f)}} \label{pfAtriv1}
\end{split}
\end{equation}
\\
\begin{flalign}
Z_B=\frac{1}{2}\int d\sigma \frac{\sinh^2{\pi(2\sigma)}}{\prod^n_{a=1}\cosh{\pi(\sigma+m_a/2)}\cosh{\pi(\sigma-m_a/2)}}\frac{1}{\cosh{\pi M_{\text{singlet}}}} \label{pfBtriv1}
\end{flalign}
Note that the mass dependence of the partition function $Z_A$ can be removed by simply shifting the integration variables $\{\sigma_{\alpha}\}, \{\tilde{\sigma}^i_{\beta}\}$ by constants. One can, therefore, ignore the masses in $Z_A$ for the proof of the duality, which is consistent with the fact that the mirror dual of the A-model does not have any FI parameters.

The dimensions of the Coulomb branches and the Higgs branches of the dual models, as well as the number of FI and mass parameters in each case are summarized in table \ref{Coulomb-Higgs}.

\begin{table}[h]
\centering
\renewcommand{\arraystretch}{1.5}
\begin{tabular}{|c|c|c|c|c|}     \hline
Model   & $\mbox{dim} M_C$ & $\mbox{dim} M_H$ & $n_{\text{FI}}$ & $n_{\text{mass}}$ \\ \hline
       A  & $2(n-1)$ & $1$ & $n+1$ & $0$ \\  \hline     
       B &  $1$ & $2(n-1)$ & $0$ & $n+1$ \\ \hline 
\end{tabular}
\caption{The dimension of the Coulomb and Higgs branches
and the number of mass and FI parameters of A and B models for the M-theory background $\mathbb{C}^2/ D_{n} \times \mathbb{C}^2$ with $k=1$}
\label{Coulomb-Higgs}
\end{table}

To prove the equality of the two partition functions and derive the mirror map, we ``integrate out" the nodes $\sigma_2$ and $\sigma_4$, in a manner described in the Appendix. The partition function reduces to :
\begin{equation}
\begin{split}
Z_A& = \frac{2^2 i^{2}}{(2!)^{n-3}\sinh{\pi \eta_2}\sinh{\pi \eta_4}}\int  \prod_{\alpha=1,3}d\sigma_{\alpha} e^{2\pi i \eta_{\alpha}\sigma_{\alpha}}\prod^{n-3}_{\beta=1}d^2\tilde{\sigma}_{\beta} \prod^{2}_{i=1}e^{2\pi i \tilde{\eta}_{\beta}\tilde{\sigma}^i_{\beta}}e^{2\pi i \eta_2 \tilde{\sigma}^1_1 }e^{2\pi i \eta_4 \tilde{\sigma}^1_{n-3} }\\
&\times \frac{1}{\prod_{i}\cosh{\pi(\sigma_1-\tilde{\sigma}^i_{1})}} \prod^{n-4}_{\beta=1} \frac{ \sinh{\pi(\tilde{\sigma}^1_{\beta}-\tilde{\sigma}^2_{\beta})}\sinh{\pi(\tilde{\sigma}^1_{\beta+1}-\tilde{\sigma}^2_{\beta+1})}}{\prod_{i,j}\cosh{\pi(\tilde{\sigma}^i_{\beta}-\tilde{\sigma}^j_{\beta+1})}}\frac{1}{\prod_{i}\cosh{\pi(\sigma_3-\tilde{\sigma}^i_{n-3})}}  \\
&\times \frac{1}{\prod_i \cosh{\pi{\sigma}^i_3}}
\end{split}
\end{equation}
On using Cauchy's determinant formula and Fourier transforming, we get
\begin{equation}
\begin{split}
Z_A& \propto \int  \prod_{\alpha=1,3}d\sigma_{\alpha}e^{2\pi i \eta_{\alpha}\sigma_{\alpha}} \prod^{n-3}_{\beta=1}d^{2}\tilde{\sigma}_{\beta}\prod^{2}_{i=1}e^{2\pi i \tilde{\eta}_{\beta}\tilde{\sigma}^i_{\beta}} \prod^{n-4}_{\beta=1}d^2\tilde{\tau}_{\beta}\; d^{2} \tau_1 d^{2}\tau_{n-3} \;d\tau \;e^{2\pi i \eta_2 \tilde{\sigma}^1_1 }e^{2\pi i \eta_4 \tilde{\sigma}^1_{n-3} } \\
&\times \left(\prod_i \frac{e^{2\pi i \tau^i_1(\sigma_{1}- \tilde{\sigma}^{i}_1)}}{\cosh{\pi \tau^i_1}} \right) \prod^{n-4}_{\beta=1}  \left(\sum_{\tilde{\rho}_{\beta}}(-1)^{\tilde{\rho}_{\beta}}\prod_i \frac{e^{2\pi i \tilde{\tau}^i_{\beta}(\tilde{\sigma}^i_{\beta}- \tilde{\sigma}^{\tilde{\rho}_{\beta}(i)}_{\beta +1})}}{\cosh{\pi \tilde{\tau}^i_{\beta}}} \right) \left(\prod_i \frac{e^{2\pi i \tau^i_{n-3}(\sigma_{3}- \tilde{\sigma}^{i}_{n-3})}}{\cosh{\pi \tau^i_{n-3}}} \right)\left(\frac{e^{2\pi i \tau\sigma_3}}{\cosh{\pi \tau}}\right)
\end{split}
\end{equation}
where we have suppressed the constant pre-factors in the formula for $Z_A$.

On integrating out $\sigma_{\alpha}$, $\tilde{\sigma}_{\beta}$, the resulting $\delta$-functions impose the following conditions on the remaining variables:
\begin{equation}
\begin{split}
&\tau^1_1 + \tau^2_1 +\eta_1=0; \tilde{\tau}^1_1 - \tau^{1}_1 +\eta_2 +\tilde{\eta}_1=0;\tilde{\tau}^{2}_1 - \tau^{2}_1 +\tilde{\eta}_1=0 \\
& \tilde{\tau}^i_2=\tilde{\tau}^{\tilde{\rho}_1^{-1}(i)}_1-\tilde{\eta}_2; \tilde{\tau}^i_3=\tilde{\tau}^{\tilde{\rho}_1^{-1}\tilde{\rho}_2^{-1}(i)}_1-\tilde{\eta}_2 -\tilde{\eta}_3;........;  \tilde{\tau}^i_{n-4}=\tilde{\tau}^{\tilde{\rho}_1^{-1}\tilde{\rho}_2^{-1}.....\tilde{\rho}_{n-5}^{-1}(i)}_1 -\tilde{\eta}_2 -\tilde{\eta}_3-.......-\tilde{\eta}_{n-4}\\
& \tau^{1}_{n-3} = - \tilde{\tau}^{\tilde{\rho}_1^{-1}\tilde{\rho}_2^{-1}.....\tilde{\rho}_{n-4}^{-1}(1)}_1+ (\eta_4 + \tilde{\eta}_2 +\tilde{\eta}_3+.....+\tilde{\eta}_{n-3} )\\
&  \tau^{2}_{n-3} =-\tilde{\tau}^{\tilde{\rho}_1^{-1}\tilde{\rho}_2^{-1}.....\tilde{\rho}_{n-4}^{-1}(2)}_1+ (\tilde{\eta}_2 +\tilde{\eta}_3+.....+\tilde{\eta}_{n-3} )\\
&  \tau + \tau^1_{n-3} + \tau^{2}_{n-3} +\eta_3=0\\
\end{split}
\end{equation}
Now, one can integrate over all the variables other than $\tilde{\tau}^i_1$, noting that all but one of the permutations $\{\tilde{\rho}_{\beta}\}$ can be trivialized, giving rise to a multiplicative factor of $(2!)^{n-5}$ in the integrand. Redefining $\sigma^i=\tilde{\tau}^i_1$, we have,
\begin{equation}
\begin{split}
Z_A &= \frac{i^{2}}{\sinh{\pi \eta_2}\sinh{\pi \eta_4}}\int d^{2} \sigma \left(\frac{\delta(\sigma^1 +\sigma^{2}+ \eta_1+\eta_2 +2\tilde{\eta}_1)}{\cosh{\pi(\sigma^1 + \eta_2 + \tilde{\eta}_1)}\cosh{\pi(\sigma^{2}+\tilde{\eta}_1)}}\right)\\
&\times \left(\frac{1}{\prod^2_{i=1} \cosh{\pi \sigma^i} \cosh{\pi (\sigma^i-\tilde{\eta}_2)}......\cosh{\pi(\sigma^i-\tilde{\eta}_2-\tilde{\eta}_3 -.....-\tilde{\eta}_{n-4})}}    \right)\\
&\times \left(\sum_{\rho}(-1)^{\rho} \frac{1}{\cosh{\pi(\sigma^{\rho(1)}  -\eta_4-\sum^{n-3}_{\beta=2}\tilde{\eta}_{\beta})}\cosh{\pi(\sigma^{\rho(2)}-\sum^{n-3}_{\beta=2}\tilde{\eta}_{\beta})}} \right)\\
&\times \frac{1}{\cosh{\pi(\sigma^1 +\sigma^{2}-\eta_3-\eta_4-2\sum^{n-3}_{\beta=2}\tilde{\eta}_{\beta})}}\\
\end{split}
\end{equation}
Changing variables to $\sigma^i \to \sigma^i -(\eta_1+\eta_2)/2 -\tilde{\eta}_1$, and noting that the first term in parenthesis needs to be antisymmetric under the exchange $\sigma^1 \leftrightarrow \sigma^2$, the above expression reduces to
\begin{equation}
\begin{split}
&Z_A =\frac{i^{2}}{\sinh{\pi \eta_2}\sinh{\pi \eta_4}} \int d^{2} \sigma \frac{\delta(\sigma^1 +\sigma^{2})}{2}\\
&\times\left(\sum_{\rho^{'}}(-1)^{\rho^{'}} \frac{1}{\cosh{\pi(\sigma^{\rho^{'}(1)}+(\eta_2-\eta_1)/2)}\cosh{\pi(\sigma^{\rho^{'}(2)}-(\eta_1+\eta_2)/2)}} \right)\\
&\times \left[\frac{1}{\prod^2_{i=1} \cosh{\pi( \sigma^i-\frac{(\eta_1+\eta_2)}{2} -\tilde{\eta}_1)} \cosh{\pi (\sigma^i-\frac{(\eta_1+\eta_2)}{2} -\tilde{\eta}_1-\tilde{\eta}_2)}......\cosh{\pi(\sigma^i-\frac{(\eta_1+\eta_2)}{2} -\sum^{n-4}_{\beta=1}\tilde{\eta}_{\beta})}} \right]\\
&\times \left(\sum_{\rho}(-1)^{\rho} \frac{1}{\cosh{\pi(\sigma^{\rho(1)} -\frac{(\eta_1+\eta_2)}{2} -\eta_4-\sum^{n-3}_{\beta=1}\tilde{\eta}_{\beta})}\cosh{\pi(\sigma^{\rho(2)}-\frac{(\eta_1+\eta_2)}{2}-\sum^{n-3}_{\beta=1}\tilde{\eta}_{\beta})}} \right)\\
&\times \frac{1}{\cosh{\pi(\sigma^1 +\sigma^{2}-\eta_1-\eta_2-\eta_3-\eta_4-2\sum^{n-3}_{\beta=1}\tilde{\eta}_{\beta})}}\\
&=i^2\int d^{2} \sigma \frac{\delta(\sigma^1 +\sigma^{2})}{2\sinh{\pi \eta_2}\sinh{\pi \eta_4}} \left(\frac{-\sinh{\pi \eta_2}\sinh{\pi(\sigma^1-\sigma^2)}}{\prod^2_{i=1}\cosh{\pi(\sigma^{i}-(\eta_1-\eta_2)/2)}\cosh{\pi(\sigma^{i}-(\eta_1+\eta_2)/2)}}\right)\\
&\times \left[\frac{1}{\prod^2_{i=1} \cosh{\pi( \sigma^i-\frac{(\eta_1+\eta_2)}{2} -\tilde{\eta}_1)} \cosh{\pi (\sigma^i-\frac{(\eta_1+\eta_2)}{2} -\tilde{\eta}_1-\tilde{\eta}_2)}......\cosh{\pi(\sigma^i-\frac{(\eta_1+\eta_2)}{2} -\sum^{n-4}_{\beta=1}\tilde{\eta}_{\beta})}} \right]\\
&\times \left(\frac{\sinh{\pi \eta_4}\sinh{\pi(\sigma^1-\sigma^2)}}{\prod^2_{i=1}\cosh{\pi(\sigma^{i} -\frac{(\eta_1+\eta_2)}{2} -\eta_4-\sum^{n-3}_{\beta=1}\tilde{\eta}_{\beta})}\cosh{\pi(\sigma^{i}-\frac{(\eta_1+\eta_2)}{2}-\sum^{n-3}_{\beta=1}\tilde{\eta}_{\beta})}} \right)\\
&\times \frac{1}{\cosh{\pi(\eta_1+\eta_2+\eta_3+\eta_4+2\sum^{n-3}_{\beta=1}\tilde{\eta}_{\beta})}}
\end{split}
\end{equation}
The above partition function can be immediately re-written as,
\begin{equation}
\begin{split}
&Z_A=\int d^{2} \sigma \frac{\delta(\sigma^1 +\sigma^{2})}{2}\sinh^2{\pi(\sigma^1-\sigma^2)}\left(\frac{1}{\prod^2_{i=1}\cosh{\pi(\sigma^{i}-(\eta_1-\eta_2)/2)}\cosh{\pi(\sigma^{i}-(\eta_1+\eta_2)/2)}}\right)\\
&\times \left[\frac{1}{\prod^2_{i=1} \cosh{\pi( \sigma^i-\frac{(\eta_1+\eta_2)}{2} -\tilde{\eta}_1)} \cosh{\pi (\sigma^i-\frac{(\eta_1+\eta_2)}{2} -\tilde{\eta}_1-\tilde{\eta}_2)}......\cosh{\pi(\sigma^i-\frac{(\eta_1+\eta_2)}{2} -\sum^{n-4}_{\beta=1}\tilde{\eta}_{\beta})}} \right]\\
&\times \left(\frac{1}{\prod^2_{i=1}\cosh{\pi(\sigma^{i} -\frac{(\eta_1+\eta_2)}{2} -\eta_4-\sum^{n-3}_{\beta=1}\tilde{\eta}_{\beta})}\cosh{\pi(\sigma^{i}-\frac{(\eta_1+\eta_2)}{2}-\sum^{n-3}_{\beta=1}\tilde{\eta}_{\beta})}} \right)\\
&\times \frac{1}{\cosh{\pi(\eta_1+\eta_2+\eta_3+\eta_4+2\sum^{n-3}_{\beta=1}\tilde{\eta}_{\beta})}}
\end{split}
\end{equation}
which proves that $Z_A=Z_B$, as we had expected, provided the FI parameters of the A-model are related to the masses of the B-model in a particular way, which can be read off by comparing the two partition functions. 

\paragraph*{\bf Mirror Map:} The number of independent mass parameters in the A-model is zero, which matches with the number of FI parameters in the B-model. The number of independent FI parameters in the A-model is $(n+1)$ which again matches with number of independent mass parameters of the B-model. The B-model masses are related to the A-model FI parameters. Then masses of the fundamental hypermultiplets are given by
\begin{equation}
\boxed{\begin{gathered}m_1= (\eta_1-\eta_2); m_2 = (\eta_2+\eta_1);m_{\beta+2}=(\eta_2+\eta_1)+2\sum^{\beta}_{\alpha=1} \tilde{\eta}_{\alpha};\\ m_{n}=(\eta_2+\eta_1) +2\eta_4+2\sum^{n-3}_{\alpha=1} \tilde{\eta}_{\alpha}\end{gathered}}
\end{equation}
where $\beta=1,2,..,n-3$. The mass of the hypermultiplet which is a singlet under $Sp(1)$ is given by
\begin{equation}
\boxed{M_{\text{singlet}} =\eta_1+\eta_2 +\eta_3 +\eta_4 + 2(\tilde{\eta}_1+....+\tilde{\eta}_{n-3})}
\end{equation}

\subsection{$k>1$ Case}
The mirror dual theories, in this case, are:

\paragraph*{\bf A-model:} $U(k)^4 \times U(2k)^{n-3}$ gauge theory with the matter content given by an extended ${D}_{n}$ quiver diagram. One of the $U(k)$ factors has one fundamental hyper.

\paragraph*{\bf B-model:} $Sp(k)$ gauge theory with $n$ fundamental hypers and one hyper in the anti-symmetric representation of $Sp(k)$.

\begin{figure}[htbp]
\begin{center}
\includegraphics[height=3.0in]{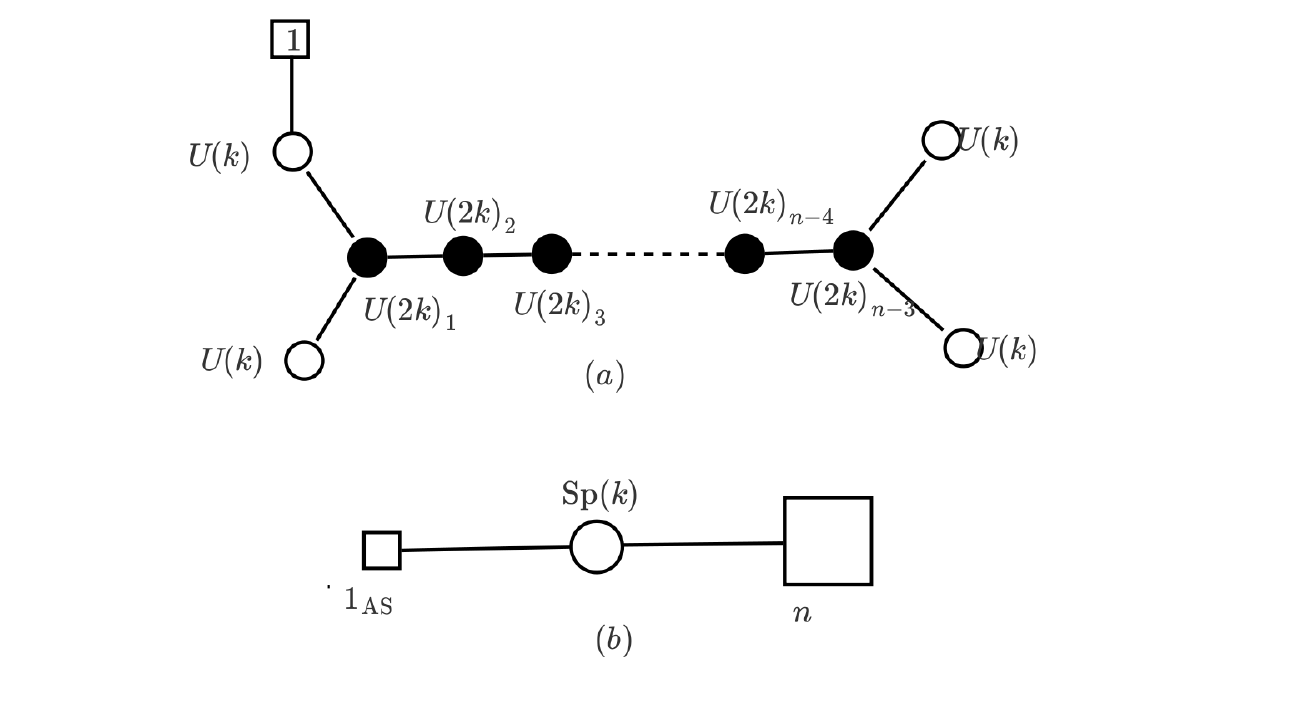}
\caption{Mirror Duals for $\Gamma_1={D}_{n}$,$\Gamma_2=\text{Trivial}$ for a generic value of k. For the special case of $k=1$, the antisymmetric hyper in (b) is replaced by a singlet.}
\label{fig3l}
\end{center}
\end{figure}

\begin{equation}
\begin{split}
Z_A&=\frac{1}{(k!)^4 (2k!)^{n-3}}\int  \prod^{4}_{\alpha=1}d^k\sigma_{\alpha} \prod^{n-3}_{\beta=1}d\tilde{\sigma}_{\beta}\prod^{k}_{i=1}e^{2\pi i \eta_{\alpha}\sigma^i_{\alpha}} \prod^{2k}_{p=1}e^{2\pi i \tilde{\eta}_{\beta}\tilde{\sigma}^p_{\beta}}\\
&\times\frac{\prod_{i<j}\sinh^2{\pi(\sigma^i_1-\sigma^j_1)}\sinh^2{\pi(\sigma^i_2-\sigma^j_2)}}{\prod_{i,p}\cosh{\pi(\sigma^i_1-\tilde{\sigma}^p_{1}+m_1)}\cosh{\pi(\sigma^i_2-\tilde{\sigma}^p_{1}+m_2)} } \frac{\prod^{n-3}_{\beta=1}\prod_{p<l}\sinh^2{\pi(\tilde{\sigma}^p_{\beta}-\tilde{\sigma}^l_{\beta})}}{\prod^{n-4}_{\beta=1}\prod_{p,l} \cosh{\pi(\tilde{\sigma}^p_{\beta}-\tilde{\sigma}^l_{\beta+1}+M_{\beta})} }\\
&\times \frac{\prod_{i<j}\sinh^2{\pi(\sigma^i_3-\sigma^j_3)}\sinh^2{\pi(\sigma^i_4-\sigma^j_4)}}{\prod_{i,p}\cosh{\pi(\sigma^i_3-\tilde{\sigma}^p_{n-3}+m_3)}\cosh{\pi(\sigma^i_4-\tilde{\sigma}^p_{n-3}+m_4)} }\frac{1}{\prod_i\cosh{\pi(\sigma^i_3+m_f)}} \label{pfAtriv}
\end{split}
\end{equation}
\begin{equation}
\begin{split}
Z_B&=\frac{1}{2^k (k!)}\int d^k\sigma \frac{1}{\prod^n_{a=1}\prod^k_{i=1}\cosh{\pi(\sigma^i+m_a/2)}\cosh{\pi(\sigma^i-m_a/2)}}\\
&\times \frac{\prod_{i<j}\sinh^2{\pi(\sigma_i-\sigma_j)}\sinh^2{\pi(\sigma_i+\sigma_j)}\prod_i \sinh^2{\pi(2\sigma_i)}}{\prod_{i<j}\cosh{\pi(\sigma^i+\sigma^j+M_{as})}\cosh{\pi(\sigma^i+\sigma^j-M_{as})}\prod_{i,j} \cosh{\pi(\sigma^i-\sigma^j-M_{as})}} \label{pfBtriv}
\end{split}
\end{equation}

The dimensions of the Coulomb branches and the Higgs branches of the dual models, as well as the number of FI and mass parameters in each case are summarized in table \ref{D-trivTable}.

\begin{table}[h]
\renewcommand{\arraystretch}{1.5}
\centering
\begin{tabular}{|c|c|c|c|c|}     \hline
Model   & $\mbox{dim} M_C$ & $\mbox{dim} M_H$ & $n_{\text{FI}}$ & $n_{\text{mass}}$ \\ \hline
       A  & $2k(n-1)$ & $k$ & $n+1$ & $0$ \\  \hline     
       B &  $k$ & $2k(n-1)$ & $0$ & $n+1$ \\ \hline 
\end{tabular}
\caption{The dimension of the Coulomb and Higgs branches
and the number of mass and FI parameters of A and B models for the M-theory background $\mathbb{C}^2/ D_{n} \times \mathbb{C}^2$ with $k>1$}
\label{D-trivTable}
\end{table}

To show that the two partition functions given above describe the same theory in the IR and to derive the corresponding mirror map, we start with the A-model.  As in the $k=1$ case described in the previous sub-section, we ``integrate out" the nodes corresponding to $\sigma_2$ and $\sigma_4$. This reduces $Z_A$ to :
\begin{equation}
\begin{split}
Z_A& = \frac{ i^{-2k} e^{-2\pi i k \eta_2 m_2 }e^{-2\pi i k \eta_4 m_4}}{(k!)^6(2k!)^{n-5}\sinh^k{\pi \eta_2}\sinh^k{\pi \eta_4}}\int  \prod_{\alpha=1,3}d^k\sigma_{\alpha} \prod^{n-3}_{\beta=1}d\tilde{\sigma}^{2k}_{\beta}\prod^{k}_{i=1}e^{2\pi i \eta_{\alpha}\sigma^i_{\alpha}}e^{2\pi i \eta_2 \tilde{\sigma}^i_1 }e^{2\pi i \eta_4 \tilde{\sigma}^i_{n-3} } \prod^{2k}_{p=1}e^{2\pi i \tilde{\eta}_{\beta}\tilde{\sigma}^p_{\beta}}\\
&\times \frac{\prod_{i<j}\sinh{\pi({\sigma}^i_1-{\sigma}^j_1)}\sinh{\pi(\tilde{\sigma}^i_1-\tilde{\sigma}^j_1)}}{\prod_{i,j}\cosh{\pi(\sigma^i_1-\tilde{\sigma}^j_{1}+m_1)}} \frac{\prod_{i<j}\sinh{\pi({\sigma}^i_1-{\sigma}^j_1)}\sinh{\pi(\tilde{\sigma}^{k+i}_1-\tilde{\sigma}^{k+j}_1)}}{\prod_{i,j}\cosh{\pi(\sigma^i_1-\tilde{\sigma}^{k+j}_{1}+m_1)}} \\
&\times \prod^{n-4}_{\beta=1} \frac{\prod_{p<l} \sinh{\pi(\tilde{\sigma}^p_{\beta}-\tilde{\sigma}^l_{\beta})}\sinh{\pi(\tilde{\sigma}^p_{\beta+1}-\tilde{\sigma}^l_{\beta+1})}}{\prod_{p,l}\cosh{\pi(\tilde{\sigma}^p_{\beta}-\tilde{\sigma}^l_{\beta+1}+M_{\beta})}}\\
&\times \frac{\prod_{i<j}\sinh{\pi({\sigma}^i_3-{\sigma}^j_3)}\sinh{\pi(\tilde{\sigma}^i_{n-3}-\tilde{\sigma}^j_{n-3})}}{\prod_{i,j}\cosh{\pi(\sigma^i_3-\tilde{\sigma}^j_{n-3}+m_3)}} \frac{\prod_{i<j}\sinh{\pi({\sigma}^i_3-{\sigma}^j_3)}\sinh{\pi(\tilde{\sigma}^{k+i}_{n-3}-\tilde{\sigma}^{k+j}_{n-3})}}{\prod_{i,j}\cosh{\pi(\sigma^i_3-\tilde{\sigma}^{k+j}_{n-3}+m_3)}}\\
&\times \frac{1}{\prod_i \cosh{\pi({\sigma}^i_3+m_f)}}
\end{split}
\end{equation}
Note that since the $\sigma^i_2$-independent part of the integrand is symmetric in $\tilde{\sigma}^p_1$, the symmetrization operation in terms of $\tilde{\sigma}^p_1$ becomes trivial as we integrate out $\sigma^i_2$ and one needs to multiply the integrand by a factor of $k!$. The same argument applies for $\tilde{\sigma}^p_{n-3}$, when we integrate out $\sigma^i_4$.

On using Cauchy's determinant formula and Fourier transforming, we get
\begin{equation}
\begin{split}
Z_A& \propto \int  \prod_{\alpha=1,3}d^k\sigma_{\alpha} \prod^{n-3}_{\beta=1}d^{2k}\tilde{\sigma}_{\beta} \prod^{n-4}_{\beta=1}d\tilde{\tau}^{2k}_{\beta} d^{2k} \tau_1 d^{2k}\tau_{n-3} d^{k}\tau \prod^{k}_{i=1}e^{2\pi i \eta_{\alpha}\sigma^i_{\alpha}}e^{2\pi i \eta_2 \tilde{\sigma}^i_1 }e^{2\pi i \eta_4 \tilde{\sigma}^i_{n-3} } \prod^{2k}_{p=1}e^{2\pi i \tilde{\eta}_{\beta}\tilde{\sigma}^p_{\beta}}\\
&\times \left(\sum_{\rho_1}(-1)^{\rho_1}\prod_i \frac{e^{2\pi i \tau^i_1(\sigma^i_1- \tilde{\sigma}^{\rho_1(i)}_1+m_1)}}{\cosh{\pi \tau^i_1}} \right)  \left(\sum_{\rho_2}(-1)^{\rho_2}\prod_i \frac{e^{2\pi i \tau^{k+i}_1(\sigma^{i}_1- \tilde{\sigma}^{k+\rho_2(i)}_1+m_1)}}{\cosh{\pi \tau^{k+i}_1}} \right)\\
&\times \prod^{n-4}_{\beta=1}  \left(\sum_{\tilde{\rho}_{\beta}}(-1)^{\tilde{\rho}_{\beta}}\prod_p \frac{e^{2\pi i \tilde{\tau}^p_{\beta}(\tilde{\sigma}^p_{\beta}- \tilde{\sigma}^{\tilde{\rho}_{\beta}(p)}_{\beta +1}+M_{\beta})}}{\cosh{\pi \tilde{\tau}^p_{\beta}}} \right)\\
&\times \left(\sum_{\rho_3}(-1)^{\rho_3}\prod_i \frac{e^{2\pi i \tau^i_{n-3}(\sigma^i_3- \tilde{\sigma}^{\rho_3(i)}_{n-3}+m_3)}}{\cosh{\pi \tau^i_{n-3}}} \right)  \left(\sum_{\rho_4}(-1)^{\rho_4}\prod_i \frac{e^{2\pi i \tau^{k+i}_{n-3}(\sigma^{i}_3- \tilde{\sigma}^{k+\rho_4(i)}_{n-3}+m_3)}}{\cosh{\pi \tau^{k+i}_{n-3}}} \right)\\
&\times \prod_i \frac{e^{2\pi i \tau^i(\sigma^i_3+m_f)}}{\cosh{\pi \tau^i}}
\end{split}
\end{equation}
where we have suppressed the constant pre-factors in the formula for $Z_A$.

From equation \eqref{pfAtriv}, it is clear that the masses of the bifundamental hypers and that of the fundamental hyper can be eliminated by simply shifting the variables $\{\sigma_{\alpha}\}, \{\tilde{\sigma}_{\beta}\}$ by constants. This shows that $Z_A$ is only a function of the FI parameters. Therefore, we set to zero all the mass terms in what follows.

On integrating out $\sigma_{\alpha}$, $\tilde{\sigma}_{\beta}$, the resulting $\delta$-functions impose the following conditions on the remaining variables:
\begin{equation}
\begin{split}
&\tau^i_1 + \tau^{k+i}_1 +\eta_1=0; \tilde{\tau}^i_1 - \tau^{\rho_1^{-1}(i)}_1 +\eta_2 +\tilde{\eta}_1=0;\tilde{\tau}^{k+i}_1 - \tau^{k+\rho_2^{-1}(i)}_1 +\tilde{\eta}_1=0 \\
& \tilde{\tau}^p_2=\tilde{\tau}^{\tilde{\rho}_1^{-1}(p)}_1-\tilde{\eta}_2; \tilde{\tau}^p_3=\tilde{\tau}^{\tilde{\rho}_1^{-1}\tilde{\rho}_2^{-1}(p)}_1-\tilde{\eta}_2 -\tilde{\eta}_3;........;  \tilde{\tau}^p_{n-4}=\tilde{\tau}^{\tilde{\rho}_1^{-1}\tilde{\rho}_2^{-1}.....\tilde{\rho}_{n-5}^{-1}(p)}_1 -\tilde{\eta}_2 -\tilde{\eta}_3-.......-\tilde{\eta}_{n-4}\nonumber\\
& \tau^{\rho^{-1}_3(i)}_{n-3} = - \tilde{\tau}^{\tilde{\rho}_1^{-1}\tilde{\rho}_2^{-1}.....\tilde{\rho}_{n-4}^{-1}(i)}_1+ (\eta_4 + \tilde{\eta}_2 +\tilde{\eta}_3+.....+\tilde{\eta}_{n-3} )\\
&  \tau^{k+\rho^{-1}_4(i)}_{n-3} =-\tilde{\tau}^{\tilde{\rho}_1^{-1}\tilde{\rho}_2^{-1}.....\tilde{\rho}_{n-4}^{-1}(k+i)}_1+ (\tilde{\eta}_2 +\tilde{\eta}_3+.....+\tilde{\eta}_{n-3} )\\
&  \tau^i + \tau^i_{n-3} + \tau^{k+i}_{n-3} +\eta_3=0\\
\end{split}
\end{equation}
Next, we integrate out all the variables other than $\tilde{\tau}^p_1$ using the $\delta$-function conditions. Note that all of four permutations $\{\rho_{\alpha}\}$ and all but one of the $n-4$ permutations $\tilde{\rho}_{\beta}$ can be trivialized and this implies multiplying the integrand by a factor of $(k!)^4 (2k!)^{n-5}$. Now, redefining $\sigma^p=\tilde{\tau}^p_1$, we have,
\begin{equation}
\begin{split}
Z_A &=\frac{i^{2k}}{(k!)^2\sinh^k{\pi \eta_2}\sinh^k{\pi \eta_4}} \int d^{2k} \sigma \prod_i \frac{\delta(\sigma^i +\sigma^{k+i}+ \eta_1+\eta_2 +2\tilde{\eta}_1)}{\cosh{\pi(\sigma^i + \eta_2 + \tilde{\eta}_1)}\cosh{\pi(\sigma^{k+i}+\tilde{\eta}_1)}}\\
&\times \left[\frac{1}{\prod_p \cosh{\pi \sigma^p} \cosh{\pi (\sigma^p-\tilde{\eta}_2)}......\cosh{\pi(\sigma^p-\tilde{\eta}_2-\tilde{\eta}_3 -.....-\tilde{\eta}_{n-4})}}    \right]\\
&\times \left(\sum_{\rho}(-1)^{\rho} \frac{1}{\prod_i \cosh{\pi(\sigma^{\rho(i)}-m)}\cosh{\pi(\sigma^{\rho(k+i)}-m^{'})}\cosh{\pi(\sigma^{\rho(i)}+\sigma^{\rho(k+i)}-M)}} \right)
\end{split}
\end{equation}
where $m=\eta_4 +\tilde{\eta}_2+\tilde{\eta}_3 +.....+\tilde{\eta}_{n-3}$, $m^{'}=\tilde{\eta}_2+\tilde{\eta}_3 +.....+\tilde{\eta}_{n-3}$, $M=\eta_3 +\eta_4 +2(\tilde{\eta}_2+....+\tilde{\eta}_{n-3})$. Note that the RHS of the above equation is completely independent of the A-model fundamental masses. This is expected since the number of independent mass parameters of the A-model is zero, thereby matching the number of FI parameters in the B-model.

The last term in parenthesis in the above equation  can be evaluated using an identity which involves a  generalization of Schur's Pfaffian identity. As described in the Appendix (equation \eqref{Id2}), one has the following identity:
\begin{equation}
\begin{split}
&\sum _{\rho} (-1)^{\rho} \frac{1}{\prod_{i} \cosh{(\sigma_{\rho(i)}-m)}\cosh{(\sigma_{\rho(k+i)}-m^{'})}\cosh{(\sigma_{\rho(i)}+\sigma_{\rho(k+i)} -M)}}  \\
=&\left(\frac{\kappa \; k!\; \sinh^k{(m-m^{'})}}{\prod_i \cosh{(\sigma_i-m)} \cosh{(\sigma_i-m^{'})} \cosh{(\sigma_{k+i}-m)}\cosh{(\sigma_{k+i}-m^{'})}}\right) \times \left(\prod_{p<l} \frac{\sinh{(\sigma_p - \sigma_l)}}{\cosh{(\sigma_p + \sigma_l-M)}}\right)
\end{split}
\end{equation}
with $\kappa=1$ for even $k=4m$ and odd $k=4m+1$ ($m=1,2,3,...$) and $\kappa=-1$ otherwise.

Therefore, using the above identity, we have
\begin{equation}
\begin{split}
Z_A &= \frac{i^{2k}}{(k!)^2\sinh^k{\pi \eta_2}\sinh^k{\pi \eta_4}} \int d^{2k} \sigma \left(\prod_i \frac{\delta(\sigma^i +\sigma^{k+i}+ \eta_1+\eta_2 +2\tilde{\eta}_1)}{\cosh{\pi(\sigma^i + \eta_2 + \tilde{\eta}_1)}\cosh{\pi(\sigma^{k+i}+\tilde{\eta}_1)}}\right)\\
&\times \left[\frac{1}{\prod_p \cosh{\pi \sigma^p} \cosh{\pi (\sigma^p-\tilde{\eta}_2)}......\cosh{\pi(\sigma^p-\tilde{\eta}_2-\tilde{\eta}_3 -.....-\tilde{\eta}_{n-4})}}    \right]\\
&\times \left(\frac{\kappa \; k! \; \sinh^k{(m-m^{'})}}{\prod_p\cosh{\pi(\sigma^p-m)} \cosh{\pi(\sigma^p-m^{'})}} \prod_{p<l} \frac{\sinh{\pi(\sigma^p-\sigma^l)}}{\cosh{\pi(\sigma^p+\sigma^l-M)}}\right)\\
\end{split}
\end{equation}
On anti-symmetrizing the first term in parenthesis under the exchange $\sigma^i \leftrightarrow \sigma^{k+i}$ (which brings in a factor of $2^{-k}$ in the integrand), we have,
\begin{equation}
\begin{split}
Z_A & =\frac{i^{2k}}{(k!)^2\sinh^k{\pi \eta_2}\sinh^k{\pi \eta_4}} \int d^{2k}\sigma \frac{\sinh^k{(-\pi\eta_2)}}{2^k} \frac{\prod_i \delta(\sigma^i +\sigma^{k+i}+ \eta_1+\eta_2 +2\tilde{\eta}_1)  \sinh{\pi (\sigma^i-\sigma^{k+i})}}{\prod_p \cosh{\pi(\sigma^p+\tilde{\eta}_1)} \cosh{\pi(\sigma^p+\tilde{\eta}_1+\eta_2)}}\\
&\times \left[\frac{1}{\prod_p \cosh{\pi \sigma^p} \cosh{\pi (\sigma^p-\tilde{\eta}_2)}......\cosh{\pi(\sigma^p-\tilde{\eta}_2-\tilde{\eta}_3 -.....-\tilde{\eta}_{n-4})}}\right]\\
&\times \left(\frac{\kappa \; k! \;\sinh^k{(\eta_4)}}{\prod_p\cosh{\pi(\sigma^p-m)} \cosh{\pi(\sigma^p-m^{'})}} \prod_{p<l} \frac{\sinh{\pi(\sigma^p-\sigma^l)}}{\cosh{\pi(\sigma^p+\sigma^l-M)}}\right)
\end{split}
\end{equation}
We redefine $\sigma^p \to \sigma^p - (\eta_1+\eta_2)/2 -\tilde{\eta}_1$ and recall that 
\begin{equation}
\prod_{p<l} \cosh{\pi(\sigma^p+\sigma^l-M)}\bigg|_{\sigma^i =-\sigma^{k+i}}=\prod_{i<j}\cosh{\pi(\sigma^i+\sigma^j+M)}\cosh{\pi(\sigma^i+\sigma^j-M)}\prod_{i,j} \cosh{\pi(\sigma^i-\sigma^j-M)}
\end{equation}
\begin{equation}
\prod_{p<l} \sinh{\pi(\sigma^p-\sigma^l)}\bigg|_{\sigma^i =-\sigma^{k+i}}=(-1)^{k(k-1)/2}\prod_{i<j} \sinh^2{\pi(\sigma^i-\sigma^j)} \sinh^2{\pi(\sigma^i+\sigma^j)} \prod_i \sinh{\pi 2\sigma^i}
\end{equation}
Note that the factor $(-1)^{k(k-1)/2}$ and the factor $\kappa$ cancel each other for any $k$.

Therefore, taking into account all the combinatorial pre-factors, we have 
\begin{equation} 
\begin{split}
&Z_A = \int \frac{d^{2k}\sigma}{2^k k!} \prod_{i<j} \sinh^2{\pi(\sigma^i-\sigma^j)} \sinh^2{\pi(\sigma^i+\sigma^j)} \prod_i \sinh^2{\pi 2\sigma^i}\\
&\times \frac{\prod_i \delta(\sigma^i +\sigma^{k+i})}{\prod_p \cosh{\pi(\sigma^p-(\eta_1+\eta_2)/2)} \cosh{\pi(\sigma^p-(\eta_1-\eta_2)/2)}}\\
&\times \left[\frac{1}{\prod_{p} \cosh{\pi( \sigma^p-\frac{(\eta_1+\eta_2)}{2} -\tilde{\eta}_1)} \cosh{\pi (\sigma^p-\frac{(\eta_1+\eta_2)}{2} -\tilde{\eta}_1-\tilde{\eta}_2)}......\cosh{\pi(\sigma^p-\frac{(\eta_1+\eta_2)}{2} -\sum^{n-4}_{\beta=1}\tilde{\eta}_{\beta})}}\right]\\
&\times \left(\frac{1}{\prod_{p}\cosh{\pi(\sigma^{p} -\frac{(\eta_1+\eta_2)}{2} -\eta_4-\sum^{n-3}_{\beta=1}\tilde{\eta}_{\beta})}\cosh{\pi(\sigma^{p}-\frac{(\eta_1+\eta_2)}{2}-\sum^{n-3}_{\beta=1}\tilde{\eta}_{\beta})}} \right)\\
&\times \left( \prod_{p<l} \frac{1}{\cosh{\pi(\sigma^p+\sigma^l-\eta_1-\eta_2-\eta_3-\eta_4-2\sum^{n-3}_{\beta=1}\tilde{\eta}_{\beta})}}\right)
\end{split}
\end{equation}
which is identical to the partition function given in equation \eqref{pfBtriv}. Therefore, $Z_A=Z_B$, provided the FI parameters in the A model are related to the masses in the B model. The corresponding mirror map is read off by comparing the two partition functions, as before.

\paragraph*{\bf Mirror Map:} The number of independent mass parameters in the A-model is zero, which matches with the number of FI parameters in the B-model. The number of independent FI parameters in the A-model is $(n+1)$ which again matches with number of independent mass parameters of the B-model. The B-model masses are related to the A-model FI parameters; the fundamental masses are given by
\begin{equation}
\boxed{\begin{gathered}m_1= (\eta_1-\eta_2); m_2 = (\eta_2+\eta_1);m_{\beta+2}=(\eta_2+\eta_1)+2\sum^{\beta}_{\alpha=1} \tilde{\eta}_{\alpha};\\ m_{n}=(\eta_2+\eta_1) +2\eta_4+2\sum^{n-3}_{\alpha=1} \tilde{\eta}_{\alpha}\end{gathered}}
\end{equation}
where $\beta=1,2,..,n-3$. The mass of the antisymmetric hypermultiplet is given by
\begin{equation}
\boxed{M_{as} =\eta_1+\eta_2 +\eta_3 +\eta_4 + 2(\tilde{\eta}_1+....+\tilde{\eta}_{n-3})}
\end{equation}

\section{$\mathbb{C}^2/ D_{n} \times \mathbb{C}^2/\mathbb{Z}_2$ Case ($n \geq 4$)}\label{D-A}
There are four important families of dual pairs in this category. The A-model for each of these families is a $D_n$ quiver with $\{n_i\}$ fundamental hypers, such that $\sum_i n_i l_i=2$, where $l_i$ is the Dynkin label of the $i$-th node in a $D_n$ quiver. Therefore,  A-models for the four families differ only in the distribution of the fundamental hypers on the quiver but their mirror duals can be significantly different from each other. The four cases are labeled by the manifest flavor symmetry groups (due to fundamental hypers) of the dual pairs - $G^A_{flavor}$ and $G^B_{flavor}$. For the A-model, the number in the subscript of an "unprimed" flavor symmetry group indicates which of the boundary nodes have the two fundamental hypers, while for a "primed" flavor symmetry group it specifies which interior node of the $D_n$ quiver contains the single fundamental hyper.  
 
\subsection{$G^A_{flavor}=U(1)_1\times U(1)_2$, $G^B_{flavor}=SO(2)\times SO(2n-2)$}\label{sec:case1}
\paragraph*{\bf A-model:} $U(k)_1 \times U(k)_2 \times U(2k)^{n-3} \times U(k)_3 \times U(k)_4$ gauge theory with bi-fundamental  matter content given by an extended ${D}_{n}$ quiver diagram. $U(k)_1$ and $U(k)_2$ gauge groups have one fundamental hyper each. The masses and the FI parameters respect the $\mathbb{Z}_2$ outer-automorphism symmetry of the extended ${D}_{n}$ quiver.

\paragraph*{\bf B-model:} $Sp(k)_1 \times Sp(k)_2$ gauge theory with one bi-fundamental hyper. In addition, one has a single hyper in the fundamental of $Sp(k)_1$ and $n-1$ hypers in the fundamental of $Sp(k)_2$.

\begin{figure}[htbp]
\begin{center}
\includegraphics[height=3.0in]{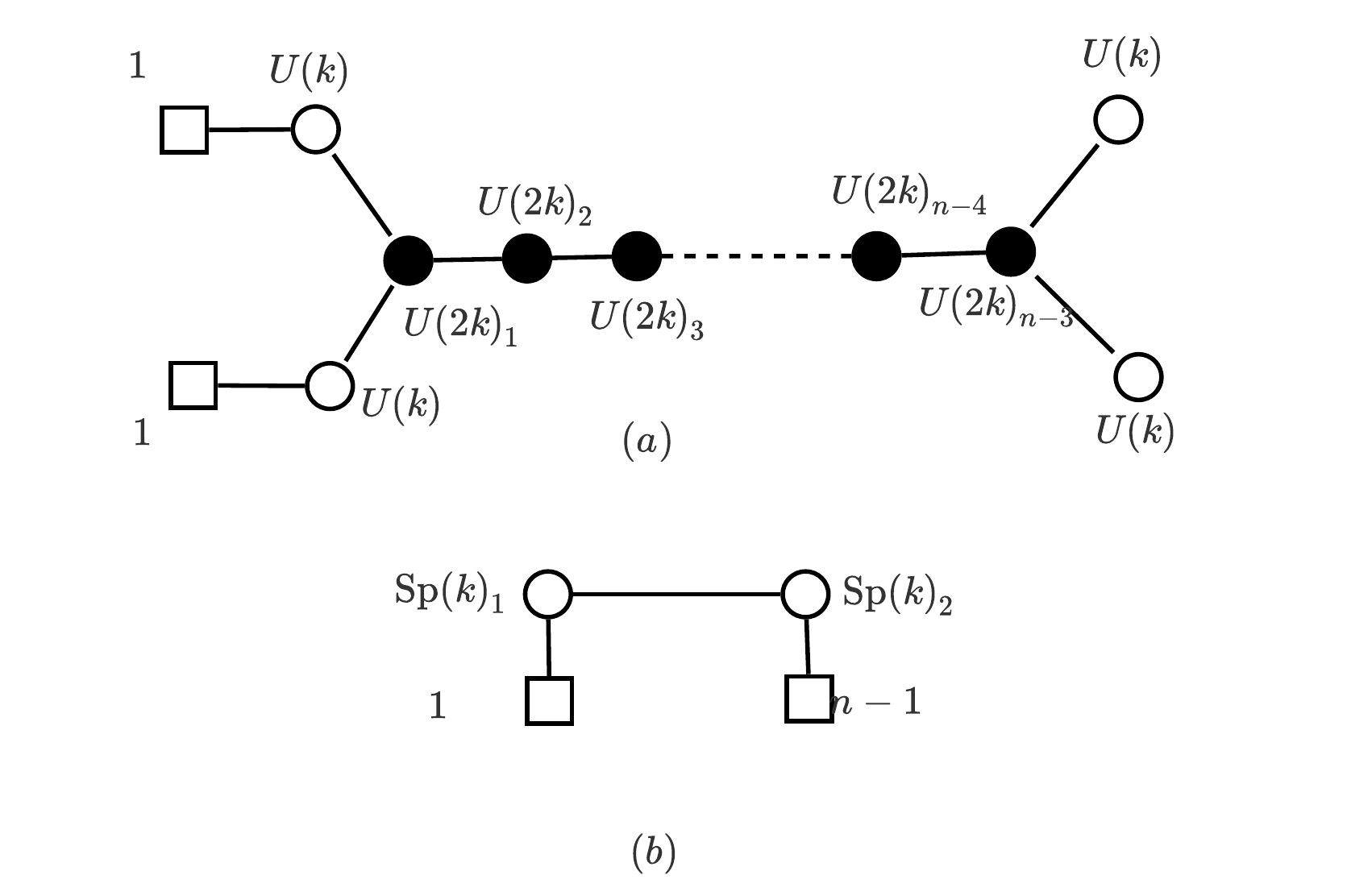}
\caption{Mirror Duals for $\Gamma_1={D}_{n}$,$\Gamma_2=\mathbb{Z}_2$. The A-model has 1 fundamental hyper each on two adjacent $U(k)$ nodes. }
\label{Z2-case1}
\end{center}
\end{figure}
The partition functions for the dual theories are as follows:
\begin{equation}
\begin{split}
Z_A&=\frac{1}{(k!)^4 (2k!)^{n-3}}\int  \prod^{4}_{\alpha=1}d^k\sigma_{\alpha} \prod^{n-3}_{\beta=1}d^{2k}\tilde{\sigma}_{\beta}\prod^{k}_{i=1}e^{2\pi i \eta_{\alpha}\sigma^i_{\alpha}} \prod^{2k}_{p=1}e^{2\pi i \tilde{\eta}_{\beta}\tilde{\sigma}^p_{\beta}}\\
&\times\frac{1}{\prod^{k}_{i=1}\cosh{\pi(\sigma_1^i+m^f)} \prod^{k}_{i=1}\cosh{\pi(\sigma_2^i+m^f)}}\\
&\times\frac{\prod_{i<j}\sinh^2{\pi(\sigma^i_1-\sigma^j_1)}\sinh^2{\pi(\sigma^i_2-\sigma^j_2)}}{\prod_{i,p}\cosh{\pi(\sigma^i_1-\tilde{\sigma}^p_{1}+m_1)}\cosh{\pi(\sigma^i_2-\tilde{\sigma}^p_{1}+m_2)} } \frac{\prod^{n-3}_{\beta=1}\prod_{p<l}\sinh^2{\pi(\tilde{\sigma}^p_{\beta}-\tilde{\sigma}^l_{\beta})}}{\prod^{n-4}_{\beta=1}\prod_{p,l} \cosh{\pi(\tilde{\sigma}^p_{\beta}-\tilde{\sigma}^l_{\beta+1}+M_{\beta})} }\\
&\times \frac{\prod_{i<j}\sinh^2{\pi(\sigma^i_3-\sigma^j_3)}\sinh^2{\pi(\sigma^i_4-\sigma^j_4)}}{\prod_{i,p}\cosh{\pi(\sigma^i_3-\tilde{\sigma}^p_{n-3}+m_3)}\cosh{\pi(\sigma^i_4-\tilde{\sigma}^p_{n-3}+m_4)} } \label{pfAm2_1}
\end{split}
\end{equation}
\begin{equation}
\begin{split}
Z_B&=\int \frac{d^k\sigma_1 d^k\sigma_2}{(2^k k!)^2} \frac{1}{\prod^{n-1}_{a=1}\prod^k_{i=1}\cosh{\pi(\sigma_2^i+m_a/2)}\cosh{\pi(\sigma_2^i-m_a/2)}\cosh{\pi(\sigma_1^i+m_n/2)}\cosh{\pi(\sigma_1^i-m_n/2)}}\\
&\times \frac{\prod_{\alpha=1,2}\prod_{i<j}\sinh^2{\pi(\sigma^i_{\alpha}-\sigma^j_{\alpha})}\sinh^2{\pi(\sigma^i_{\alpha}+\sigma^j_{\alpha})}\prod_i \sinh^2{\pi(2\sigma^i_{\alpha})}}{\prod_{i,j}\cosh{\pi(\sigma_1^i+\sigma_2^j+M_{bif})}\cosh{\pi(\sigma_1^i+\sigma_2^j-M_{bif})}\cosh{\pi(\sigma_1^i-\sigma_2^j+M_{bif})} \cosh{\pi(\sigma_1^i-\sigma_2^j-M_{bif})}} \label{pfBm2_1}
\end{split}
\end{equation}
Note that the number of independent mass parameters in the A-model --- after taking into account the $\mathbb{Z}_2$ discrete gauge symmetry (which imposes $m_1=m_2, \eta_1=\eta_2$ on the one hand and $m_3=m_4$, $\eta_3=\eta_4$ on the other, in addition to the two fundamental hypers having equal masses ) --- is zero. One can see this directly, from the partition function of A, by shifting the variables $\{\sigma_{\alpha}\},\{\tilde{\sigma}_{\beta}\}$ by constants. This matches the number of FI parameters in the B-model.

The number of independent FI parameters in the A-model is $(n-3) +2=n-1$, after taking into account the $\mathbb{Z}_2$ discrete gauge symmetry. As we will see, this again matches the total number of mass parameters in the B-model.

The dimensions of the Coulomb branches and the Higgs branches of the dual models, as well as the number of FI and mass parameters in each case are summarized in table \ref{tab:D-Atable}.

\begin{table}[h]
\centering
\renewcommand{\arraystretch}{1.5}
\begin{tabular}{|c|c|c|c|c|}     \hline
Model   & $\mbox{dim} M_C$ & $\mbox{dim} M_H$ & $n_{\text{FI}}$ & $n_{\text{mass}}$ \\ \hline
       A  & $2k(n-1)$ & $2k$ & $n-1$ & $0$ \\  \hline     
       B &  $2k$ & $2k(n-1)$ & $0$ & $n-1$ \\ \hline 
\end{tabular}
\caption{The dimension of the Coulomb and Higgs branches
and the number of mass and FI parameters of A and B models for the M-theory background $\mathbb{C}^2/ D_{n} \times \mathbb{C}^2/\mathbb{Z}_2$ with generic $k$ for Case 1}
\label{tab:D-Atable}
\end{table}

To show the equality of $Z_A$ and $Z_B$, one uses a ``doubling trick" for the two boundaries of the quiver A: define $\tilde{\sigma}^p_0=(\sigma^i_1, \sigma^i_2)$ and $\tilde{\sigma}^p_{n-3}=(\sigma^i_3, \sigma^i_4)$. $Z_A$ (modulo the numerical pre-factors) can now be re-written as,
\begin{equation}
\begin{split}
Z_A& \propto \int   \prod^{n-2}_{\beta=0}d^{2k}\tilde{\sigma}_{\beta} \prod^{2k}_{p=1}e^{2\pi i \tilde{\eta}_{\beta}\tilde{\sigma}^p_{\beta}} \frac{1}{\prod^{2k}_{p=1}\cosh{\pi\tilde{\sigma}_0^p}}\left(\frac{\prod_{i<j}\sinh{\pi(\tilde{\sigma}^i_0-\tilde{\sigma}^j_0)}\sinh{\pi(\tilde{\sigma}^{k+i}_0-\tilde{\sigma}^{k+j}_0)}}{\prod_{i,j}\sinh{\pi(\tilde{\sigma}^i_0-\tilde{\sigma}^{k+j}_0)}}\right)\\
&\times\prod^{n-3}_{\beta=0} \frac{\prod_{p<l}\sinh{\pi(\tilde{\sigma}^p_{\beta}-\tilde{\sigma}^l_{\beta})}\sinh{\pi(\tilde{\sigma}^p_{\beta+1}-\tilde{\sigma}^l_{\beta+1})}}{\prod_{p,l} \cosh{\pi(\tilde{\sigma}^p_{\beta}-\tilde{\sigma}^l_{\beta+1})} }\left(\frac{\prod_{i<j}\sinh{\pi(\tilde{\sigma}^i_{n-2}-\tilde{\sigma}^j_{n-2})}\sinh{\pi(\tilde{\sigma}^{k+i}_{n-2}-\tilde{\sigma}^{k+j}_{n-2})}}{\prod_{i,j}\sinh{\pi(\tilde{\sigma}^i_{n-2}-\tilde{\sigma}^{k+j}_{n-2})}}\right)\\ \label{intpfAm2_1}
\end{split}
\end{equation}
As described in the appendix, the terms in parenthesis (which can be thought of as boundary contributions from a Type IIB perspective), can be re-written, using equations \eqref{Cauchysinh},\eqref{FTcosec}, as follows:
\begin{flalign}
&\frac{\prod_{i<j}\sinh{\pi(\tilde{\sigma}^i_0-\tilde{\sigma}^j_0)}\sinh{\pi(\tilde{\sigma}^{k+i}_0-\tilde{\sigma}^{k+j}_0)}}{\prod_{i,j}\sinh{\pi(\tilde{\sigma}^i_0-\tilde{\sigma}^{k+j}_0)}}=(-1)^{k(k-1)/2}\sum_{\rho} (-1)^{\rho} \frac{1}{\prod_i \sinh{\pi(\tilde{\sigma}^i_0-\tilde{\sigma}^{k+\rho(i)}_0)}}& \nonumber\\
&=(-1)^{k(k-1)/2} i^k\int d^k\tau\left(\sum_{\rho} (-1)^{\rho} \prod_i \tanh{\pi \tau^i}e^{2\pi i \tau^i (\tilde{\sigma}^i_0-\tilde{\sigma}^{k+\rho(i)}_0)}\right)
\end{flalign}
The boundary term involving $\tilde{\sigma}^p_{n-2}$ can be re-written in precisely the same way. Therefore, introducing auxiliary variables through Fourier transformations and using the identity (\ref{Cauchycosh}), $Z_A$ can be expressed as,
\begin{equation}
\begin{split}
&Z_A= i^{2k}\int  \frac{d^k\tau d^k \tau^{'}}{(k!)^4 (2k!)^{n-3}} \prod^{n-1}_{\beta=0}d^{2k}\tilde{\sigma}_{\beta} \prod^{n-2}_{\beta=0}d^{2k}\tilde{\tau}_{\beta} \prod^{n-1}_{\beta=0}\prod^{2k}_{p=1}e^{2\pi i {\zeta}_{\beta}\tilde{\sigma}^p_{\beta}}\left(\sum_{\rho} (-1)^{\rho} \prod_i \tanh{\pi \tau^i}e^{2\pi i \tau^i (\tilde{\sigma}^i_0-\tilde{\sigma}^{k+\rho(i)}_0)}\right)\\
&\times\frac{1}{2k!} \left(\sum_{\rho_0}(-1)^{\rho_0} \frac{e^{2\pi i \tilde{\tau}^p_0(\tilde{\sigma}^p_0-\tilde{\sigma}^{\rho_0(p)}_1)}}{\cosh{\pi \tilde{\sigma}^p_1}}\right)\prod^{n-2}_{\beta=1} \left(\sum_{\rho_{\beta}}(-1)^{\rho_{\beta}} \frac{e^{2\pi i \tilde{\tau}^p_{\beta}(\tilde{\sigma}^p_{\beta}-\tilde{\sigma}^{\rho_{\beta}(p)}_{\beta+1})}}{\cosh{\pi \tilde{\tau}^p_{\beta}}}\right)\\
&\times \left(\sum_{\rho{'}} (-1)^{\rho{'}} \prod_i \tanh{\pi \tau'^{i}}e^{2\pi i \tau'^{i} (\tilde{\sigma}^i_{n-1}-\tilde{\sigma}^{k+\rho{'}(i)}_{n-1})}\right)
\end{split}
\end{equation}
Note that we have introduced an extra pair of $(\tilde{\sigma}, \tilde{\tau})$ variables in the above action, such that we have $n$ pairs of variables $(\tilde{\sigma}_{\beta}, \tilde{\tau}_{\beta})$ as opposed to $n-1$ such pairs in equation \eqref{intpfAm2_1}. The FI parameters $\{\zeta_{\beta}\}$ are related to the original ones as $\zeta_0=0,\zeta_1=\eta_1=\eta_2, \zeta_{\beta}=\tilde{\eta}_{\beta-1} (\beta=2,3,...,n-2),  \zeta_{n-1}=\eta_3=\eta_4$.

Next, we integrate out the variables  $\{\tilde{\sigma}_{\beta}\}$ and use the resulting $\delta$-function conditions to integrate out $n-3$ of the variables $\{\tilde{\tau}_{\beta}\}$, thereby trivializing $n-3$ sums over permutations. The integrand therefore gets a multiplicative factor $(2k!)^{n-3}$.
\begin{equation}
\begin{split}
Z_A&= \int \frac{i^{2k}d^{2k}\tilde{\tau}_0 d^{2k}\tilde{\tau}_1 d^k\tau^{'}}{(k!)^4 (2k!)} \left(\sum_{\rho}(-1)^{\rho} \prod_i \tanh{(-\pi \tilde{\tau}^i_0)} \delta(\tilde{\tau}^{k+i}_0+\tilde{\tau}^{\rho^{-1}(i)}_0)\right)\\
&\times \left(\sum_{\rho_0} (-1)^{\rho_0} \frac{1}{\prod_p \cosh{\pi (\tilde{\tau}^p_1-\tilde{\tau}^{\rho^{-1}_0(p)}_0 +\eta_1)}} \right)\\
&\times \frac{1}{\prod_p \cosh{\pi \tilde{\tau}^p_1} \cosh{\pi(\tilde{\tau}^p_1 - \tilde{\eta}_1)}  \cosh{\pi(\tilde{\tau}^p_1 - \tilde{\eta}_1-\tilde{\eta}_2)}...........\cosh{\pi(\tilde{\tau}^p_1 - \tilde{\eta}_1-\tilde{\eta}_2-...-\tilde{\eta}_{n-3})}}\\
&\times \sum_{\rho^{'}, \tilde{\rho}} (-1)^{\rho^{'}+\tilde{\rho}}\prod_i \tanh{\pi\tau^{'i}}\delta(\tau^{'i}-\tilde{\tau}^{\tilde{\rho}^{-1}(i)}_1+\eta+\eta_3)\delta(\tau^{'\rho^{'-1}(i)}+\tilde{\tau}^{\tilde{\rho}^{-1}(k+i)}_1-\eta -\eta_3)
\end{split}
\end{equation}
where $\eta=\sum^{n-3}_{\beta=1}\tilde{\eta}_{\beta}$. While $\rho$ and $\rho^{'}$ are permutations over $k$ objects, $\rho_0$ and $\tilde{\rho}$ are permutations over $2k$ objects.

Now, redefining $\tilde{\tau}^i_1 \to \tilde{\tau}^{\rho{'}\circ\tilde{\rho}(i)}_1+\eta+\eta_3$, $\tilde{\tau}^{k+i}_1 \to \tilde{\tau}^{\tilde{\rho}(k+i)}_1+\eta+\eta_3$ and $\tilde{\tau}^i_0 \to \tilde{\tau}^{\rho(i)}_0$, one can see that all but one of the permutations can be trivialized by change of variables --- the one that survives is a permutation of $2k$ objects. The integrand, as a result, gets a multiplicative factor of $(k!)^2 2k!$.

Therefore, $Z_A$ is reduced to,
\begin{equation}
\begin{split}
Z_A&=\int \frac{d^{2k}\tilde{\tau}_0 d^{2k}\tilde{\tau}_1}{(2^k k!)^2} \frac{\prod_i \sinh{\pi 2\tilde{\tau}^i_0}}{\prod_p \cosh{\pi \tilde{\tau}^p_0}} \frac{\prod_{p<l} \sinh{\pi(\tilde{\tau}^p_0-\tilde{\tau}^l_0)}\sinh{\pi(\tilde{\tau}^p_1-\tilde{\tau}^l_1)}}{\prod_{p,l} \cosh{\pi(\tilde{\tau}^p_1-\tilde{\tau}^p_0+\eta_1+\eta+\eta_3)} }  \frac{\prod_i \sinh{\pi 2\tilde{\tau}^i_1}}{\prod_i \cosh{\pi \tilde{\tau}^p_1}} \\
& \times \prod^{n-2}_{\beta=1}\frac{1}{\prod_p\cosh{\pi(\tilde{\tau}^p_1+\eta_{\beta} +\eta_3)}} \delta(\tilde{\tau}^i_0+\tilde{\tau}^{k+i}_0) \delta(\tilde{\tau}^i_1+\tilde{\tau}^{k+i}_1)\\
&=Z_B
\end{split}
\end{equation}
where we have defined $\eta_{\beta}=\sum^{\beta}_{i=1} \tilde{\eta}_i$ for $\beta=1,2,...,n-3$(after a relabeling of the A-model FI parameters as $\tilde{\eta}_{\beta} \to \tilde{\eta}_{n-\beta -2}$) and $\eta_{n-3}=0$. The mirror map can now be obtained by comparing the parameters (FI parameters on one side and masses on the other) in the two expressions.

\paragraph*{\bf Mirror Map:} The bi-fundamental mass in the B-model is related to the FI parameters of the A-model as follows: 
\begin{equation}
\boxed{M_{bif}= \eta_1+\eta + \eta_3; \; \; \; \eta=\sum^{n-3}_{\beta=1}\tilde{\eta}_{\beta}}
\end{equation}
The fundamental mass associated with the $Sp(k)_1$ gauge factor is zero. One of the fundamental masses associated with $Sp(k)_2$ is also zero. The remaining $n-2$ non-zero fundamental masses are given by
\begin{equation}
\boxed{m_{a} = 2(\eta_3 + \sum^{a}_{i=1}\tilde{\eta}_i)\;  (a=1,2,...., n-3), \; m_{n-2}=2\eta_3}
\end{equation}

\subsection{$G^A_{flavor}=U(2)_4$, $G^B_{flavor}=SO(4) \times SO(2n)$}\label{sec:case2}
The mirror dual theories in this case:

\paragraph*{\bf A-model:} $U(k)_1 \times U(k)_2 \times U(2k)^{n-3} \times U(k)_3 \times U(k)_4$ gauge theory with bi-fundamental  matter content given by an extended ${D}_{n}$ quiver diagram. $U(k)_4$ has 2 fundamental hypers with equal masses.

\paragraph*{\bf B-model:} The gauge group consists of $2k$ copies of $SU(2)$ --- $k$ of which have $n$ fundamental hypers each, while the other $k$ $SU(2)$s have 2 fundamental hypers each.The $k$ sets of $n$ hypers have the same $n$ independent masses, while the $k$ sets of 2 fundamental hypers are all massless.

\begin{figure}[htbp]
\begin{center}
\includegraphics[height=3.0in]{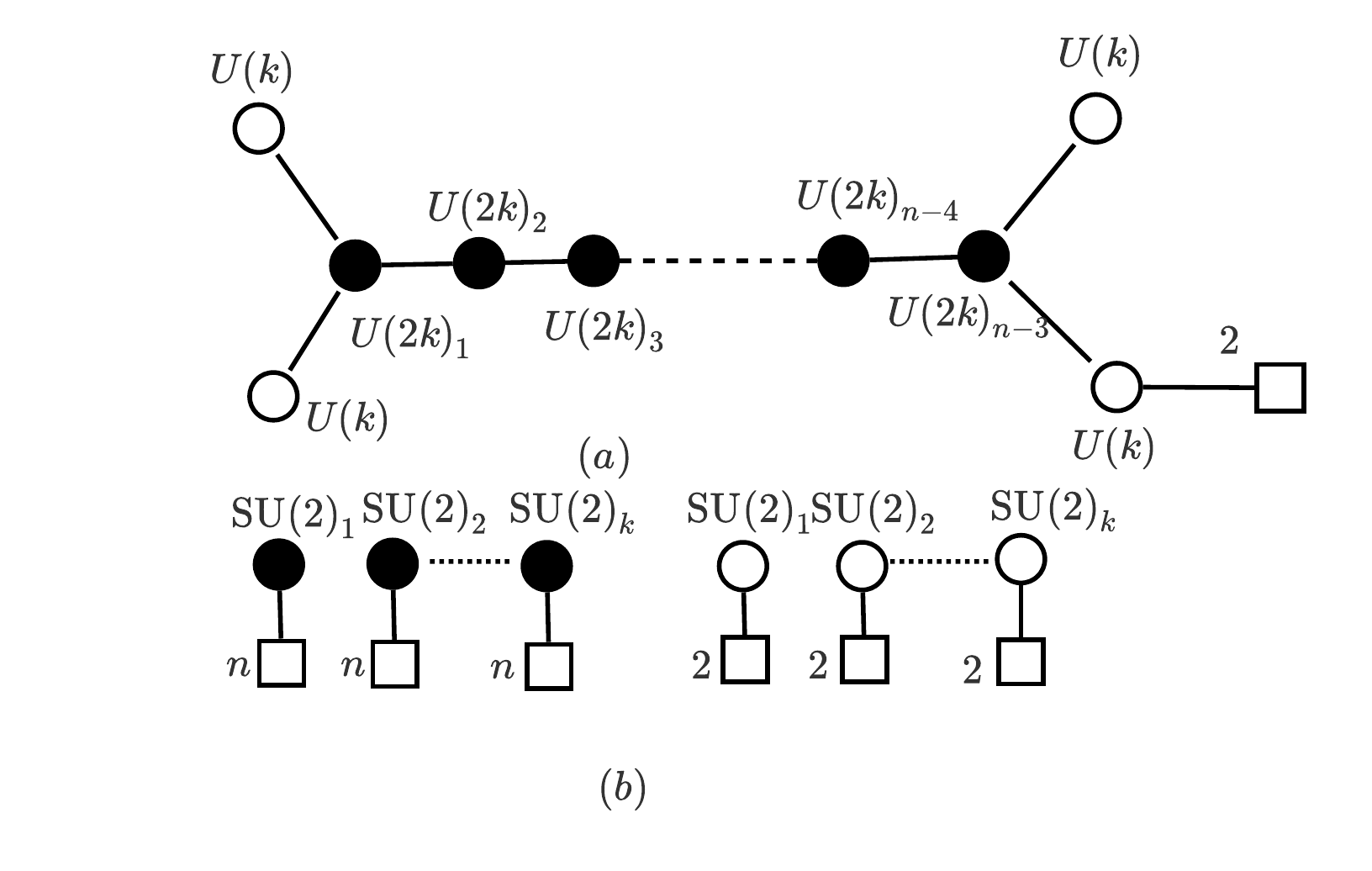}
\caption{Mirror Duals for $\Gamma_1={D}_{n}$,$\Gamma_2=\mathbb{Z}_2$. The A-model has 2 fundamental hypers on one of the $U(k)$ nodes.}
\label{Z2-case2}
\end{center}
\end{figure}

For this family of dual pairs, we work out the equality of the partition functions explicitly for the case $k=1$. The generalization of the proof to $k>1$ theories is straightforward; therefore, we simply state the mirror map in this case.

For $k=1$, the mirror dual theories have the following partition functions:
\begin{equation}
\begin{split}
Z_A&=\frac{1}{(2!)^{n-3}}\int  \prod^{4}_{\alpha=1}d\sigma_{\alpha}e^{2\pi i \eta_{\alpha}\sigma_{\alpha}} \prod^{n-3}_{\beta=1}d^2\tilde{\sigma}_{\beta} \prod^{2}_{i=1}e^{2\pi i \tilde{\eta}_{\beta}\tilde{\sigma}^i_{\beta}} \frac{1}{\cosh{\pi(\sigma_4+m_1^f)} \cosh{\pi(\sigma_4+m_2^f)}}\\
&\times\frac{1}{\prod_{i}\cosh{\pi(\sigma_1-\tilde{\sigma}^i_{1}+m_1)}\cosh{\pi(\sigma_2-\tilde{\sigma}^i_{1}+m_2)} } \frac{\prod^{n-3}_{\beta=1}\sinh^2{\pi(\tilde{\sigma}^1_{\beta}-\tilde{\sigma}^2_{\beta})}}{\prod^{n-4}_{\beta=1}\prod_{i,j} \cosh{\pi(\tilde{\sigma}^i_{\beta}-\tilde{\sigma}^j_{\beta+1}+M_{\beta})} }\\
&\times \frac{1}{\prod_{i}\cosh{\pi(\sigma_3-\tilde{\sigma}^i_{n-3}+m_3)}\cosh{\pi(\sigma_4-\tilde{\sigma}^i_{n-3}+m_4)} } \label{pfAm2_2}
\end{split}
\end{equation}

\begin{equation}
\begin{split}
Z_B=\int \frac{d^2\sigma_1}{2} \frac{d^2\sigma_2}{2} \frac{\prod^2_{\alpha=1}\delta(\sigma^1_{\alpha}+\sigma^2_{\alpha})}{\prod^{n}_{a=1}\prod^2_{i=1}\cosh{\pi(\sigma_1^i+m^{(1)}_a)}}\frac{\prod_{\alpha=1,2}\sinh^2{\pi(\sigma^1_{\alpha}-\sigma^2_{\alpha})}}{\prod_{b=1,2}\prod_{i}\cosh{\pi(\sigma^i_2 +m^{(2)}_b)}} \label{pfBm2_1}
\end{split}
\end{equation}
In the partition function of the A-model, all bi-fundamental masses (and one of the fundamental masses) can be set to zero by shifting integration variables by constants. Therefore, in addition to the FI parameters, it is sufficient to consider $Z_A$ as an explicit function of only the difference of the two fundamental masses. We later show that the masses have to be the same for the proposed duality to hold, which naturally implies that $Z_A$ does not depend on any of the mass parameters. Also, one needs to impose the following condition on the FI parameters of the A-model : $\sum^4_{\alpha=1} \eta_{\alpha} + 2\sum^{n-3}_{\beta=1} \tilde{\eta}_{\beta}=0$, for the duality to hold.

The counting evidence for the duality is summarized in table \ref{D-Atable2}.

\begin{table}[h]
\centering
\renewcommand{\arraystretch}{1.5}
\begin{tabular}{|c|c|c|c|c|}     \hline
Model   & $\mbox{dim} M_C$ & $\mbox{dim} M_H$ & $n_{\text{FI}}$ & $n_{\text{mass}}$ \\ \hline
       A  & $2k(n-1)$ & $2k$ & $n$ & $0$ \\  \hline     
       B &  $2k$ & $2k(n-1)$ & $0$ & $n$ \\ \hline 
\end{tabular}
\caption{The dimension of the Coulomb and Higgs branches
and the number of mass and FI parameters of A and B models for the M-theory background $\mathbb{C}^2/ D_{n} \times \mathbb{C}^2/\mathbb{Z}_2$ with generic $k$ for Case 2.}
\label{D-Atable2}
\end{table}

To prove the proposed duality, we consider the partition function $Z_A$ and ``integrate out" the two nodes corresponding to the variables $\sigma_2$ and $\sigma_4$, we get,
\begin{equation}
\begin{split}
Z_A&= \frac{2^2 i^2}{(2!)^{n-3}\sinh{\pi \eta_2}\sinh{\pi \eta_4}}\int  \prod_{\alpha=1,3}d\sigma_{\alpha} e^{2\pi i \eta_{\alpha}\sigma_{\alpha}}\prod^{n-3}_{\beta=1}d^2\tilde{\sigma}_{\beta} \prod^{2}_{i=1}e^{2\pi i \tilde{\eta}_{\beta}\tilde{\sigma}^i_{\beta}}e^{2\pi i \eta_2 \tilde{\sigma}^1_1 }e^{2\pi i \eta_4 \tilde{\sigma}^1_{n-3} }\\
&\times \frac{1}{\prod_{i}\cosh{\pi(\sigma_1-\tilde{\sigma}^i_{1})}} \prod^{n-4}_{\beta=1} \frac{ \sinh{\pi(\tilde{\sigma}^1_{\beta}-\tilde{\sigma}^2_{\beta})}\sinh{\pi(\tilde{\sigma}^1_{\beta+1}-\tilde{\sigma}^2_{\beta+1})}}{\prod_{i,j}\cosh{\pi(\tilde{\sigma}^i_{\beta}-\tilde{\sigma}^j_{\beta+1})}}\frac{1}{\prod_{i}\cosh{\pi(\sigma_3-\tilde{\sigma}^i_{n-3})}}  \\
&\times \frac{1}{ \sinh{\pi(\tilde{\sigma}^1_{n-3}-m^1_f)}\sinh{\pi(\tilde{\sigma}^1_{n-3}-m^2_f)}}
\end{split}
\end{equation}
On using Cauchy's determinant formula (equation \eqref{Cauchycosh}) and Fourier transforming, we get
\begin{equation}
\begin{split}
Z_A& \propto \int  \prod_{\alpha=1,3}d\sigma_{\alpha}e^{2\pi i \eta_{\alpha}\sigma_{\alpha}} \prod^{n-3}_{\beta=1}d^{2}\tilde{\sigma}_{\beta}\prod^{2}_{i=1}e^{2\pi i \tilde{\eta}_{\beta}\tilde{\sigma}^i_{\beta}} \prod^{n-4}_{\beta=1}d^2\tilde{\tau}_{\beta}\; d^{2} \tau_0 d^{2}\tau_{n-3} \;d^2\tau \;e^{2\pi i \eta_2 \tilde{\sigma}^1_1 }e^{2\pi i \eta_4 \tilde{\sigma}^1_{n-3} } \\
&\times \left(\prod_i \frac{e^{2\pi i \tau^i_1(\sigma_{1}- \tilde{\sigma}^{i}_1)}}{\cosh{\pi \tau^i_1}} \right) \prod^{n-4}_{\beta=1}  \left(\sum_{\tilde{\rho}_{\beta}}(-1)^{\tilde{\rho}_{\beta}}\prod_i \frac{e^{2\pi i \tilde{\tau}^i_{\beta}(\tilde{\sigma}^i_{\beta}- \tilde{\sigma}^{\tilde{\rho}_{\beta}(i)}_{\beta +1})}}{\cosh{\pi \tilde{\tau}^i_{\beta}}} \right) \left(\prod_i \frac{e^{2\pi i \tau^i_{n-3}(\sigma_{3}- \tilde{\sigma}^{i}_{n-3})}}{\cosh{\pi \tau^i_{n-3}}} \right)\\
&\times \left(i \tanh{\pi \tau^1}e^{2\pi i \tau^1(\tilde{\sigma}^{1}_{n-3}-m^1_f)}\right)\left(i \tanh{\pi \tau^1}e^{2\pi i \tau^2(\tilde{\sigma}^{1}_{n-3}-m^2_f)}\right)
\end{split}
\end{equation}
where we have suppressed the constant pre-factors in the formula for $Z_A$.

On integrating over $\sigma_1$ and $\tilde{\sigma}^i_{1}$, we respectively obtain the following $\delta$-function conditions:
\begin{eqnarray}
\tau^1_0 +\tau^2_0 +\eta_1=0\\
\tau^1_0=\tilde{\tau}^1_1 +\tilde{\eta}_1 +\eta_2,\; \tau^1_0=\tilde{\tau}^2_1 +\tilde{\eta}_1
\end{eqnarray}
which on integrating over $\tau^i_0$ can be combined to give the condition,
\begin{equation}
\tilde{\tau}^1_1+\tilde{\tau}^2_1+2\tilde{\eta}_1+\eta_1+\eta_2=0 \label{Bd1}
\end{equation}
Similarly, on integrating over $\tilde{\sigma}^i_{\beta}$ and $\sigma_3$, we get
\begin{eqnarray}
\tilde{\tau}^i_{\beta}=\tilde{\tau}^{\tilde{\rho}^{-1}_{1}..... \tilde{\rho}^{-1}_{\beta -1}(i)}_{1} -\sum^{\beta}_{a=2} \tilde{\eta}_{a} \; \; (\beta=2,3,....,n-4)\\
{\tau}^1_{n-3}=-\tilde{\tau}^{\tilde{\rho}^{-1}_{1}..... \tilde{\rho}^{-1}_{n-4}(1)}_{1} +\eta_4 +\sum^{n-3}_{a=2} \tilde{\eta}_{a} + \tau_1 +\tau_2 \label{Bd2.1} \\
{\tau}^2_{n-3}=-\tilde{\tau}^{\tilde{\rho}^{-1}_{1}..... \tilde{\rho}^{-1}_{n-4}(2)}_{1} +\sum^{n-3}_{a=2} \tilde{\eta}_{a}\label{Bd2.2}\\
\tau^1_{n-3} +\tau^2_{n-3} +\eta_3=0 \label{Bd2.3}
\end{eqnarray}
Integrating over $\tau^i_{n-3}$, equations \eqref{Bd2.1}-\eqref{Bd2.3} and \eqref{Bd1} would imply,
\begin{equation}
\tau_1+\tau_2=\sum^4_{\alpha=1} \eta_{\alpha} + 2\sum^{n-3}_{\beta=1} \tilde{\eta}_{\beta}=0
\end{equation}
In the last equation, we imposed the condition $\sum^4_{\alpha=1} \eta_{\alpha} + 2\sum^{n-3}_{\beta=1} \tilde{\eta}_{\beta}=0$ on the FI parameters of the A-model. This reduces the total number of independent FI parameters in the A-model to $n$, as noted in table \ref{D-Atable3}. Therefore, imposing the $\delta$-function conditions and trivializing appropriate sums over permutations as before, $Z_A$ reduces to
\begin{equation}
\begin{split}
&Z_A =  \int \frac{d^2 \tau d^{2} \tilde{\tau}_1\delta(\tilde{\tau}_1^1 +\tilde{\tau}_1^{2})\delta(\tau^1+\tau^2)}{\sinh{\pi \eta_2}\sinh{\pi \eta_4}} \left(\sum_{\rho^{'}}(-1)^{\rho^{'}} \frac{1}{\cosh{\pi(\tilde{\tau}_1^{\rho^{'}(1)}+(\eta_2-\eta_1)/2)}\cosh{\pi(\tilde{\tau}_1^{\rho^{'}(2)}-(\eta_1+\eta_2)/2)}} \right)\\
&\times \left[\frac{1}{\prod^2_{i=1} \cosh{\pi( \tilde{\tau}_1^i-\frac{(\eta_1+\eta_2)}{2} -\tilde{\eta}_1)} \cosh{\pi (\tilde{\tau}_1^i-\frac{(\eta_1+\eta_2)}{2} -\tilde{\eta}_1-\tilde{\eta}_2)}......\cosh{\pi(\tilde{\tau}_1^i-\frac{(\eta_1+\eta_2)}{2} -\sum^{n-4}_{\beta=1}\tilde{\eta}_{\beta})}} \right]\\
&\times \left(\sum_{\rho}(-1)^{\rho} \frac{1}{\cosh{\pi(\tilde{\tau}_1^{\rho(1)} -\frac{(\eta_1+\eta_2)}{2} -\eta_4-\sum^{n-3}_{\beta=1}\tilde{\eta}_{\beta})}\cosh{\pi(\tilde{\tau}_1^{\rho(2)}-\frac{(\eta_1+\eta_2)}{2}-\sum^{n-3}_{\beta=1}\tilde{\eta}_{\beta})}} \right)\\
&\times \frac{-\sinh^2{\pi(\tau^1-\tau^2)}}{2^2\prod_i \cosh^2{\pi \tau^i} } e^{-2\pi i \tau^1 m^1_f -2\pi i \tau^2 m^2_f}
\end{split}
\end{equation}
Finally, summing over the permutations using equation \eqref{Cauchycosh}, we obtain
\begin{equation}
\begin{split}
&Z_A =  \frac{1}{4}\int d^2 \tau d^{2} \tilde{\tau}_1 \delta(\tau^1+\tau^2)\delta(\tilde{\tau}_1^1 +\tilde{\tau}_1^{2})\left( \frac{\sinh{\pi(\tilde{\tau}_1^1 -\tilde{\tau}_1^{2})}}{\prod_i\cosh{\pi(\tilde{\tau}_1^{i}+(\eta_2-\eta_1)/2)}\cosh{\pi(\tilde{\tau}_1^{i}-(\eta_1+\eta_2)/2)}} \right)\\
&\times \left[\frac{1}{\prod^2_{i=1} \cosh{\pi( \tilde{\tau}_1^i-\frac{(\eta_1+\eta_2)}{2} -\tilde{\eta}_1)} \cosh{\pi (\tilde{\tau}_1^i-\frac{(\eta_1+\eta_2)}{2} -\tilde{\eta}_1-\tilde{\eta}_2)}......\cosh{\pi(\tilde{\tau}_1^i-\frac{(\eta_1+\eta_2)}{2} -\sum^{n-4}_{\beta=1}\tilde{\eta}_{\beta})}} \right]\\
&\times \left( \frac{\sinh{\pi(\tilde{\tau}_1^1 -\tilde{\tau}_1^{2})}}{\prod_i\cosh{\pi(\tilde{\tau}_1^{i} -\frac{(\eta_1+\eta_2)}{2} -\eta_4-\sum^{n-3}_{\beta=1}\tilde{\eta}_{\beta})}\cosh{\pi(\tilde{\tau}_1^{i}-\frac{(\eta_1+\eta_2)}{2}-\sum^{n-3}_{\beta=1}\tilde{\eta}_{\beta})}} \right)\\
&\times \frac{\sinh^2{\pi(\tau^1-\tau^2)}}{\prod_i \cosh^2{\pi \tau^i} }e^{-2\pi i \tau^1 m^1_f -2\pi i \tau^2 m^2_f}\\
&= Z_B,\; \;\;\;\mbox{iff} \;\;\; m^1_f=m^2_f
\end{split}
\end{equation}
The duality holds only if the two fundamental masses for the A-model are equal to each other, which, in turn, implies that the A-model does not have any independent mass parameters. This is expected since the B-model does not have any FI parameters.

\paragraph*{\bf Mirror Map:} The mirror map, relating FI parameters of the A-model to the mass parameters of the B-model, can now be obtained by comparing the parameters appearing in the two formul\ae. The $n$ fundamental masses for the first SU(2)s are given by
\begin{equation}
\boxed{\begin{gathered}
m^{(1)}_a=\frac{\eta_1+\eta_2}{2}+\sum^{a}_{\beta=1}\tilde{\eta}_{\beta}\;(a=1,2,..,n-3);m^{(1)}_{n-2}=\eta_4+\frac{\eta_1+\eta_2}{2}+\sum^{n-3}_{\beta=1}\tilde{\eta}_{\beta};\\
m^{(1)}_{n-1}=\frac{\eta_1+\eta_2}{2};m^{(1)}_{n}=\frac{\eta_1-\eta_2}{2}
\end{gathered}} \label{map2.2}
\end{equation}
The 2 fundamental hypers of the second SU(2)s are massless.
\begin{equation}
\boxed{m^{(2)}_1=0, m^{(2)}_2=0}
\end{equation}

For $k>1$, one has $k$ copies of $n$ hypers transforming in the fundamental of $k$ different $SU(2)$s and another $k$ copies of $2$ hypers transforming in the fundamental of another set of $k$ $SU(2)$s.
The $k$ copies of $n$ hypers have the same set of masses while the $k$ copies of $2$ hypers are all massless, so that we still have $n$ independent mass parameters in the B-model to match the $n$ independent FI parameters of the A-model. The mirror map for the case of general $k$ is exactly the same as in the $k=1$ case.

\subsection{$G^A_{flavor}=U(1)_1\times U(1)_3$, $G^B_{flavor}=U(n)$}\label{sec:case3}
\paragraph*{\bf A-model:} $U(k)_1 \times U(k)_2 \times U(2k)^{n-3} \times U(k)_3 \times U(k)_4$ gauge theory with bi-fundamental  matter content given by an extended ${D}_{n}$ quiver diagram. $U(k)_1$ and $U(k)_3$ gauge groups have one fundamental hyper each.

\paragraph*{\bf B-model:} $U(2k)$ gauge theory with $n$ fundamental hypers and two hypers in the antisymmetric representation of $U(2k)$. For $k=1$, the $U(2)$ gauge theory has $n$ fundamental hypers and two hypers which are singlets of the gauge group. \\
In this subsection, we present the computation for the generic $k>1$ case; the special case of $k=1$ can be handled in a similar fashion.

\begin{figure}[htbp]
\begin{center}
\includegraphics[height=3.0in]{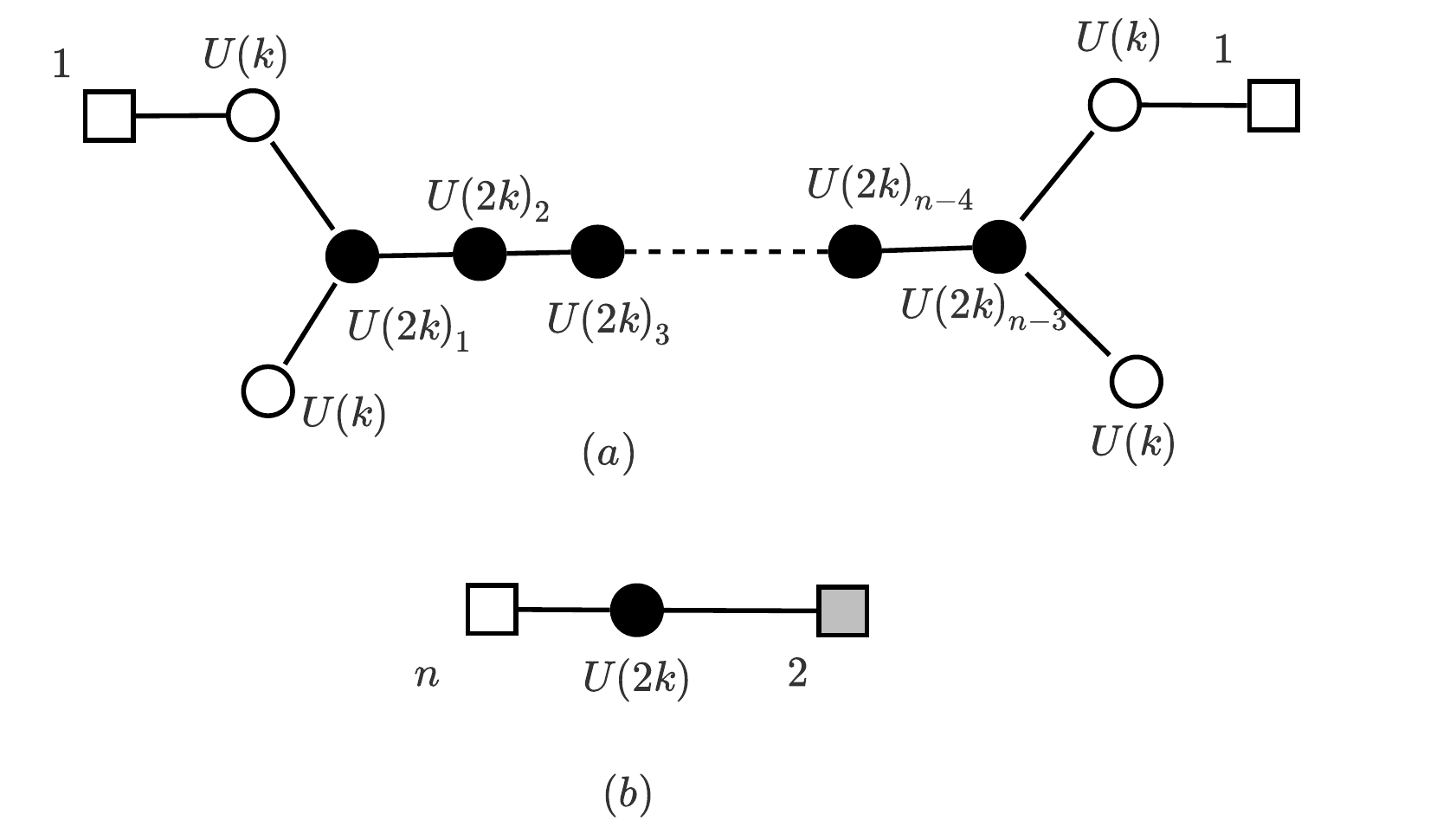}
\caption{Mirror Duals for $\Gamma_1={D}_{n}$,$\Gamma_2=\mathbb{Z}_2$. The shaded box in (b) denotes 2 antisymmetric hypers for $k>1$. For $k=1$, one has 2 singlets instead.}
\label{Z2-case3}
\end{center}
\end{figure}
\begin{equation}
\begin{split}
Z_A&=\frac{1}{(k!)^4 (2k!)^{n-3}}\int  \prod^{4}_{\alpha=1}d^k\sigma_{\alpha} \prod^{n-3}_{\beta=1}d^{2k}\tilde{\sigma}_{\beta}\prod^{k}_{i=1}e^{2\pi i \eta_{\alpha}\sigma^i_{\alpha}} \prod^{2k}_{p=1}e^{2\pi i \tilde{\eta}_{\beta}\tilde{\sigma}^p_{\beta}}\\
&\times\frac{1}{\prod^{k}_{i=1}\cosh{\pi(\sigma_1^i+m_1^f)} \prod^{k}_{i=1}\cosh{\pi(\sigma_3^i+m_3^f)}}\\
&\times\frac{\prod_{i<j}\sinh^2{\pi(\sigma^i_1-\sigma^j_1)}\sinh^2{\pi(\sigma^i_2-\sigma^j_2)}}{\prod_{i,p}\cosh{\pi(\sigma^i_1-\tilde{\sigma}^p_{1}+m_1)}\cosh{\pi(\sigma^i_2-\tilde{\sigma}^p_{1}+m_2)} } \frac{\prod^{n-3}_{\beta=1}\prod_{p<l}\sinh^2{\pi(\tilde{\sigma}^p_{\beta}-\tilde{\sigma}^l_{\beta})}}{\prod^{n-4}_{\beta=1}\prod_{p,l} \cosh{\pi(\tilde{\sigma}^p_{\beta}-\tilde{\sigma}^l_{\beta+1}+M_{\beta})} }\\
&\times \frac{\prod_{i<j}\sinh^2{\pi(\sigma^i_3-\sigma^j_3)}\sinh^2{\pi(\sigma^i_4-\sigma^j_4)}}{\prod_{i,p}\cosh{\pi(\sigma^i_3-\tilde{\sigma}^p_{n-3}+m_3)}\cosh{\pi(\sigma^i_4-\tilde{\sigma}^p_{n-3}+m_4)} } \label{pfAm2_1}
\end{split}
\end{equation}
\begin{equation}
\begin{split}
Z_B&=\int \frac{d^{2k}\sigma}{2k!}  \prod^{2k}_{p=1} e^{2\pi i \eta \sigma^p}\frac{\prod_{p<l}\sinh^2{\pi(\sigma^p-\sigma^l)}}{\prod_{a=1,2}\prod_{p<l}\cosh{\pi(\sigma^p+\sigma^l-M^{as}_{a})}\prod^{n}_{b=1}\prod^{2k}_{p=1}\cosh{\pi(\sigma^p+M^f_b)}}
\end{split}
\end{equation}
All the bi-fundamental masses in $Z_A$ can be set to zero by shifting integration variables by appropriate constants, as before. In addition, one can get rid of one of the fundamental masses, which shows that the total number of independent mass parameters in the A-model is one. However, in the following computation, we retain both fundamental masses --- $m^f_1$ and $m^f_3$ --- and show that only a particular linear combination of the masses is physically relevant.

The counting evidence for the duality summarized in table \ref{D-Atable3}.

\begin{table}[h]
\centering
\renewcommand{\arraystretch}{1.5}
\begin{tabular}{|c|c|c|c|c|}     \hline
Model   & $\mbox{dim} M_C$ & $\mbox{dim} M_H$ & $n_{\text{FI}}$ & $n_{\text{mass}}$ \\ \hline
       A  & $2k(n-1)$ & $2k$ & $n+1$ & $1$ \\  \hline     
       B &  $2k$ & $2k(n-1)$ & $1$ & $n+1$ \\ \hline 
\end{tabular}
\caption{The dimension of the Coulomb and Higgs branches
and the number of mass and FI parameters of A and B models for the M-theory background $\mathbb{C}^2/ D_{n} \times \mathbb{C}^2/\mathbb{Z}_2$ with generic $k$ for Case 3.}
\label{D-Atable3}
\end{table}

To begin with, we ``integrate out" the variables $\sigma_2$ and $\sigma_4$ corresponding to the $U(k)$ nodes which do not have any fundamental hyper. The partition function of the A model can then be written as,
\begin{equation}
\begin{split}
Z_A&=\frac{i^{-2k}e^{-2\pi i k \eta_2 m_2 }e^{-2\pi i k \eta_4 m_4}}{(k!)^6 (2k!)^{n-5}\sinh^k{\pi \eta_2}\sinh^k{\pi \eta_4}}\int  \prod_{\alpha=1,3}d^k\sigma_{\alpha} \prod^{n-3}_{\beta=1}d\tilde{\sigma}^{2k}_{\beta}\prod^{k}_{i=1}e^{2\pi i \eta_{\alpha}\sigma^i_{\alpha}}e^{2\pi i \eta_2 \tilde{\sigma}^i_1 }e^{2\pi i \eta_4 \tilde{\sigma}^i_{n-3} } \prod^{2k}_{p=1}e^{2\pi i \tilde{\eta}_{\beta}\tilde{\sigma}^p_{\beta}}\\
&\times \frac{\prod_{i<j}\sinh{\pi({\sigma}^i_1-{\sigma}^j_1)}\sinh{\pi(\tilde{\sigma}^i_1-\tilde{\sigma}^j_1)}}{\prod_{i,j}\cosh{\pi(\sigma^i_1-\tilde{\sigma}^j_{1})}} \frac{\prod_{i<j}\sinh{\pi({\sigma}^i_1-{\sigma}^j_1)}\sinh{\pi(\tilde{\sigma}^{k+i}_1-\tilde{\sigma}^{k+j}_1)}}{\prod_{i,j}\cosh{\pi(\sigma^i_1-\tilde{\sigma}^{k+j}_{1})}} \\
&\times \prod^{n-4}_{\beta=1} \frac{\prod_{p<l} \sinh{\pi(\tilde{\sigma}^p_{\beta}-\tilde{\sigma}^l_{\beta})}\sinh{\pi(\tilde{\sigma}^p_{\beta+1}-\tilde{\sigma}^l_{\beta+1})}}{\prod_{p,l}\cosh{\pi(\tilde{\sigma}^p_{\beta}-\tilde{\sigma}^l_{\beta+1})}}\\
&\times \frac{\prod_{i<j}\sinh{\pi({\sigma}^i_3-{\sigma}^j_3)}\sinh{\pi(\tilde{\sigma}^i_{n-3}-\tilde{\sigma}^j_{n-3})}}{\prod_{i,j}\cosh{\pi(\sigma^i_3-\tilde{\sigma}^j_{n-3})}} \frac{\prod_{i<j}\sinh{\pi({\sigma}^i_3-{\sigma}^j_3)}\sinh{\pi(\tilde{\sigma}^{k+i}_{n-3}-\tilde{\sigma}^{k+j}_{n-3})}}{\prod_{i,j}\cosh{\pi(\sigma^i_3-\tilde{\sigma}^{k+j}_{n-3})}}\\
&\times \frac{1}{\prod_i \cosh{\pi({\sigma}^i_1+m^f_1)}\prod_i \cosh{\pi({\sigma}^i_3+m^f_3)}}
\end{split}
\end{equation}
As in previous cases, one can introduce auxiliary variables through Fourier transformations in the action such that
\begin{equation}
\begin{split}
Z_A& \propto \int  \prod_{\alpha=1,3}d^k\sigma_{\alpha} \prod^{n-3}_{\beta=1}d^{2k}\tilde{\sigma}_{\beta} \prod^{n-4}_{\beta=1}d\tilde{\tau}^{2k}_{\beta} d^{2k} \tau_1 d^{2k}\tau_{n-3} d^{k}\tau d^k \tau' \prod^{k}_{i=1}e^{2\pi i \eta_{\alpha}\sigma^i_{\alpha}}e^{2\pi i \eta_2 \tilde{\sigma}^i_1 }e^{2\pi i \eta_4 \tilde{\sigma}^i_{n-3} } \prod^{2k}_{p=1}e^{2\pi i \tilde{\eta}_{\beta}\tilde{\sigma}^p_{\beta}}\\
&\times \left(\sum_{\rho_1}(-1)^{\rho_1}\prod_i \frac{e^{2\pi i \tau^i_1(\sigma^i_1- \tilde{\sigma}^{\rho_1(i)}_1+m_1)}}{\cosh{\pi \tau^i_1}} \right)  \left(\sum_{\rho_2}(-1)^{\rho_2}\prod_i \frac{e^{2\pi i \tau^{k+i}_1(\sigma^{i}_1- \tilde{\sigma}^{k+\rho_2(i)}_1+m_1)}}{\cosh{\pi \tau^{k+i}_1}} \right)\\
&\times \prod^{n-4}_{\beta=1}  \left(\sum_{\tilde{\rho}_{\beta}}(-1)^{\tilde{\rho}_{\beta}}\prod_p \frac{e^{2\pi i \tilde{\tau}^p_{\beta}(\tilde{\sigma}^p_{\beta}- \tilde{\sigma}^{\tilde{\rho}_{\beta}(p)}_{\beta +1}+M_{\beta})}}{\cosh{\pi \tilde{\tau}^p_{\beta}}} \right)\\
&\times \left(\sum_{\rho_3}(-1)^{\rho_3}\prod_i \frac{e^{2\pi i \tau^i_{n-3}(\sigma^i_3- \tilde{\sigma}^{\rho_3(i)}_{n-3}+m_3)}}{\cosh{\pi \tau^i_{n-3}}} \right)  \left(\sum_{\rho_4}(-1)^{\rho_4}\prod_i \frac{e^{2\pi i \tau^{k+i}_{n-3}(\sigma^{i}_3- \tilde{\sigma}^{k+\rho_4(i)}_{n-3}+m_3)}}{\cosh{\pi \tau^{k+i}_{n-3}}} \right)\\
&\times \prod_i \frac{e^{2\pi i \tau^i(\sigma^i_1+m^f_1)}}{\cosh{\pi \tau^i}} \times \prod_i \frac{e^{2\pi i \tau'^i(\sigma^i_3+m^f_3)}}{\cosh{\pi \tau'^i}}
\end{split}
\end{equation}
Integrating over the variables $\{\tilde{\sigma}^p_{\beta}\}, \sigma^i_1$ and $\sigma^i_3$ in the action and imposing the resulting $\delta$-function constraints, the partition function for the A-model can be written as,
\begin{equation}
\begin{split}
&Z_A= \int \frac{i^{-2k}d^{2k} \tilde{\tau}_1\prod_p e^{2\pi i (m^f_3-m^f_1)\tilde{\tau}^p_1}}{(k!)^2 (2k!)\sinh^k{\pi \eta_2}\sinh^k{\pi \eta_4}}\\
&\times\left(\sum_{\rho'}\frac{(-1)^{\rho'}}{\prod_i \cosh{\pi (\tilde{\tau}^{\rho'(i)}_1+\tilde{\tau}^{\rho'(k+i)}_1+\eta_1+\eta_2+\eta_3+\eta_4+2(\tilde{\eta}_1+....+\tilde{\eta}_{n-3}))}}\right.\\
&\left.\times \frac{1}{\prod_i \cosh{\pi (\tilde{\tau}^{\rho'(i)}_1+\eta_2+(\eta_3+\eta_4)/2+(\tilde{\eta}_1+....+\tilde{\eta}_{n-3}))}\cosh{\pi (\tilde{\tau}^{\rho'(k+i)}_1+(\eta_3+\eta_4)/2+(\tilde{\eta}_1+....+\tilde{\eta}_{n-3}))}}\right)\\
&\times \frac{1}{\prod_p \cosh{\pi(\tilde{\tau}^p_1+\frac{\eta_3+\eta_4}{2}+\tilde{\eta}_2+....+\tilde{\eta}_{n-3})}}..............\frac{1}{\prod_p \cosh{\pi(\tilde{\tau}^p_1+\frac{\eta_3+\eta_4}{2}+\tilde{\eta}_{n-3})}}\\
&\times \left(\sum_{\rho} (-1)^{\rho} \frac{1}{\prod_i \cosh{\pi(\tilde{\tau}^{\rho(i)}_1+\tilde{\tau}^{\rho({k+i})}_1)}\cosh{\pi(\tilde{\tau}^{\rho(i)}_1+\frac{\eta_3-\eta_4}{2})}\cosh{\pi(\tilde{\tau}^{\rho({k+i})}_1+\frac{\eta_3+\eta_4}{2})}}\right)\\
\end{split}
\end{equation}
The two sums over permutations can be computed using the identity given in equation \eqref{Id2}. In addition, relabeling the FI parameters of the A-model as $\tilde{\eta}_{\beta} \to \tilde{\eta}_{n-\beta -2}$, we get
\begin{equation}
\begin{split}
&Z_A=\int \frac{d^{2k} \tilde{\tau}_1}{2k!}\prod_p e^{2\pi i (m^f_3-m^f_1)\tilde{\tau}^p_1}\left(\frac{\prod_{p<l} \sinh{\pi(\tilde{\tau}^p_1-\tilde{\tau}^l_1))}}{\prod_{p<l}\cosh{\pi(\tilde{\tau}^p_1+\tilde{\tau}^l_1+\eta_1+\eta_2+\eta_3+\eta_4+2(\tilde{\eta}_1+....+\tilde{\eta}_{n-3})}}\right)\\
&\times \frac{1}{\prod_p \cosh{\pi (\tilde{\tau}^{p}_1+\eta_2+(\eta_3+\eta_4)/2+(\tilde{\eta}_1+....+\tilde{\eta}_{n-3}))}\cosh{\pi (\tilde{\tau}^{p}_1+(\eta_3+\eta_4)/2+(\tilde{\eta}_1+....+\tilde{\eta}_{n-3}))}}\\
&\times \frac{1}{\prod_p \cosh{\pi(\tilde{\tau}^p_1+\frac{\eta_3+\eta_4}{2}+\tilde{\eta}_1)}}..............\frac{1}{\prod_p \cosh{\pi(\tilde{\tau}^p_1+\frac{\eta_3+\eta_4}{2}+\tilde{\eta}_1+.......+\tilde{\eta}_{n-4})}}\\
&\times \frac{\prod_{p<l} \sinh{\pi(\tilde{\tau}^p_1-\tilde{\tau}^l_1)}}{\prod_{p<l}\cosh{\pi(\tilde{\tau}^p_1+\tilde{\tau}^l_1)}\prod_p\cosh{\pi(\tilde{\tau}^{p}_1+\frac{\eta_3-\eta_4}{2})}\cosh{\pi(\tilde{\tau}^{p}_1+\frac{\eta_3+\eta_4}{2}})}=Z_B
\end{split}
\end{equation}

\paragraph*{\bf Mirror Map:}
  The FI parameter $\eta$ of the gauge group $U(2k)$ in the B-model is related to the fundamental masses as follows:
\begin{equation}
\boxed{\eta= m^f_3-m^f_1}
\end{equation}
The fundamental masses of the B-model are related to the FI parameters of the A-model in the following manner:
\begin{eqnarray}
\boxed{M^f_b=\frac{\eta_3+\eta_4}{2}+\sum^b_{\beta=1}\tilde{\eta}_{\beta}, \; b=1,2,....., n-3}\\
\boxed{M^f_{n-2}=\eta_2+\frac{\eta_3+\eta_4}{2}+\sum^{n-3}_{\beta=1}\tilde{\eta}_{\beta},\;M^f_{n-1}=\frac{\eta_3+\eta_4}{2}, \; M^f_{n}=\frac{\eta_3-\eta_4}{2}}
\end{eqnarray}
The masses for hypers in the anti-symmetric representation of $U(2k)$ are given by
\begin{equation}
\boxed{M^{as}_1=\eta_1+\eta_2+\eta_3+\eta_4+2\sum^{n-3}_{\beta=1}\tilde{\eta}_{\beta},\;M^{as}_2=0 }
\end{equation}
The number of independent mass parameters in the A-model, therefore, matches the number of FI parameters in the B-model; the only FI parameter is related to the difference of the two fundamental masses in the A-model. The number of independent FI parameters in the A-model is $n+1$, which again matches the number non-zero mass parameters in the B-model, as expected.

\subsection{$G^A_{flavor}=U(1)^{'}_{\alpha}$, $G^B_{flavor}=SO(2\alpha+2)\times SO(2n-2\alpha -2)$}\label{sec:case4}
\textbf{A-model:} $U(k)_1 \times U(k)_2 \times U(2k)^{n-3} \times U(k)_3 \times U(k)_4$ gauge theory with bi-fundamental  matter content given by an extended ${D}_{n}$ quiver diagram. There is one fundamental hyper associated to $U(2k)_{\alpha}$ where $1\leq \alpha \leq n-3 $.The masses and the FI parameters respect the $\mathbb{Z}_2$ outer-automorphism symmetry of the extended ${D}_{n}$ quiver, as in Case 1. 

\paragraph*{\bf B-model:} $Sp(k)_1 \times Sp(k)_2$ gauge theory with $\alpha+1$ hypers in the fundamental of $Sp(k)_1$ and $n-\alpha -1$ hypers in the fundamental of $Sp(k)_2$.

\begin{figure}[htbp]
\begin{center}
\includegraphics[height=3.0in]{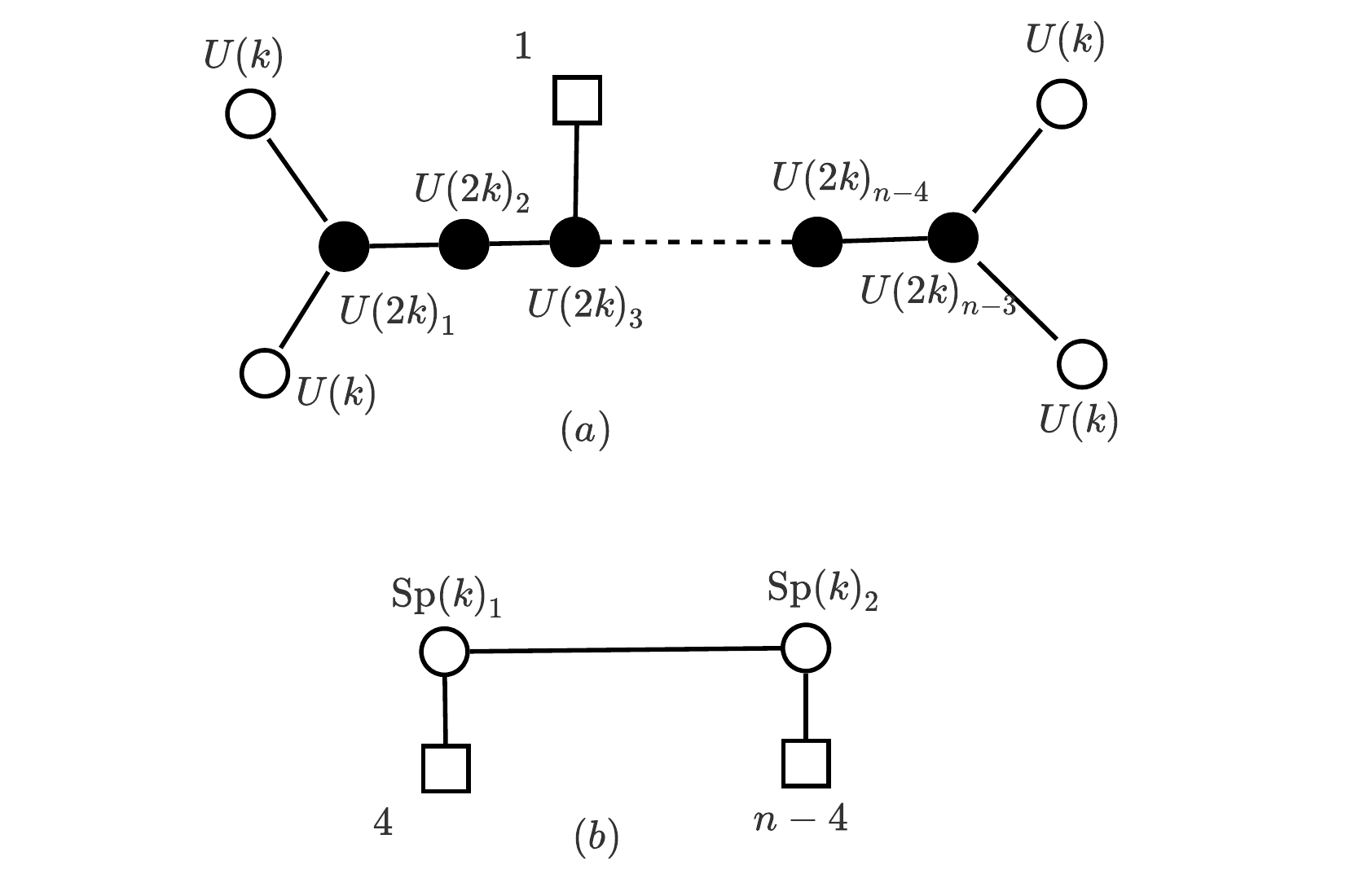}
\caption{Mirror Duals for $\Gamma_1={D}_{n}$,$\Gamma_2=\mathbb{Z}_2$. The A-model has a single fundamental hyper on one of the $U(2k)$ nodes- $\alpha=3$ in this figure.}
\label{Z2-case4}
\end{center}
\end{figure}
\begin{equation}
\begin{split}
Z_A&=\frac{1}{(k!)^4 (2k!)^{n-3}}\int  \prod^{4}_{\alpha=1}d^k\sigma_{\alpha} \prod^{n-3}_{\beta=1}d^{2k}\tilde{\sigma}_{\beta}\prod^{k}_{i=1}e^{2\pi i \eta_{\alpha}\sigma^i_{\alpha}} \prod^{2k}_{p=1}e^{2\pi i \tilde{\eta}_{\beta}\tilde{\sigma}^p_{\beta}} \frac{1}{\prod^{2k}_{p=1}\cosh{\pi(\tilde{\sigma}_{\alpha}^p+m^f)}}\\
&\times\frac{\prod_{i<j}\sinh^2{\pi(\sigma^i_1-\sigma^j_1)}\sinh^2{\pi(\sigma^i_2-\sigma^j_2)}}{\prod_{i,p}\cosh{\pi(\sigma^i_1-\tilde{\sigma}^p_{1}+m_1)}\cosh{\pi(\sigma^i_2-\tilde{\sigma}^p_{1}+m_2)} } \frac{\prod^{n-3}_{\beta=1}\prod_{p<l}\sinh^2{\pi(\tilde{\sigma}^p_{\beta}-\tilde{\sigma}^l_{\beta})}}{\prod^{n-4}_{\beta=1}\prod_{p,l} \cosh{\pi(\tilde{\sigma}^p_{\beta}-\tilde{\sigma}^l_{\beta+1}+M_{\beta})} }\\
&\times \frac{\prod_{i<j}\sinh^2{\pi(\sigma^i_3-\sigma^j_3)}\sinh^2{\pi(\sigma^i_4-\sigma^j_4)}}{\prod_{i,p}\cosh{\pi(\sigma^i_3-\tilde{\sigma}^p_{n-3}+m_3)}\cosh{\pi(\sigma^i_4-\tilde{\sigma}^p_{n-3}+m_4)} } \label{pfAm2_1}
\end{split}
\end{equation}

\begin{equation}
\begin{split}
Z_B=&\int \frac{d^k\sigma_1 d^k\sigma_2/(2^k k!)^2 }{\prod^k_{i=1}\prod^{\alpha+1}_{a=1}\cosh{\pi(\sigma_1^i+m_a/2)}\cosh{\pi(\sigma_1^i-m_a/2)}\prod^{n-\alpha-1}_{b=1}\cosh{\pi(\sigma_1^i+m_b/2)}\cosh{\pi(\sigma_1^i-m_b/2)}}\\
&\times \frac{\prod_{\alpha=1,2}\prod_{i<j}\sinh^2{\pi(\sigma^i_{\alpha}-\sigma^j_{\alpha})}\sinh^2{\pi(\sigma^i_{\alpha}+\sigma^j_{\alpha})}\prod_i \sinh^2{\pi(2\sigma^i_{\alpha})}}{\prod_{i,j}\cosh{\pi(\sigma_1^i+\sigma_2^j+M_{bif})}\cosh{\pi(\sigma_1^i+\sigma_2^j-M_{bif})}\cosh{\pi(\sigma_1^i-\sigma_2^j+M_{bif})} \cosh{\pi(\sigma_1^i-\sigma_2^j-M_{bif})}} \label{pfBm2_1}
\end{split}
\end{equation}
The counting evidence for the duality is summarized in table \ref{tab:D-Atable4}.

\begin{table}[h]
\centering
\renewcommand{\arraystretch}{1.5}
\begin{tabular}{|c|c|c|c|c|}     \hline
Model   & $\mbox{dim} M_C$ & $\mbox{dim} M_H$ & $n_{\text{FI}}$ & $n_{\text{mass}}$ \\ \hline
       A  & $2k(n-1)$ & $2k$ & $n-1$ & $0$ \\  \hline     
       B &  $2k$ & $2k(n-1)$ & $0$ & $n-1$ \\ \hline 
\end{tabular}
\caption{The dimension of the Coulomb and Higgs branches
and the number of mass and FI parameters of A and B models for the M-theory background $\mathbb{C}^2/ D_{n} \times \mathbb{C}^2/\mathbb{Z}_2$ with generic $k$ for Case 4.}
\label{tab:D-Atable4}
\end{table}

On using the ``doubling trick" (as described in \S\ref{sec:case1} for \textbf{Case 1}), introducing a set of auxiliary variables via Fourier transform and using the identity (\ref{Cauchycosh}), $Z_A$ can be expressed as,
\begin{equation}
\begin{split}
&Z_A = \int  \frac{i^{2k}d^k\tau d^k \tau^{'}}{(k!)^4(2k!)^{n-3}} \prod^{n-1}_{\beta=0}d^{2k}\tilde{\sigma}_{\beta} \prod^{n-2}_{\beta=0}d^{2k}\tilde{\tau}_{\beta} \prod^{n-1}_{\beta=0}\prod^{2k}_{p=1}e^{2\pi i {\zeta}_{\beta}\tilde{\sigma}^p_{\beta}}\left(\sum_{\rho} (-1)^{\rho} \prod_i \tanh{\pi \tau^i}e^{2\pi i \tau^i (\tilde{\sigma}^i_0-\tilde{\sigma}^{k+\rho(i)}_0)}\right)\\
&\times \prod^{n-2}_{\beta=0} \left(\sum_{\rho_{\beta}}(-1)^{\rho_{\beta}} \frac{e^{2\pi i \tilde{\tau}^p_{\beta}(\tilde{\sigma}^p_{\beta}-\tilde{\sigma}^{\rho_{\beta}(p)}_{\beta+1})}}{I_{\beta}(\tilde{\tau}_{\beta},\tilde{\sigma}_{\beta})}\right)\left(\sum_{\rho{'}} (-1)^{\rho{'}} \prod_i \tanh{\pi \tau'^{i}}e^{2\pi i \tau'^{i} (\tilde{\sigma}^i_{n-1}-\tilde{\sigma}^{k+\rho{'}(i)}_{n-1})}\right)
\end{split}
\end{equation}
where we have defined $\tilde{\sigma}^p_0=(\sigma^i_1,\sigma^i_2)$ and $\tilde{\sigma}^p_{n-1}=(\sigma^i_3,\sigma^i_4)$ and the FI parameters $\{\zeta_{\beta}\}$ as $\zeta_0=\eta_1=\eta_2$, $\zeta_{\beta}=\tilde{\eta}_{\beta} \; (\beta <\alpha)$, $\zeta_{\alpha}=0$, $\zeta_{\beta}=\tilde{\eta}_{\beta -1} \; (\alpha<\beta < n-1)$, $\zeta_{n-1}=\eta_3=\eta_4$. The function $I_{\beta}(\tilde{\tau}_{\beta},\tilde{\sigma}_{\beta})$ is defined as,
\begin{equation}
I_{\beta}(\tilde{\tau}_{\beta},\tilde{\sigma}_{\beta})=
\left\{
	\begin{array}{ll}
		\cosh{\pi \tilde{\tau}^p_{\beta}}  & \mbox{if } \beta=0,1,...,\alpha-1 \\
		\cosh{\pi \tilde{\sigma}^p_{\beta}} & \mbox{if } \beta=\alpha\\
		\cosh{\pi \tilde{\tau}^p_{\beta}}  & \mbox{if } \beta=\alpha+1,\alpha+2,....,n-2
	\end{array}
\right.
\end{equation}
On integrating over the variables $\{\tilde{\sigma}_{\beta}\}$ and imposing the resulting $\delta$-function conditions, one can reduce partition function to the following form,
 \begin{equation}
\begin{split}
Z_A&=\int \frac{d^{2k}\tilde{\tau}_0 d^{2k}\tilde{\tau}_{\alpha}}{(2^k k!)^2} \frac{\prod_i \sinh{\pi 2\tilde{\tau}^i_0}}{\prod_p \cosh{\pi \tilde{\tau}^p_0}} \frac{\prod_{p<l} \sinh{\pi(\tilde{\tau}^p_0-\tilde{\tau}^l_0)}\sinh{\pi(\tilde{\tau}^p_{\alpha}-\tilde{\tau}^l_{\alpha})}}{\prod_{p,l} \cosh{\pi(\tilde{\tau}^p_{\alpha}-\tilde{\tau}^p_0+\sum^{n-1}_{\beta=0}\zeta_{\beta})} }  \frac{\prod_i \sinh{\pi 2\tilde{\tau}^i_{\alpha}}}{\prod_i \cosh{\pi \tilde{\tau}^p_{\alpha}}} \\
& \times \prod^{\alpha-1}_{a=0}\frac{1}{\prod_p\cosh{\pi(\tilde{\tau}^p_{0}-\sum^a_{\beta=0}\zeta_{\beta})}} \prod^{n-\alpha -2}_{b=1}\frac{1}{\prod_p\cosh{\pi(\tilde{\tau}^p_{\alpha}+\sum^b_{\beta=1}\zeta_{n-\beta})}}\\
&\times \delta(\tilde{\tau}^i_0+\tilde{\tau}^{k+i}_0)\delta(\tilde{\tau}^i_{\alpha}+\tilde{\tau}^{k+i}_{\alpha})\\
&=Z_B
\end{split}
\end{equation}

\paragraph*{\bf Mirror Map:}
The number of independent mass parameters in the A-model is zero, which matches with the number of FI parameters in the B-model. The $(\alpha +1)$ masses of the fundamental hypers of $Sp(k)_1$ are given in terms of the A-model FI parameters as follows,
\begin{equation}
\boxed{m^{(1)}_1=2\eta_1,\; m^{(1)}_{a}=2\eta_1+2\sum^{a-1}_{\beta=1}\tilde{\eta}_{\beta} \; \; (a=2,3,....,\alpha), \; \; m^{(1)}_{\alpha +1}=0 }
\end{equation}
Similarly, the $(n-\alpha -1)$ masses of the fundamental hypers of $Sp(k)_2$ are given in terms of the A-model FI parameters as follows,
\begin{equation}
\boxed{m^{(2)}_1=2\eta_3,\; m^{(2)}_{b}=2\eta_3+2\sum^{b-1}_{\beta=1}\tilde{\eta}_{n-\beta-2} \; \; (b=2,3,....,n-\alpha-2), \; \; m^{(2)}_{n-\alpha-1}=0 }
\end{equation}
The bi-fundamental mass parameter is given by
\begin{equation}
\boxed{M_{bif}=\eta_1 +\eta_3 + \sum^{n-3}_{\beta=1} \tilde{\eta}_{\beta}}
\end{equation}
Therefore, the number of independent FI parameters in the A-model- $(n-3)+2=n-1$ -matches the number of non-zero independent mass parameters in the B-model, as expected.

\section{$\mathbb{C}^2/ D_{n} \times \mathbb{C}^2/\mathbb{Z}_m$ ($m=2s,s>1$,$n\geq 4$)}\label{D-AgenEven}
\begin{figure}[htbp]
\begin{center}
\includegraphics[height=3.0in]{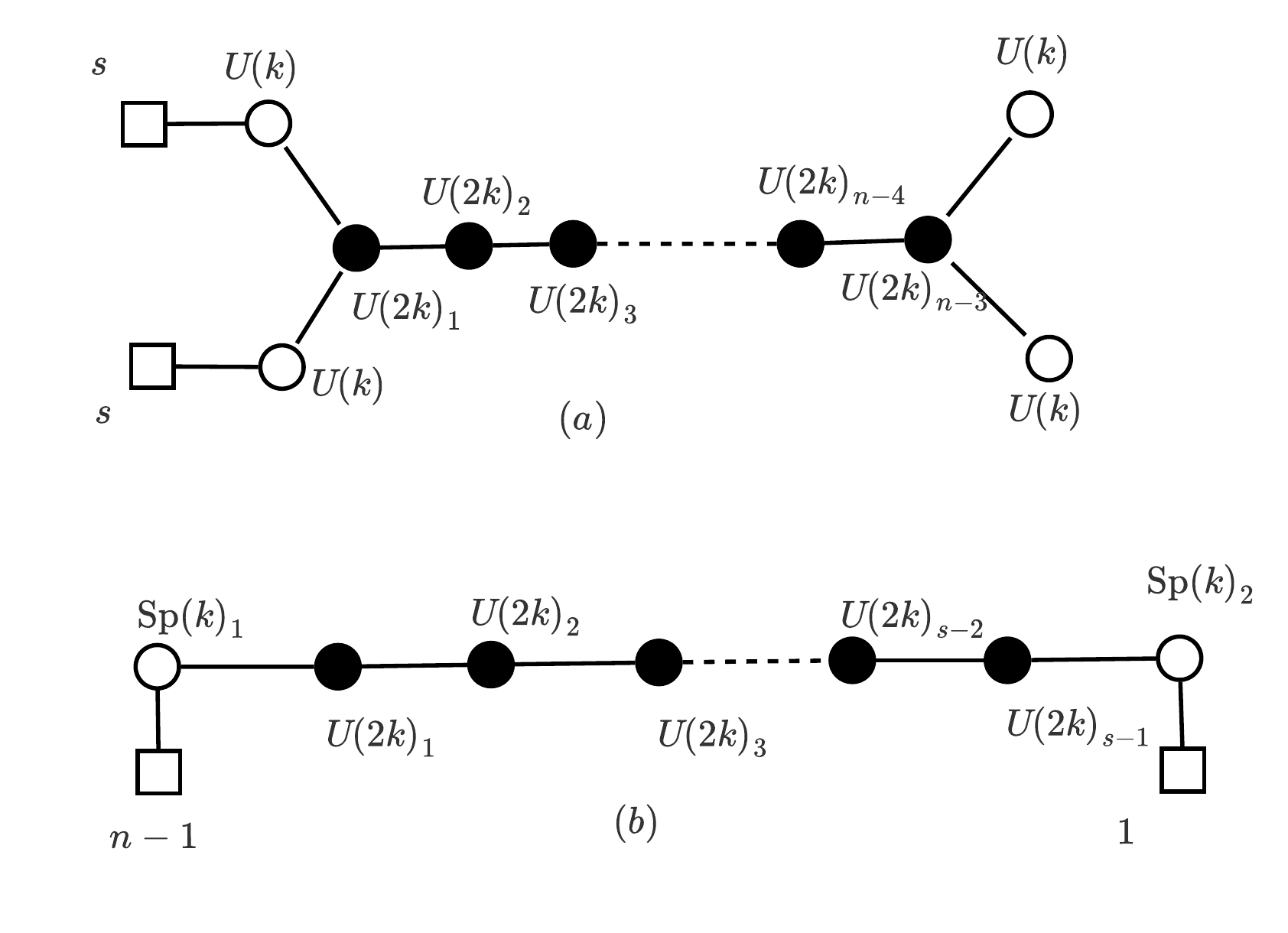}
\caption{A mirror pair of $\mathbb{D}_{n-2} \times \mathbb{Z}_{m}$ type, $m=2s$ even}
\label{ADeven}
\end{center}
\end{figure}

For this class of dual theories, we discuss the case of a particular family of dual theories which  was addressed in \cite{Dey:2011pt}. The computation proceeds on the lines of \textbf{Case 1} in \S\ref{sec:case1}. We briefly review the results here. The dual theories, in this case, are:

\paragraph*{\bf A-model:} A $U(k)^2 \times U(2k)^{n-3} \times U(k)^2$ gauge theory with bi-fundamental hypers and two sets of $s$ hypers in the fundamental of $U(k)_1$ and $U(k)_2$ respectively (figure \ref{ADeven} (a)). The masses and the FI parameters respect the $\mathbb{Z}_2$ outer-automorphism symmetry of the extended ${D}_{n}$ quiver.

\paragraph*{\bf B-model:} A $Sp(k) \times U(2k)^{s-1} \times Sp(k)$ with bi-fundamental hypers and $(n-1)$ hypers in the fundamental of $Sp(k)_1$ and one hyper in the fundamental of $Sp(k)_2$ (figure \ref{ADeven} (b)).

As explained in \cite{Dey:2011pt}, the total number of independent mass parameters in the A-model is $(s-1)$ which precisely matches the number of FI parameters in the B-model. Also, the number of independent FI parameters in the A-model is $(n-1)$ (the FI parameters for $U(k)_1$ and $U(k)_2$ are equal, same is true for $U(k)_3$ and $U(k)_4$) which again matches with the number of independent mass parameters in the B-model --- $(n-2)$ fundamental masses and one bifundamental mass.

The dimensions of the Coulomb and the Higgs branches of the dual theories, and the number of masses and FI parameters of each, are summarized in table \ref{tab:D-Ageneric}.

\begin{table}[h]
\centering
\renewcommand{\arraystretch}{1.5}
\begin{tabular}{|c|c|c|c|c|}     \hline
Model   & $\mbox{dim} M_C$ & $\mbox{dim} M_H$ & $n_{\text{FI}}$ & $n_{\text{mass}}$ \\ \hline
       A  & $2k(n-1)$ & $2sk$ & $n-1$ & $s-1$ \\  \hline     
       B &  $2sk$ & $2k(n-1)$ & $s-1$ & $n-1$ \\ \hline 
\end{tabular}
\caption{The dimension of the Coulomb and Higgs branches
and the number of mass and FI parameters of A and B models for the M-theory background $\mathbb{C}^2/ D_{n} \times \mathbb{C}^2/\mathbb{Z}_m$ with generic $k$ for the case in figure (\ref{ADeven}).}
\label{tab:D-Ageneric}
\end{table}

\paragraph*{\bf Mirror Map:}The $(s-1)$ FI parameters of the B-model can be expressed in terms of the fundamental masses of the A-model as follows:
\begin{equation}
\boxed{\tilde{\eta}_{\beta}=m^f_{\beta +1}-m^f_{\beta}, \beta=1,2,...,s-1}
\end{equation}
The $n-1$ independent mass parameters of the B-model can also be expressed in terms of the FI parameters of the A-model. The only non-zero bifundamental mass is given by
\begin{equation}
\boxed{m_{bif}=\eta_1+\eta_3 +(\eta^{'}_1 + \eta^{'}_2 +....... +\eta^{'}_{n-3})}
\end{equation}
where the primed parameters are associated with the $U(2k)$ nodes while the unprimed parameters are associated with the $U(k)$ nodes of the A model.

The fundamental masses, as functions of the A-model FI parameters, are given by
\begin{equation}
\boxed{\begin{gathered}M^f_a=(\eta + \eta_3 - \sum^{a}_{n-3} \eta^{'}_{\beta})=(\sum^{a-1}_{1} \eta^{'}_{\beta} + \eta_3)  \; \;(a=1,2,...,n-3);\\ M^f_{n-2}=(\eta + \eta_3); M^f_{n-1}=0,M^f_{n}=0\end{gathered}}
\end{equation}

\section{$ \mathbb{C}^2/ D_{n}\times \mathbb{C}^2/\mathbb{Z}_m$ ($m=2s +1$;$n \geq 4$) }\label{D-AgenOdd}
An interesting example of a family of dual pairs in this category is as follows:

\paragraph*{\bf A-model:}$U(k)^2 \times U(2k)^{n-3} \times U(k)^2$ gauge theory with the matter content given by a ${D}_{n}$ quiver diagram. Two of the adjacent $U(k)$ factors (which we label as 1 and 2 respectively) have $s$ fundamental hypers each, while one of the other $U(k)$ factors has one fundamental hyper.

\paragraph*{\bf B-model:}$Sp(k)\times U(2k)^{\frac{m-1}{2}}$ gauge theory with bi-fundamentals and one hyper in the antisymmetric representation of $U(2k)_{\frac{m-1}{2}}$. The $Sp(k)$ gauge group has one fundamental hyper while $U(2k)_{\frac{m-1}{2}}$ has $(n-1)$ fundamental hypers.

\begin{figure}[htbp]
\begin{center}
\includegraphics[height=3.0in]{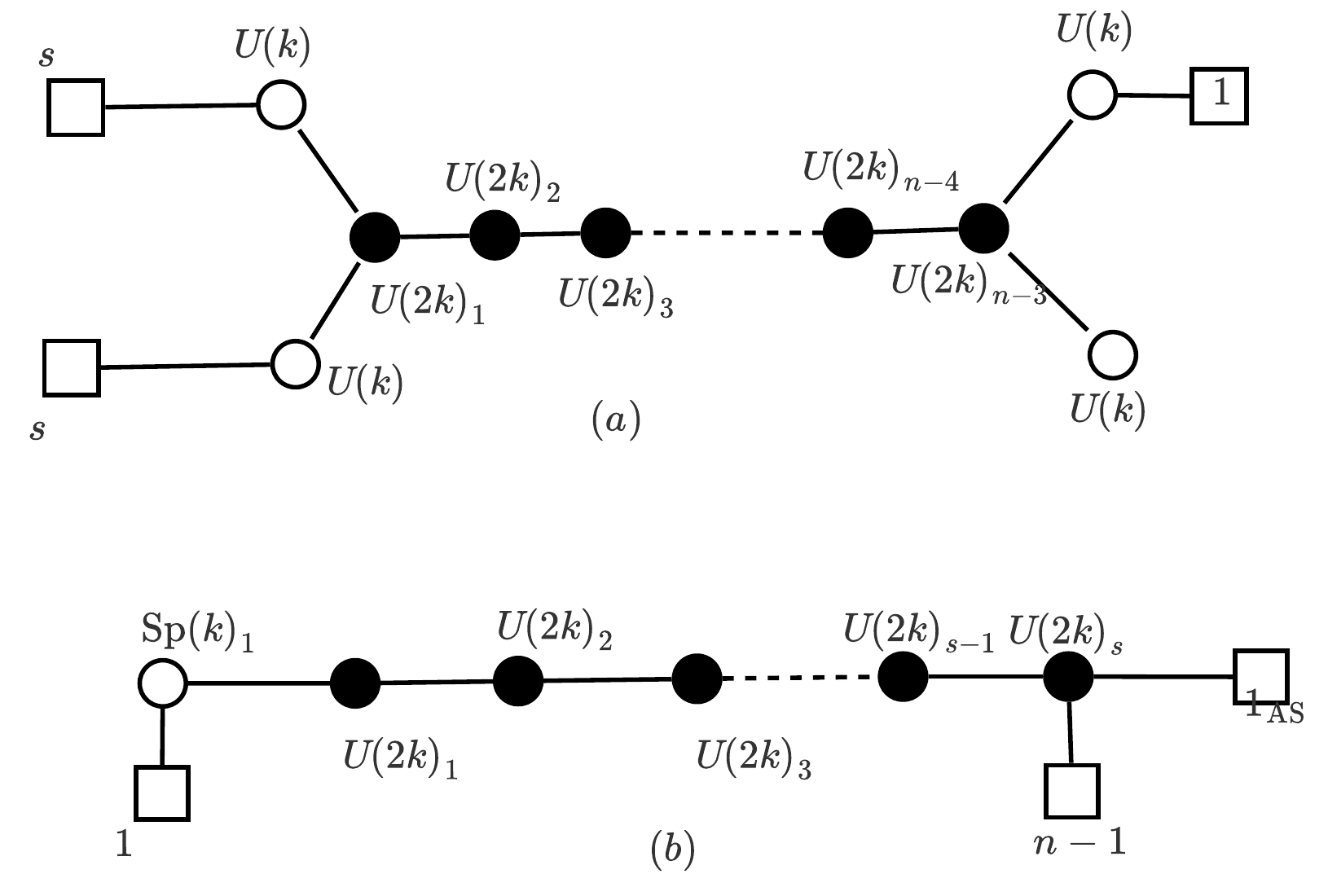}
\caption{A mirror pair of ${D}_{n-2} \times \mathbb{Z}_{m}$ type, $m=2s+1$ odd}
\label{ADodd}
\end{center}
\end{figure}

\noindent The partition function of the A model is:
\begin{equation}
\begin{split}
Z_A&=\frac{1}{(k!)^4 (2k!)^{n-3}}\int  \prod^{4}_{\alpha=1}d^k\sigma_{\alpha} \prod^{n-3}_{\beta=1}d\tilde{\sigma}_{\beta}\prod^{k}_{i=1}e^{2\pi i \eta_{\alpha}\sigma^i_{\alpha}} \prod^{2k}_{p=1}e^{2\pi i \tilde{\eta}_{\beta}\tilde{\sigma}^p_{\beta}}\\
&\times\frac{1}{\prod^{s}_{a=1}\prod_i\cosh{\pi(\sigma_1^i+m^f_a)} \cosh{\pi(\sigma_2^i+m^f_a)}}\\
&\times\frac{\prod_{i<j}\sinh^2{\pi(\sigma^i_1-\sigma^j_1)}\sinh^2{\pi(\sigma^i_2-\sigma^j_2)}}{\prod_{i,p}\cosh{\pi(\sigma^i_1-\tilde{\sigma}^p_{1}+m_1)}\cosh{\pi(\sigma^i_2-\tilde{\sigma}^p_{1}+m_2)} } \frac{\prod^{n-3}_{\beta=1}\prod_{p<l}\sinh^2{\pi(\tilde{\sigma}^p_{\beta}-\tilde{\sigma}^l_{\beta})}}{\prod^{n-4}_{\beta=1}\prod_{p,l} \cosh{\pi(\tilde{\sigma}^p_{\beta}-\tilde{\sigma}^l_{\beta+1}+M_{\beta})} }\\
&\times \frac{\prod_{i<j}\sinh^2{\pi(\sigma^i_3-\sigma^j_3)}\sinh^2{\pi(\sigma^i_4-\sigma^j_4)}}{\prod_{i,p}\cosh{\pi(\sigma^i_3-\tilde{\sigma}^p_{n-3}+m_3)}\cosh{\pi(\sigma^i_4-\tilde{\sigma}^p_{n-3}+m_4)} }\frac{1}{\prod_i\cosh{\pi(\sigma^i_3+m^f_{s+1})}} \label{pfAodd}
\end{split}
\end{equation}
As in the example with $m$ even, the masses of the fundamental hypers of $U(k)_1$ and $U(k)_2$ are pair-wise equal. Also the bifundamental masses associated with these nodes are also taken to be equal, i.e. $m_1=m_2$. The FI parameters associated with $U(k)_1$ and $U(k)_2$ are also equal, i.e. $\eta_1=\eta_2$. Therefore, the total number of independent FI parameters in the A-model is $(n-3)+2+1=n$. Also, by appropriately shifting the integration variables by constants, one can set all the bi-fundamental masses and one of the fundamental masses to zero. Therefore, the total number of independent mass parameters in the A-model is $s$.

The partition function of the B-model is:
\begin{equation}
\begin{split}
Z_B&=\frac{1}{(2^k k!) (2k!)^{s}}\int d^k\sigma \prod^{s}_{\beta=1}d^{2k}{\sigma}^{'}_{\beta} \prod^{2k}_{p=1}e^{2\pi i {\zeta}_{\beta}{\sigma}^{'p}_{\beta}} \frac{1}{\Pi_{i}\cosh{\pi(\sigma^i+M^f_1)}\cosh{\pi(\sigma^i-M^f_1)}}\\
\times&\frac{\prod_{i<j}\sinh^2{\pi(\sigma^i-\sigma^j)}\sinh^2{\pi(\sigma^i+\sigma^j)}\prod_i \sinh^2{\pi(2\sigma^i)}}{\prod_{i,p}\cosh{\pi(\sigma^i-{\sigma}^{'p}_{1}+m^{bif}_{1})}\cosh{\pi(\sigma^i+{\sigma}^{'p}_{1}-m^{bif}_{1})}}\\
\times & \frac{\prod^{s}_{\beta=1}\prod_{p<l}\sinh^2{\pi({\sigma}^{'p}_{\beta}-{\sigma}^{'l}_{\beta})}}{\prod^{s-1}_{\beta=1}\prod_{p,l} \cosh{\pi({\sigma}^{'p}_{\beta}-{\sigma}^{'l}_{\beta+1}+m^{bif}_{\beta+1})} } \frac{1}{\prod^{n-1}_{a=1} \prod_{p}\cosh{\pi(\sigma_s^{'p}+M^f_a)}\prod_{p<l} \cosh{\pi(\sigma_s^{'p} + \sigma_s^{'l} +M_{as})}} \label{pfBodd}
\end{split}
\end{equation}
The number of independent FI parameters in the B-model is evidently $s$. By suitably shifting the integration variables by constants in $Z_B$, one can check that the number of mass parameters is $n$.

We summarize the counting evidence for the duality in table \ref{D-Atablekodd}.

\begin{table}[h]
\centering
\renewcommand{\arraystretch}{1.5}
\begin{tabular}{|c|c|c|c|c|}     \hline
Model   & $\mbox{dim} M_C$ & $\mbox{dim} M_H$ & $n_{\text{FI}}$ & $n_{\text{mass}}$ \\ \hline
       A  & $2k(n-1)$ & $(2s+1)k$ & $n$ & $s$ \\  \hline     
       B &  $(2s+1)k$ & $2k(n-1)$ & $s$ & $n$ \\ \hline 
\end{tabular}
\caption{The dimension of the Coulomb and Higgs branches
and the number of mass and FI parameters of A and B models for the M-theory background $\mathbb{C}^2/ D_{n} \times \mathbb{C}^2/\mathbb{Z}_m$ with generic $k$ for the case in figure (\ref{ADodd}).}
\label{D-Atablekodd}
\end{table}

To show that the above partition functions correspond to mirror dual theories, we apply the ``doubling trick" for the boundary nodes $U(k)_1$ and $U(k)_2$ and ``integrate out" the node $U(k)_4$.

First, define $\tilde{\sigma}_0^{p} = (\sigma^i_1,\sigma^i_2)$ and the corresponding fundamental mass as $M_0=m_1=m_2$ and the FI parameter as $\tilde{\eta}_0 =\eta_1=\eta_2$. With this definition, one can write the boundary contribution of the nodes $U(k)_1$ and $U(k)_2$ as ,
\begin{equation}
\begin{split}
&\frac{\prod^{2}_{\alpha=1}\prod^{k}_{i=1}e^{2\pi i \eta_{\alpha}\sigma^i_{\alpha}}}{\prod^{s}_{i,a=1}\cosh{\pi(\sigma_1^i+m^f_a)} \prod^{l}_{i;a=1}\cosh{\pi(\sigma_2^i+m^f_a)}}\frac{\prod_{i<j}\sinh^2{\pi(\sigma^i_1-\sigma^j_1)}\sinh^2{\pi(\sigma^i_2-\sigma^j_2)}}{\prod_{i,p}\cosh{\pi(\sigma^i_1-\tilde{\sigma}^p_{1}+m_1)}\cosh{\pi(\sigma^i_2-\tilde{\sigma}^p_{1}+m_2)} }\\
=&\frac{\prod^{2k}_{p=1}e^{2\pi i \tilde{\eta}_{0}\tilde{\sigma}^p_{0}}}{\prod^{s}_{p,a=1}\cosh{\pi(\tilde{\sigma}_0^p+m^f_a)}} \frac{\prod_{p<l}\sinh{\pi(\tilde{\sigma}^p_{0}-\tilde{\sigma}^l_{0})}}{\prod_{p,l} \cosh{\pi(\tilde{\sigma}^p_{0}-\tilde{\sigma}^l_{1}+M_{0})} }\frac{\prod_{i<j}\sinh{\pi(\sigma^i_1-\sigma^j_1)}\sinh{\pi(\sigma^i_2-\sigma^j_2)}}{\prod_{i,j}\sinh{\pi(\sigma^i_1-{\sigma}^j_{2})}}\\
=&\frac{\prod^{2k}_{p=1}e^{2\pi i \tilde{\eta}_{0}\tilde{\sigma}^p_{0}}}{\prod^{s}_{p,a=1}\cosh{\pi(\tilde{\sigma}_0^p+m^f_a)}} \frac{\prod_{p<l}\sinh{\pi(\tilde{\sigma}^p_{0}-\tilde{\sigma}^l_{0})}}{\prod_{p,l} \cosh{\pi(\tilde{\sigma}^p_{0}-\tilde{\sigma}^l_{1}+M_{0})} } \left(\sum _{\rho} (-1)^{\rho} \frac{1}{\prod_{i}\sinh{(\tilde{\sigma}^i_0 -\tilde{\sigma}_0^{k+\rho(i)})}}\right)
\end{split}
\end{equation}
where in the last step we have used the result of equation \eqref{Cauchysinh}. On the other boundary, we ``integrate out" the node corresponding to $\sigma_4$ as before. This reduces $Z_A$ to,
\begin{equation}
\begin{split}
Z_A&=\frac{(-1)^k i^{(2k-1)k}}{(k!)^5 (2k!)^{n-4}\sinh^k{\pi \eta_4}}\int  d^k\sigma_{3} \prod^{n-3}_{\beta=0}d^{2k}\tilde{\sigma}_{\beta} \prod^{2k}_{p=1}e^{2\pi i \tilde{\eta}_{\beta}\tilde{\sigma}^p_{\beta}} \left(\sum _{\rho} (-1)^{\rho} \frac{1}{\prod_{i}\sinh{(\tilde{\sigma}^i_0 -\tilde{\sigma}_0^{k+\rho(i)})}}\right)\\
&\times \frac{\prod^{2k}_{p=1}e^{2\pi i \tilde{\eta}_{0}\tilde{\sigma}^p_{0}}}{\prod^{s}_{p,a=1}\cosh{\pi(\tilde{\sigma}_0^p+m^f_a)}}  \prod^{n-4}_{\beta=0} \frac{\prod_{p<l} \sinh{\pi(\tilde{\sigma}^p_{\beta}-\tilde{\sigma}^l_{\beta})}\sinh{\pi(\tilde{\sigma}^p_{\beta+1}-\tilde{\sigma}^l_{\beta+1})}}{\prod_{p,l}\cosh{\pi(\tilde{\sigma}^p_{\beta}-\tilde{\sigma}^l_{\beta+1})}}\\
&\times \frac{\prod_{i<j}\sinh{\pi({\sigma}^i_3-{\sigma}^j_3)}\sinh{\pi(\tilde{\sigma}^i_{n-3}-\tilde{\sigma}^j_{n-3})}}{\prod_{i,j}\cosh{\pi(\sigma^i_3-\tilde{\sigma}^j_{n-3})}} \frac{\prod_{i<j}\sinh{\pi({\sigma}^i_3-{\sigma}^j_3)}\sinh{\pi(\tilde{\sigma}^{k+i}_{n-3}-\tilde{\sigma}^{k+j}_{n-3})}}{\prod_{i,j}\cosh{\pi(\sigma^i_3-\tilde{\sigma}^{k+j}_{n-3})}}\\
&\times \frac{1}{\prod_i \cosh{\pi({\sigma}^i_3+m^f_{s+1})}}\prod^{k}_{i=1}e^{2\pi i \eta_4 \tilde{\sigma}^i_{n-3} }
\end{split}
\end{equation}
where we have explicitly set the bi-fundamental masses to zero. Now, on using Cauchy's determinant formula and Fourier transforming, we get
\begin{equation}   
\begin{split}
Z_A& \propto \int  d^k\sigma_{3} \prod^{n-3}_{\beta=0}d^{2k}\tilde{\sigma}_{\beta}  \prod^{n-4}_{\beta=0}d^{2k}\tilde{\tau}_{\beta} d^{2k} \tau_{n-3} d^k \tau \prod^{s+1}_{a=1} d^{2k}x_a d^{2k}z_a d^kx\prod^{n-3}_{\beta=0}\prod^{2k}_{p=1}e^{2\pi i \tilde{\eta}_{\beta}\tilde{\sigma}^p_{\beta}}\\
&\times \left(\sum _{\rho} (-1)^{\rho} \prod_i i\tanh{(\pi x^i)} e^{2\pi i x^i(z^i_{s+1}- z^{k+\rho(i)}_{s+1})}\right)\\
&\times \frac{1}{(2k!)^{s+1}}\left(\sum_{\{\rho_a\}}(-1)^{\{\rho_a\}} \frac{\prod^{1}_{a=s+1}\prod_p e^{2\pi i x^p_a(z^p_a -z^{\rho_{a-1}(p)}_{a-1})} }{\prod^{s}_{a=1} \prod_p \cosh{\pi (z^p_a + m_a) }} \right)\\
&\times \prod^{n-4}_{\beta=0}  \left(\sum_{\tilde{\rho}_{\beta}}(-1)^{\tilde{\rho}_{\beta}}\prod_p \frac{e^{2\pi i \tilde{\tau}^p_{\beta}(\tilde{\sigma}^p_{\beta}- \tilde{\sigma}^{\tilde{\rho}_{\beta}(p)}_{\beta +1})}}{\cosh{\pi \tilde{\tau}^p_{\beta}}} \right)\\
&\times \left(\sum_{\rho_3}(-1)^{\rho_3}\prod_i \frac{e^{2\pi i \tau^i_{n-3}(\sigma^i_3- \tilde{\sigma}^{\rho_3(i)}_{n-3})}}{\cosh{\pi \tau^i_{n-3}}} \right)  \left(\sum_{\rho_4}(-1)^{\rho_4}\prod_i \frac{e^{2\pi i \tau^{k+i}_{n-3}(\sigma^{i}_3- \tilde{\sigma}^{k+\rho_4(i)}_{n-3})}}{\cosh{\pi \tau^{k+i}_{n-3}}} \right)\\
&\times \prod_i \frac{e^{2\pi i \tau^i(\sigma^i_3+m^f_{s+1})}}{\cosh{\pi \tau^i}} \prod^{k}_{i=1}e^{2\pi i \eta_4 \tilde{\sigma}^i_{n-3} }
\end{split}
\end{equation}
where we have suppressed the constant pre-factors as before. In the above equation, $z^p_0=\tilde{\sigma}^p_0$ while $(-1)^{\{\rho_a\}}= (-1)^{\rho_1 +.... + \rho_{s+1}}$.
 
Integrating over the variables $z_a, x$ and $\tilde{\sigma}_{\beta}$, the $\delta$-functions impose the following conditions on the remaining variables,
\begin{equation}   
\begin{split}
x^i_{s+1} + x^{k+\rho(i)}_{s+1}=0\\
\tilde{\tau}^p_0= x^{\rho^{-1}_0(p)}_1-\tilde{\eta}_0 \\
 \tilde{\tau}^p_1={x}^{\rho^{-1}_0\tilde{\rho}_1^{-1}(p)}_1-\tilde{\eta}_0 -\tilde{\eta}_1\\
 \tilde{\tau}^p_2={x}^{\rho^{-1}_0\tilde{\rho}_1^{-1}\tilde{\rho}_2^{-1}(p)}_1 - \tilde{\eta}_0 -\tilde{\eta}_1 -\tilde{\eta}_2\\
............................ .......................\\
 \tilde{\tau}^p_{n-4}={x}^{\rho^{-1}_0 \tilde{\rho}_1^{-1}\tilde{\rho}_2^{-1}.....\tilde{\rho}_{n-5}^{-1}(p)}_1 - \tilde{\eta}_0 -\tilde{\eta}_1-\tilde{\eta}_2 -\tilde{\eta}_3-.......-\tilde{\eta}_{n-4}\\
\tau^{\rho^{-1}_3(i)}_{n-3} = - {x}^{\rho^{-1}_0\tilde{\rho}_1^{-1}\tilde{\rho}_2^{-1}.....\tilde{\rho}_{n-4}^{-1}(i)}_1+ ( \tilde{\eta}_0 +\tilde{\eta}_1+ \tilde{\eta}_2 +\tilde{\eta}_3+.....+\tilde{\eta}_{n-3} +\eta_4)\\
 \tau^{k+\rho^{-1}_4(i)}_{n-3} =-{x}^{\rho^{-1}_0\tilde{\rho}_1^{-1}\tilde{\rho}_2^{-1}.....\tilde{\rho}_{n-4}^{-1}(k+i)}_1+ (\tilde{\eta}_0 +\tilde{\eta}_1+\tilde{\eta}_2 +\tilde{\eta}_3+.....+\tilde{\eta}_{n-3} )\\
 \tau^i + \tau^i_{n-3} + \tau^{k+i}_{n-3} +\eta_3=0\\
 \end{split}
\end{equation}
Define $m=\tilde{\eta}_0+\tilde{\eta}_1+....+\tilde{\eta}_{n-3}+\eta_4, m^{'}=\tilde{\eta}_0+\tilde{\eta}_1+....+\tilde{\eta}_{n-3}$ and $M=\eta_3+\eta_4 +2 (\tilde{\eta}_0 + \tilde{\eta}_1 +..... + \tilde{\eta}_{n-3})$. Also, define $s$ constants $\{l_a\}$, such that $l_1=m^f_{s+1}-m^f_1$, $l_a=m^f_{a-1}-m^f_a$ for $a=2,3,....,s$. Now, implementing the above $\delta$-function conditions, we have
\begin{equation}   
\begin{split}
Z_A& =\frac{(-1)^k i^{2k^2}}{(k!)^3 (2k!)^s\sinh^k{\pi \eta_4}} \int \prod^{s+1}_{a=1} d^{2k}x_a \left(\sum _{\rho} (-1)^{\rho} \prod_i \tanh{(\pi x^i_{s+1})} {\delta(x^i_{s+1}+ x^{k+\rho(i)}_{s+1})}\right)\\
&\times \prod^{2}_{a=s+1}\left( \sum_{\rho_a} (-1)^{\rho_a} \frac{1}{\prod_p \cosh{\pi(x^p_a- x^{\rho_a(p)}_{a-1})}}\right)\prod^{s}_{a=1} \prod_p e^{2\pi i l_a x^p_a} \\
&\times \frac{1}{\prod_p \cosh{\pi(x^p_1-\tilde{\eta}_0)}} \frac{1}{\prod_p \cosh{\pi(x^p_1-\tilde{\eta}_0-\tilde{\eta}_1)}}.......\frac{1}{\prod_p \cosh{\pi(x^p_1-\tilde{\eta}_0-\tilde{\eta}_1-....-\tilde{\eta}_{n-4})}}\\
&\times \left(\sum_{\rho}(-1)^{\rho} \frac{1}{\prod_i \cosh{\pi(x_1^{\rho(i)}-m)}\cosh{\pi(x_1^{\rho(k+i)}-m^{'})}\cosh{\pi(x_1^{\rho(i)}+x_1^{\rho(k+i)}-M)}} \right) \\
& = \frac{\kappa}{2^k k! (2k!)^s\sinh^k{\pi \eta_4}}\int \prod^{s+1}_{a=1} d^{2k}x_a \prod^{s}_{a=1} \prod_p e^{2\pi i l_a x^p_a} \left(\prod_i  \delta(x^i_{s+1}+ x^{k+i}_{s+1}) \frac{\sinh{\pi 2x^i_{s+1}}}{\cosh^2{\pi x^i_{s+1}}}   \right)\\
&\times \frac{\prod_{p<l}\sinh{\pi(x^p_{s+1}-x^l_{s+1})}\sinh{\pi(x^p_{s}-x^l_{s})}}{\prod_{p,l}\cosh{\pi (x^p_{s+1}-x^l_{s})}}....... \frac{\prod_{p<l}\sinh{\pi(x^p_{2}-x^l_{2})}\sinh{\pi(x^p_{1}-x^l_{1})}}{\prod_{p,l}\cosh{\pi (x^p_{2}-x^l_{1})}}\\
&\times \frac{1}{\prod_p \cosh{\pi(x^p_1-\tilde{\eta}_0)}} \frac{1}{\prod_p \cosh{\pi(x^p_1-\tilde{\eta}_0-\tilde{\eta}_1)}}.......\frac{1}{\prod_p \cosh{\pi(x^p_1-\tilde{\eta}_0-\tilde{\eta}_1-....-\tilde{\eta}_{n-4})}}\\
&\times \frac{1}{\prod_p \cosh{\pi(x^p_1-m)}} \frac{1}{\prod_p\cosh{\pi(x^p_1-m^{'})}} \frac{\sinh^k{(m-m^{'})}\prod_{p<l} \sinh{\pi(x^p_1-x^l_1)}}{\prod_{p<l} \cosh{\pi(x^p_1+x^l_1-M)}}
\end{split}
\end{equation}
where we have used equation \eqref{Id2} in the last step.

Rearranging the above expression and integrating over the variables $x^{k+i}_{s+1}$, we have
\begin{equation}   
\begin{split}
Z_A& =\frac{1}{2^k k! (2k!)^s} \int \prod^{s}_{a=1} d^{2k}x_a d^k x_{s+1}  \frac{\prod_{i<j}\sinh^2{\pi(x^i_{s+1}-x^j_{s+1})}\sinh^2{\pi(x^i_{s+1}+x^j_{s+1})}\prod_i \sinh^2{\pi(2x^i_{s+1})}}{\prod_i\cosh^2{\pi x^i_{s+1}} \prod_{i,p}\cosh{\pi(x^i_{s+1}-x^{p}_{s})}\cosh{\pi(x^i_{s+1}+{x}^{p}_{s})}}\\
&\times \frac{\prod^{1}_{a=s}\prod_{p<l}\sinh^2{\pi(x^p_a-x^l_a)}}{\prod^{2}_{a=s}\prod_{p,l} \cosh{\pi(x^p_a-x^l_{a-1})}}\times \frac{1}{\prod_p \cosh{\pi(x^p_1-\tilde{\eta}_0)}}.......\frac{1}{\prod_p \cosh{\pi(x^p_1-\tilde{\eta}_0-\tilde{\eta}_1-....-\tilde{\eta}_{n-4})}}\\
&\times \frac{1}{\prod_p \cosh{\pi(x^p_1-m)}} \frac{1}{\prod_p\cosh{\pi(x^p_1-m^{'})}} \frac{1}{\prod_{p<l} \cosh{\pi(x^p_1+x^l_1-M)}} \times \prod^{s}_{a=1} \prod_p e^{2\pi i {\zeta}_a x^p_a}
\end{split}
\end{equation}
which is evidently identical to $Z_B$ as given in equation \eqref{pfBodd} after the transformation $x^p_a \to -x^p_a$ for $a=1,2,..,s$. The corresponding mirror map can now be read off by comparing the two expressions.

\paragraph*{\bf Mirror Map:}
The A model has $s$ independent mass parameters, which exactly matches the number of FI parameters in the B model. The number of independent FI parameters in the A model is $(n-3)+2+1=n$, which again matches the number of independent mass parameters in the B-model.

Now, FI parameters $\{\zeta_a\}$ of model B, in terms of the fundamental masses of model A, are given by
\begin{equation}
\boxed{\zeta_1= m^f_1- m^f_{s+1} , \zeta_a=m^f_{a}-m^f_{a-1} (a=2,3,....,s).} 
\end{equation}
The bi-fundamental masses of model B are all zero. The fundamental masses of model B are obtained in terms of the FI parameters of model A:
\begin{equation}
\boxed{M^f_1=0, M^f_2=\eta_1,M^f_i =\eta_1+ \sum^{i-2}_{\beta=1} \tilde{\eta}_{\beta} \;  (i=3,..., n-1), M^f_n=\sum^{n-3}_{\beta=1} \tilde{\eta}_{\beta} +\eta_1+\eta_4}
\end{equation}
Finally, the mass of the antisymmetric tensor hypermultiplet is given by
\begin{equation}
\boxed{M_{as}= 2 \sum^{n-3}_{\beta=0} \tilde{\eta}_{\beta} +\eta_3 +\eta_4}
\end{equation}

\section*{Acknowledgements} 
This material is based upon work supported by the National Science Foundation under Grant Number PHY-0969020.

\appendix
\section{Appendices}\label{Appendix}
\subsection{Schur's Pfaffian Identity and Other Useful Identities}\label{SchurPfaffian}
The most elementary form of Schur's Pfaffian identity  is given as follows: Consider a $2k \times 2k$ matrix $A$ with entries $A_{ij}= \frac{x_i - x_j}{x_i + x_j}$ (such that $x_i \neq \pm x_j$ if $i \neq j$). Then the Pfaffian of the matrix A is
\begin{equation}
\Pf A = \prod _{1\leq i< j \leq 2k}\frac{x_i - x_j}{x_i + x_j}
\end{equation}
The above identity can be generalized in many different ways. One such generalization \cite{Okada:2004Pf} that will be useful for our investigation of mirror symmetry is
\begin{equation}
\Pf \left(\frac{x_i - x_j}{g(x_i, x_j)} \right)_{1\leq i,j \leq 2k} =  \prod _{1\leq i< j \leq 2k}\frac{x_i - x_j}{g(x_i,x_j)} \label{SPI}
\end{equation}
where $g(x,y)= 1+xy $.

In proving mirror duality for pairs of theories arising from a M-theory background involving a product of A-type ALEs (or equivalently, in the Hanany-Witten picture, theories for which the compact direction is $S^1$), one needs to use the following version of the Cauchy's determinant identity \cite{Kapustin:2010xq}:
\begin{equation}
\sum _{\rho} (-1)^{\rho} \frac{1}{\prod_{i}\cosh{(x_i -y_{\rho(i)})}}=\frac{\prod_{i <j} \sinh{(x_i-x_j)} \sinh{(y_i -y_j)}}{\prod_{i,j}\cosh{(x_i -y_j)}} \label{Cauchycosh}
\end{equation}
where $\rho$ is a permutation of the set $\{1,2,...,k-1,k\}$ and $i,j=1,2,3,...,k$.

A related version of the identity (proved in the next subsection) which plays a crucial role in proving mirror symmetry for dual theories associated to D-type ALEs in the M-theory picture ( resulting in the appearance of orbifold/orientifold 5-planes on the boundary of the compact direction wrapped by the D3 branes) is:
\begin{equation}
\sum _{\rho} (-1)^{\rho} \frac{1}{\prod_{i}\sinh{(x_i +y_{\rho(i)})}}=\frac{\prod_{i <j} \sinh{(x_i-x_j)} \sinh{(y_i -y_j)}}{\prod_{i,j}\sinh{(x_i +y_j)}} \label{Cauchysinh}
\end{equation}

In cases involving D-type ALEs, one encounters more complicated identities, which can be derived from certain generalizations of the Schur's Pfaffian identity. An important identity that we encounter in the case of dual theories coming from the M-theory background $\mathcal{M}= \mathbb{R}^{2,1} \times \mathbb{C}^2/ D_{n} \times \mathbb{C}^2/\mathbb{Z}_m$, we have,
\begin{equation}
\begin{split}
&\sum _{\rho} (-1)^{\rho} \frac{1}{\prod_{i} \cosh{(x_{\rho(i)}-m)}\cosh{(x_{\rho(k+i)}-m^{'})}\cosh{(x_{\rho(i)}+x_{\rho(k+i)} -M)}}  \\
=&\left(\frac{\kappa \;k! \; \sinh^k{(m-m^{'})}}{\prod_i \cosh{(x_i-m)} \cosh{(x_i-m^{'})} \cosh{(x_{k+i}-m)}\cosh{(x_{k+i}-m^{'})}}\right) \times \Pf\left[\frac{\sinh{(x_p - x_l)}}{\cosh{(x_p + x_l-M)}} \right] \\ 
=&\left(\frac{\kappa \; k!\; \sinh^k{(m-m^{'})}}{\prod_i \cosh{(x_i-m)} \cosh{(x_i-m^{'})} \cosh{(x_{k+i}-m)}\cosh{(x_{k+i}-m^{'})}}\right) \times \left(\prod_{p<l} \frac{\sinh{(x_p - x_l)}}{\cosh{(x_p + x_l-M)}}\right) \label{Id2}
\end{split}
\end{equation}
where $\rho$ now denotes the set of all permutations of the integers $\{1,2,....,k,k+1,...,2k-1,2k\}$ with $p,l =1,2,..,2k$ and $i,j = 1,2,...,k$. The constant $\kappa$ depends on $k$ : $\kappa=1$ for even $k=4m$ and odd $k=4m+1$ ($m=1,2,3,...$) and $\kappa=-1$ otherwise.

The last equality simply uses the Pfaffian identity given in equation \eqref{SPI}. We prove this identity in the next subsection.

\subsection{Proof of the Identities}\label{sec:proofs}
Let us consider the identity (\ref{Cauchysinh}) first. A simple generalization of the Cauchy determinant formula (see, for example, equation (15) of \cite{Okada:2004Pf}) is:
\begin{equation}
\mbox{det}\left(\frac{1}{1-u_i v_j}\right)= \frac{\prod_{i<j}(u_i-u_j)(v_i-v_j)}{\prod_{i,j}(1-u_i v_j)} \label{Cauchygen}
\end{equation}
Substituting $u_i=e^{2x_i}$,$v_i=e^{2y_i}$ and multiplying the numerator and the denominator by the factor $e^{-\sum_{i<j} (x_i +x_j)}e^{-\sum_{i<j}(y_i +y_j)} e^{-\sum_i(x_i+y_i)}$, we have,
\begin{equation}
\frac{\prod_{i<j}(u_i-u_j)(v_i-v_j)}{\prod_{i,j}(1-u_i v_j)}=\frac{\prod_{i <j} \sinh{(x_i-x_j)} \sinh{(y_i -y_j)}}{\prod_{i,j}\sinh{(x_i +y_j)}}\times \frac{(-1)^k e^{-\sum_i(x_i +y_i)}}{2^k}
\end{equation}
On the other hand, the determinant can be rewritten as,
\begin{equation}
\mbox{det}\left(\frac{1}{1-u_i v_j}\right)=\frac{(-1)^k e^{-\sum_i(x_i +y_i)}}{2^k} \sum_{\rho} (-1)^{\rho} \frac{1}{\prod_i \sinh{(x_i +y_{\rho(i)})}}
\end{equation}
On equating the two sides in (\ref{Cauchygen}), one arrives at the desired identity.\\ 

Now, we prove the identity (\ref{Id2}), which , as indicated above, is related to a particular form of Schur's Pfaffian identity. The LHS of identity (\ref{Id2}) can be re-written in the following manner,
\begin{equation}
\begin{split}
&\mbox{LHS}=\sum _{\rho} (-1)^{\rho} \frac{1}{\prod_{i} \cosh{(x_{\rho(i)}-m)}\cosh{(x_{\rho(k+i)}-m^{'})}\cosh{(x_{\rho(i)}+x_{\rho(k+i)} -M)}}  \\
&=\sum _{{\rho}_1,{\rho}_2} (-1)^{{\rho}_1} \frac{1}{\prod_i\cosh{(x_{\rho_1(i)}+x_{\rho_1(k+i)} -M)}}\left[\frac{(-1)^{\rho_2}}{\prod_i\cosh{(x_{\rho_2\circ \rho_1(i)}-m)}\cosh{(x_{\rho_2\circ \rho_1(k+i)}-m'})}\right]
\end{split}
\end{equation}
where $\rho_1$ denotes a subset of the permutations which produce inequivalent variations of the term
$ \frac{1}{\prod_i\cosh{(x_{i}+x_{k+i} -M)}}$ while $\rho_2$ is the set of all possible permutations $x_{\rho_1(i)} \leftrightarrow x_{\rho_1(k+i)}$ for a given $\rho_1$. In other words, $\rho_1$ corresponds 
to the set of distinct ways in which one can form $k$ pairs from the set of $2k$ integers while $\rho_2$, for a given choice of $k$ pairs, corresponds to the set of all possible permutations involving exchanges within each of those $k$ pairs. The number of ways  in which $k$ pairs can be drawn from the set of $2k$ integers is $\prod^{k-1}_{r=0}{2k-2r \choose 2}= \frac{(2k)!}{2^k} $. For every given choice of such $k$ pairs, there exists $2^k$ permutations corresponding to exchange of objects within pairs. Therefore, the total number of permutations is $\prod^{k-1}_{r=0}{2k-2r \choose 2} 2^k= 2k !$, as expected for a set of $2k$ integers.

Now, since the expression in square brackets is simply the anti-symmetrization of the term $\frac{1}{\prod_i\cosh{(x_{\rho_1(i)}-m)}\cosh{(x_{ \rho_1(k+i)}-m'})}$ under all possible permutations of the form $x_{\rho_1(i)} \leftrightarrow x_{\rho_1(k+i)}$, we have,
\begin{equation}
\begin{split}
&\sum _{{\rho}_2}\frac{(-1)^{\rho_2}}{\prod_i\cosh{(x_{\rho_2\circ \rho_1(i)}-m)}\cosh{(x_{\rho_2\circ \rho_1(k+i)}-m'})}\\
&=\frac{\sinh^k{(m-m')}\prod_i\sinh{(x_{\rho_1(i)}-x_{\rho_1(k+i)})}}{\prod_i \cosh{(x_{i}-m)}\cosh{(x_{k+i}-m})\cosh{(x_{i}-m')}\cosh{(x_{k+i}-m'})}
\end{split}
\end{equation}
where we have used Cauchy's determinant formula for each pair of variables $(x_{\rho_1(i)},x_{\rho_1(k+i)})$. The original expression can now be written as,
\begin{equation}
\begin{split}
&\mbox{LHS}=\sum _{\rho} (-1)^{\rho} \frac{1}{\prod_{i} \cosh{(x_{\rho(i)}-m)}\cosh{(x_{\rho(k+i)}-m^{'})}\cosh{(x_{\rho(i)}+x_{\rho(k+i)} -M)}}\\
&=\frac{\sinh^k{(m-m')}}{\prod_i \cosh{(x_{i}-m)}\cosh{(x_{k+i}-m})\cosh{(x_{i}-m')}\cosh{(x_{k+i}-m'})}\\ 
&\times \left(\sum _{{\rho}_1} (-1)^{{\rho}_1} \frac{\prod_i\sinh{(x_{\rho_1(i)}-x_{\rho_1(k+i)})}}{\prod_i\cosh{(x_{\rho_1(i)}+x_{\rho_1(k+i)} -M)}}\right)\\
\end{split}
\end{equation}
Note that
$$\sum _{{\rho}_1} (-1)^{{\rho}_1} \frac{\prod_i\sinh{(x_{\rho_1(i)}-x_{\rho_1(k+i)})}}{\prod_i\cosh{(x_{\rho_1(i)}+x_{\rho_1(k+i)} -M)}}=\kappa \;k!\; \Pf\left[\frac{\sinh{(x_p - x_l)}}{\cosh{(x_p + x_l-M)}}\right]$$
with $p,l=1,2,...,2k-1,2k$, where $\kappa=1$ for even $k=4m$ and odd $k=4m+1$ ($m=1,2,3,...$) and $\kappa=-1$ otherwise.

The above equation, therefore, implies,
\begin{equation}
\begin{split}
\mbox{LHS}&=\frac{\kappa \;k! \; \sinh^k{(m-m')}}{\prod_i \cosh{(x_{i}-m)}\cosh{(x_{k+i}-m})\cosh{(x_{i}-m')}\cosh{(x_{k+i}-m'})}\Pf\left[\frac{\sinh{(x_p - x_l)}}{\cosh{(x_p + x_l-M)}}\right]\\
&=\frac{\kappa\;k! \; \sinh^k{(m-m')}}{\prod_i \cosh{(x_{i}-m)}\cosh{(x_{k+i}-m})\cosh{(x_{i}-m')}\cosh{(x_{k+i}-m'})}\left(\prod_{p<l} \frac{\sinh{(x_p - x_l)}}{\cosh{(x_p + x_l-M)}}\right)
\end{split}
\end{equation}
where, for the final equality, we have used the generalized Pfaffian identity given in equation \eqref{SPI}.

\subsection{Useful Fourier Transforms}\label{sec:FTs}
There are two useful fourier transforms used in manipulating the partition functions. One of them is the transform of the hyperbolic secant:
\begin{equation}
\int \frac{e^{2\pi i x z}}{\cosh{\pi z}} dx = \frac{1}{\cosh{\pi x}}
\end{equation}
The other transform which is extremely useful for manipulating boundary contributions in theories with orbifold 5-planes is that hyperbolic cosec:
\begin{equation}
\int \frac{e^{2\pi i x z}}{\sinh{\pi z}} dx = i\tanh{\pi x} \label{FTcosec}
\end{equation}

\subsection{Integrating Boundary Nodes in a Quiver}\label{sec:boundaryNodes}
Consider a $\mathcal{N}=4$ $U(N_c)$ gauge theory in three dimensions with $N_f (>N_c)$ fundamental hypers. The partition function on $S^3$ for this theory is
\begin{equation}
Z^{N_c}_{N_f}= \binom{N_f}{N_c}\left(\frac{i^{(N_f-1)}2 e^{\pi \eta}}{1+ (-1)^{(N_f-1)} e^{2\pi \eta}}\right)^{N_c}\left(\prod^{N_c}_{j=1} e^{2\pi i \eta m_j}\right) \left(\prod^{N_c}_{j=1} \prod^{N_f}_{k=N_c +1} \sinh{\pi(m_j -m_k)} \right)^{-1} \arrowvert_{\{m_j\}}
\end{equation}
where $\{m_p\}$ ($p=1,2,...,N_f$) are the masses and $\eta$ is the FI parameter. The symbol $\arrowvert_{\{m_j\}}$  denotes symmetrization w.r.t. the masses $m_j$ (see \cite{Kapustin:2010mh} and Appendix \ref{Appendix}), with $j=1,2,...,N_c$.

In a given quiver, consider a boundary node corresponding to a gauge group $U(k)$ with a fundamental hyper that also transforms in the fundamental of a gauge group of rank $2k$. Let $\{\sigma^i\}=diag(\sigma^1,\sigma^2,...., \sigma^k)$ denote an element in the Cartan of $U(k)$ and $\{\tilde{\sigma}^p\}=diag(\tilde{\sigma}^1,\tilde{\sigma}^2,...., \tilde{\sigma}^{2k})$ denote an element in the Cartan of $U(2k)$. Let $Z[\sigma]$ be the $\sigma$-dependent part of the partition function of the quiver on $S^3$. Therefore, from the above equation, we have
\begin{eqnarray}
Z[\sigma]&=&\int d^k \sigma \prod_i e^{2\pi i \eta \sigma^i} \frac{\prod_{i<j}\sinh^2{\pi(\sigma^i -\sigma^j)}}{\prod_{i,p}\cosh{\pi (\sigma^i -\tilde{\sigma}^p +m)} } \nonumber \\
&=&\binom{2k}{k} \left(\frac{i^{(2k-1)}e^{\pi \eta}}{1+ (-1)^{(2k-1)} e^{2\pi \eta}}\right)^k \left(\prod^{k}_{i=1} e^{2\pi i \eta (\tilde{\sigma}^i -m)}\right)  \left(\prod^{k}_{i=1} \prod^{2k}_{j=k+1} 2\sinh{\pi(\tilde{\sigma}^i -\tilde{\sigma}^{j})} \right)^{-1} \arrowvert_{\{\tilde{\sigma}^i\}} \nonumber \\
\end{eqnarray}
``Integrating out a node" in a quiver will therefore be equivalent to inserting a function like above in the integrand for the $S^3$ partition function. If the rest of the integrand is symmetric in $\{\tilde{\sigma}^p\}$, then the integral automatically picks out the symmetric part of the function and as such the symmetrization operation is redundant.

\bibliographystyle{utphys}
\bibliography{mirror2}

\end{document}